\documentclass[12pt]{article}
\usepackage{amsmath}
\usepackage{amsfonts}
\usepackage{amssymb}
\usepackage{bbm}
\newcommand{\be}{\begin{equation}}
\newcommand{\ee}{\end{equation}}
\newcommand{\ba}{\begin{eqnarray}}
\newcommand{\ea}{\end{eqnarray}}
\newcommand{\nn}{\nonumber}
\newcommand{\kl}{\langle}
\newcommand{\kr}{\rangle}

\newcommand{\Tr}{{\rm{Tr}}}

\newcommand{\ha}{\hat{A}}

\newcommand{\ps}{\bar{\psi}}
\newcommand{\hp}{\hat{\psi}}
\renewcommand{\theequation}{\arabic{section}.\arabic{equation}}

\renewcommand{\arraystretch}{2.5}

\newcommand{\grgl}{\:\hbox to -0.2pt{\lower2.5pt\hbox{$\sim$}\hss}
           {\raise3pt\hbox{$>$}}\:}
\newcommand{\klgl}{\:\hbox to -0.2pt{\lower2.5pt\hbox{$\sim$}\hss}
           {\raise3pt\hbox{$<$}}\:}
\newcommand{\qslash}{\mathbin{q\mkern-10mu\big/}}
\renewcommand{\arraystretch}{1.0}
\begin{document}
\begin{titlepage}

\quad\\
\vspace{1.8cm}
\begin{center}
{\bf\LARGE Spinors in euclidean field theory,}\\
\bigskip
{\bf\LARGE complex structures and discrete symmetries}\\
\vspace{1cm}
C. Wetterich\\
\bigskip
Institut  f\"ur Theoretische Physik\\
Universit\"at Heidelberg\\
Philosophenweg 16, D-69120 Heidelberg\\
\vspace{3cm}
\date{}
\begin{abstract}
We discuss fermions for arbitrary dimensions and signature of the metric, with special emphasis on euclidean space. Generalized Majorana spinors are defined for $d=2,3,4,8,9$ mod $8$, independently of the signature.  These objects permit a consistent analytic continuation of Majorana spinors in Min\-kowski space to euclidean signature.  Compatibility of charge conjugation with complex conjugation requires for euclidean signature a new complex structure which involves a reflection in euclidean time. The possible complex structures for Minkowski and euclidean signature can be understood in terms of a modulo two periodicity in the signature. The concepts of a real action and hermitean observables depend on the choice of the complex structure. For a real action the expectation values of all hermitean multi-fermion observables are real. This holds for arbitrary signature, including euclidean space. In particular, a chemical potential is compatible with a real action for the euclidean theory. We also discuss the discrete symmetries of parity, time reversal and charge conjugation for arbitrary dimension and signature. 
\end{abstract}
\parbox[t]{\textwidth}{}
\end{center}\end{titlepage}

\newpage
\section{Introduction}
\label{Introduction}
\setcounter{equation}{0}
Analytic continuation from Minkowski space to euclidean space is a central tool in quantum field theory. Many non-perturbative computations, as lattice gauge theory for quantum chromodynamics, are directly performed in euclidean space-time. If the euclidean functional integral obeys Osterwalder-Schrader positivity \cite{OS}, and if a continuous analytic continuation of the action from euclidean space to Minkowski space can be realized, the scattering amplitudes of the physical world in Minkowski space can be obtained from the scattering amplitudes computed in euclidean space. More precisely, the $n$-point functions in momentum space can be computed by analytic continuation of the momenta. For all momenta within the range where the $n$-point functions remain analytic, these analytically continued values coincide with the physical $n$-point functions as computed from the functional integral in Minkowski space \cite{ZJ}. Euclidean and Minkowski space differ by the signature of the metric. Analytic continuation from euclidean to Minkowski signature affects the properties of the generalized Lorentz transformations. The standard Lorentz transformations $SO(1,d-1)$ for Minkowski signature become the (euclidean) rotations $SO(d)$ for euclidean signature. 

For fermions, the mapping between Minkowski space and euclidean space is, however, not without problems. The basic issue is a modulo two periodicity in signature for the reality properties of representations of the generalized Lorentz group. While the two fundamental spinor representations $2_L,_R$ of the Lorentz group in four dimensions are complex, the fundamental spinors $2_1,2_2$ are (pseudo)-real for the orthogonal group $SO(4)$. Complex conjugation maps the spinor representations of $SO(1,3)$ into each other, $2_L\leftrightarrow 2_R$, while the $SO(4)$-spinors $2_1$ and $2_2$ are mapped into themselves.

This simple property has important consequences if one tries to define charge conjugation and Majorana spinors in terms of complex conjugation. Often, Majorana spinors are associated to real $2^{\left[\frac d2\right]}$-component spinor representations of the generalized $d$-dimensional Lorentz group (with $\left[\frac d2\right]=(d-1)/2$ for odd). For Minkowski signature, they exist for $d=2,3,4,8,9$ mod $8$ \cite{W1}. (In four dimensions, a four-component Majorana  spinor can be associated to a two-component complex Weyl spinor - the four real components can be obtained as linear combinations of the real and imaginary parts of the Weyl spinors.) A similar prescription for euclidean signature would allow Majorana spinors for $d=2,6,7,8,9$  \cite{W1}. No euclidean Majorana spinor would exist for $d=4$ according to this definition.

It is therefore sometimes advocated that the number of spinor degrees has to be doubled if four-dimensional Majorana spinors are described with euclidean signature. From the point of view of analytic continuation a different number of spinor components for euclidean and Minkowski signature would obviously not be a very satisfactory situation. Continuity is not compatible with a sudden jump of the number of degrees of freedom. In this paper we advocate that the correct implementation of Majorana spinors in quantum field theory should not be based on the reality properties of the representations of the generalized Lorentz group. 

Majorana spinors are eigenstates of a suitable charge conjugation operator. We present a definition of charge conjugation for which physical Majorana spinors can be implemented whenever the symmetric product of two identical $2^{\left[\frac d2\right]}$-component spinors contains a vector with respect to the generalized Lorentz transformations. This is the case for $d=2,3,4,8,9$ mod $8$, independently of the signature. Imposing the Majorana constraint reduces the number of independent degrees of freedom by a factor $1/2$. This is the same for Minkowski and euclidean signature. To every model in Minkowski space corresponds then a euclidean model with the same number of spinor degrees of freedom. No ``doubling of degrees of freedom'' is necessary for Majorana spinors in four-dimensional euclidean space. What has to be questioned, however, is the relation of the discrete symmetries like charge conjugation, to the notion of complex conjugation. In this paper we present a detailed discussion of the complex structure that is compatible with analytic continuation.

We discuss quantum field theories for fermions in the context of Grassmann functional integrals \cite{ZJ}. Dirac spinors are based on a Grassmann algebra that is constructed from two sets of independent Grassmann variables $\psi_\gamma(x)$ and $\bar\psi_\gamma(x)$, where the number of components $\gamma$ equals $2^{\left[\frac d2\right]}$. This basic setting does not need the notion of a complex conjugation that relates $\psi$ and $\bar\psi$. For Majorana spinors the Grassmann variables $\bar\psi$ are no longer independent - they can be expressed as linear combinations of the Grassmann variables $\psi$. Correspondingly, the functional integral involves only an integration over $\psi$, in contrast to the integration over $\psi$ and $\bar\psi$ for Dirac spinors. The number of independent Grassmann variables for Majorana spinors is only half the number for Dirac spinors. Dirac spinors can be interpreted as two independent Majorana spinors. 

For Dirac spinors both $\psi$ and $\bar\psi$ belong to the same representation of the generalized Lorentz group (which may be reducible). From the point of view of a consistent implementation of generalized Lorentz transformations no obstacle prevents the identification of $\bar\psi_\gamma(x)$ with a suitable linear combination of variables $\psi_\gamma(x)$. The restrictions arise from the requirement that for physical Majorana spinors it must be possible to construct a Lorentz invariant kinetic term from a single Majorana spinor. This requires that the symmetric product of two identical Majorana spinors contains a vector of the generalized Lorentz transformations. This is not the case for $d=5,6,7$ mod $8$. 

For all other dimensions, $d=2,3,4,8,9$ mod $8$, we will construct generalized Majorana spinors which make euclidean and Minkowski signature compatible. They are similar to the definition of Majorana spinors by Nicolai \cite{NIC} (see also ref. \cite{OSA}) and generalize this concept. Our setting also leads to a generalized concept of ``reality'' of the action, closely related to the old observation that hermiticity for the action in Minkowski space corresponds to Osterwalder-Schrader positivity in euclidean space \cite{OS}. The notion of ``reality'' for spinors is no longer linked to the representation theory of the generalized Lorentz group, but rather to eigenstates of a generalized complex conjugation. Similar remarks actually apply to the chiral antisymmetric tensor representation of rank $d/2$ (for even dimensions), where the reality properties also jump with a modulo two periodicity in the signature \cite{CT}. 

We present a unified picture of the discrete symmetries parity, time reversal and charge conjugation for arbitrary dimension and signature. The basic formulation involves the independent Grassmann variables $\psi$ and $\bar\psi$ for Dirac spinors and is independent of the complex structure. If we express charge conjugation in terms of complex conjugation it reflects the property that complex conjugation in euclidean space involves a reflection in euclidean time and the zero-component of momentum $q_0$. (With $t$ the Minkowski time and $\tau=it$ the euclidean time, the complex conjugation $(it)^*=-it=-\tau$ indeed results in $\tau\to-\tau$.) As a byproduct of our investigation, we develop a consistent notation for spinors and discrete symmetries in arbitrary $d$ and for arbitrary signature. 

Our definition of the action and the symmetries for euclidean spinors is such that a continuous analytic continuation from euclidean to Minkowski signature is always possible. The Greens functions in Minkowski space can then be computed from the analytically continued euclidean Greens functions. Besides analytic continuation, euclidean spinors are also of interest in thermal field theories. In this case the euclidean time is on a torus with circumference $1/T$, reflecting the temperature $T$. Our formulation of euclidean spinors implements this concept for arbitrary dimensions. The properties of the spinor representations of the two-dimensional rotation group (two euclidean dimensions) are particularly interesting in this respect, since symmetries could forbid mass terms in this case. Finally, euclidean spinors in higher dimensions are crucial if analytic continuation is used as a tool for unified theories. This applies, in particular, to proposals that the difference between space and time could originate from spontaneous symmetry breaking \cite{CWPO}.

In sect. 2 we start with basic notions of spinor fields in the functional integral approach to quantum field theory. We formulate analytic continuation in sect. 3 in terms of an analytic continuation of the vielbein for fixed Grassmann variables and coordinates.(This is briefly compared to other versions of analytic continuation \cite{MEH1}, \cite{MEH}.) We proceed in sect. 4 to a formulation of Majorana spinors for euclidean signature which is compatible with analytic continuation. We discuss the case of four dimensions rather explicitly in order to give a first account of the concepts involved. 

We then proceed to general signature and general $d$. The discussion of sects. \ref{Complex conjugation for Minkowski signature}-\ref{Hermiteanspinor} concerns the complex structure. The usual notion of complex conjugation in Minkowski space is discussed in sect. \ref{Complex conjugation for Minkowski signature}. We interpret this as a mapping $\theta_M$ in the space of spinor components. In sect. \ref{Modulo two periodicity in the signature} we introduce the analogous mapping $\theta$ in euclidean space. It includes a reflection of the time coordinate. We discuss an important modulo two periodicity in the signature, by which the role of $\theta$ and $\theta_M$ is switched as the number of time-like dimensions increases by one unit. In particular, the map $\theta$ for euclidean signature induces the same type of map between the Weyl spinors as $\theta_M$ does for Minkowski signature.

In sect. \ref{Complexconjugation} we describe a generalized complex conjugation for spinors with euclidean signature. It is based on the map $\theta$ and therefore involves an additional reflection of the zero-component of the momentum or a reflection of euclidean time. This is used in sect. \ref{Realfermionicactions} in order to define a real action as a Grassmann element that is invariant under the transformation $\theta$ or $\theta_M$. (The action of $\theta$ and $\theta_M$ in the Grassmann algebra includes a total reordering of all Grassmann variables (transposition).) The functional integral with a real action yields a real result. We generalize this discussion to the possible presence of bosonic fields in addition to the fermionic spinors. In sect. \ref{Hermiteanspinor} we discuss hermitean spinor bilinears which can be used in order to construct a real action and for providing observables that have a real expectation value. We show that a chemical potential is compatible with a real action for the euclidean theory. The expectation values of ``real observables'' can be computed with real weight factors in presence of a nonvanishing chemical potential. 

In sects.  \ref{Generalizedcharge}, \ref{Generalized Majorana} we turn to the proper definition of physical Majorana and Majorana-Weyl spinors for arbitrary signature, and in particular for euclidean signature. Physical Majorana spinors do not coincide with the real spinor representations of the rotation group $SO(d)$. We define in sect. \ref{Generalizedcharge} a generalized charge conjugation which maps the spinor $\psi$ to its conjugate spinor $\bar\psi$. If this map defines an involution ${\cal C}_W$ we can define in sect. \ref{Generalized Majorana} the generalized Majorana spinors as the eigenstates of ${\cal C}_W$ with eigenvalues $+1$ or $-1$. For Majorana spinors the Grassmann variables $\psi$ and $\bar\psi$ in the functional integral are no longer independent, $\bar\psi$ can be expressed in terms of $\psi$ \cite{NIC}. The generalized Majorana constraint reduces the number of degrees of freedom by a factor two, as expected for Majorana spinors. For Minkowski signature the generalized Majorana spinors coincide with the algebraic notion of Majorana spinors. We define physical Majorana spinors by the constraint that an invariant kinetic term in the action should be allowed. Physical Majorana spinors exist for $d=2,3,4,8,9$ mod $8$, and physical Majorana-Weyl spinors are allowed for $d=2$ mod $8$. The existence of physical Majorana spinors therefore depends only on the dimension $d$, and not on the signature $s$. In particular, analytic continuation between Minkowski and euclidean signature is always possible for physical Majorana and Majorana-Weyl spinors. We also address in sect. \ref{Generalized Majorana} the compatibility of the Majorana constraint with the complex structure. 

In sect. \ref{Parityandtimereversal} we extend our discussion to the appropriate definitions of parity and time reversal which are compatible with analytic continuation. These symmetries can be defined consistent with the Majorana constraint and with analytic continuation. In sect. \ref{Continuousinternalsymmetries} we address possible continuous internal symmetries acting on the spinors. For $N$ species of Dirac spinors the kinetic term is invariant under the group $SL(2 N,{\mathbbm C})$ of regular complex $2N\times 2N$ matrices. The chiral transformations $U(N)\times U(N)$ are a subgroup of this group. While the chiral transformations are compatible with the complex structure, this does not hold for the generalized $SL(2N,{\mathbbm C})$ transformations. We present our conclusions in sect. \ref{Conclusions}. 

Detailed conventions for four dimensions are collected in appendix A and a short general discussion of possible complex structures for a real or complex Grassmann algebra is given in appendix B. The appendix C summarizes various properties of the Clifford algebra which are needed in this paper.

\section{Spinor degrees of freedom}
\label{Spinor degrees of freedom}
\setcounter{equation}{0}

Let us consider an arbitrary number of dimensions $d$
with arbitrary
signature $s$, where the diagonal metric $\eta_{mn}$ has $d-s$ eigenvalues $+1$
and $s$ eigenvalues $-1$. Later we will concentrate on the
euclidean case $s=0$ and compare it with a Minkowski signature
$s=1$ (or $s=d-1$).
We start with Dirac spinors
described by associated elements of a Grassmann algebra
$\psi$ and $\bar\psi$. This Grassmann algebra is generated by two sets of independent Grassmann variables $\psi_u$ and $\bar\psi_v$ which fulfill the usual anticommutation relations
\ba\label{2.A}
 \{\psi_u,\psi_v\}=\{\psi_u,\bar
\psi_v\}=\{\bar\psi_u,\bar\psi_v\}=0.
\ea
We may choose $\psi_u \equiv\psi^a_\gamma(x)$ or $\psi_u\equiv\psi^a_\gamma(q)$ in a coordinate or momentum representation. Here $\gamma$ are the ``Lorentz-indices'' on which the generalized Lorentz group $SO(s,d-s)$ acts, while the index $a$ denotes further possible internal degrees of freedom or different species of Dirac spinors. 

Integration and differentiation with
Grassmann variables obey the usual rules
\ba\label{2.1}
&&\int d\psi_ug=\frac{\partial}{\partial\psi_u}g\quad,\quad
\int d\psi_u\psi_v=\delta_{uv},\quad \int d\psi_u\bar\psi_v=
\int d\psi_u=0\nonumber\\
&&\{\frac{\partial}{\partial\psi_u},\frac{\partial}{\partial\psi_v}\}
=0,\quad \{d\psi_u,d\psi_v\}=0,
\ea
such that
\be\label{2.1a}
\int\prod_{u'}(d\psi_{u'}d\bar\psi_{u'})\exp(\bar\psi_uA_{uv}
\psi_v)=\det\ A.
\ee
Elements of the Grassmann algebra are sums of products of Grassmann variables with complex coefficients. Spinors are elements of the Grassmann algebra which are linear in $\psi_u$ or $\bar\psi_v$. In particular, for a Dirac spinor one has $\psi=\sum_u a_u\psi_u,\bar\psi=\sum_vb_v\bar\psi_v$, with $\psi^2=0,\bar\psi^2=0,\{\psi,\bar\psi\}=0$. For example, a spinor wave function in position space can be expressed in terms of the Grassmann variables $\psi^a_\gamma(q)$ by 
\be\label{2.B}
\psi^a_\gamma(x)=\sum_q\exp (iq_\mu x^\mu)\psi^a_\gamma(q).
\ee

We can consider $\psi^a_\gamma(x)$ as new Grassmann variables obeying eqs. \eqref{2.A}, \eqref{2.1} such that eq. \eqref{2.B} amounts to a change of basis for the Grassmann algebra. Indeed, every regular linear transformation $A$ defines a new set of Grassmann variables,
\be\label{2.C}
\varphi_u=A_{uv}\psi_v~,~
\frac{d}{d\varphi_w}=A^{-1}_{vw}\frac{d}{d\psi_v}~,~
\frac{d}{d\varphi_w}\varphi_u=\delta_{wu}.
\ee
Every complete set of spinors can be taken as a complete set of Grassmann variables. The products of Grassmann variables from a complete set define a basis for the Grassmann algebra. We observe that
the Grassmann algebra $G$ is defined over the complex
numbers, i.e. $\lambda g$ is defined for all $g\in G$ and
$\lambda\in {\cal C}$, but we do not assume a priori the existence
of a complex conjugation within $G$, i.e. $g^*$ (or $\psi^*)$ is not
necessarily defined. 

We define the ``functional measure''
\ba\label{2.2A}
\int {\cal D}\psi {\cal D}\bar{\psi}=\int \prod\limits_{u'}(d\psi_{u'}d\bar{\psi}_{u'})
\ea
and the partition function
\be\label{2.2AA}
Z=\int{\cal D}\psi{\cal D}\bar{\psi}\exp \big(-S_E[\psi,\bar{\psi}]\big).
\ee
Here the euclidean action $S_E$ is a polynomial of an even number of Grassman variables.
(For $s=1, S_E$ will be related to the ``Minkowski action'' $S_M$ by $S_E=-iS_M$.) The action
may contain appropriate source terms such that Z becomes the generating functional for
Greens functions in the standard way.

A Dirac spinor has $2^{[\frac{d}{2}]}$ components labeled by the spinor index 
$\gamma$ with $[\frac{d}{2}]=\frac{d}{2}$ for
$d$ even and $[\frac{d}{2}]=\frac{d-1}{2}$ for $d$ odd. The spinor  $\psi_\gamma$
transforms under generalized infinitesimal Lorentz transformations
as\footnote{Here $\epsilon_{mn}=-\epsilon_{nm}=\epsilon^*_{mn}$
and the index $m$
runs from 0 to $d-1$ in order to be close to standard
Minkowski space notation. For an euclidean signature $s=0$
the Lorentz transformations correspond to standard $SO(d)$
rotations. We do not write explicitly the Lorentz transformation of coordinates or momenta.}
\be\label{2.2}
\delta\psi_\gamma=-\frac{1}{2}\epsilon_{mn}(\Sigma^{mn})_\gamma^{\
\delta}\psi_\delta.
\ee
Here the generators $\Sigma^{mn}$ of the group $SO(s,d-s)$ are related to the Clifford algebra by
\be\label{2.3}
\Sigma^{mn}=-\frac{1}{4}[\gamma^m,\gamma^n]\quad,\quad
\{\gamma^m,\gamma^n\}=2\eta^{mn},
\ee
with $(\gamma^m)^\dagger=\gamma^m$ for $\eta^{mm}=1$ and
$(\gamma^m)^\dagger=-\gamma^m$ for $\eta^{mm}=-1$. (Hermitean generators can be obtained from $\Sigma^{mn}$ by suitable multiplication of factors $i$.) We postulate that an infinitesimal Lorentz
transformation of $\bar\psi$ reads
\be\label{2.6}
\delta\bar\psi=\frac{1}{2}\epsilon_{mn}
\bar\psi\Sigma^{mn}.\ee
Then the bilinears $\bar\psi\psi,\bar\psi\gamma^m
\psi,\bar\psi\Sigma^{mn}
\psi$ etc. transform as Lorentz scalars, vectors, second rank
antisymmetric tensors and so on. A Lorentz-invariant action can be constructed from these
building blocks.

In even dimensions Dirac spinors are reducible representations. They may be decomposed
into Weyl spinors by use of the $d$-dimensional generalization of the $\gamma^5$-matrix
$\bar{\gamma}$, which is defined as
\be\label{2.4}
\bar\gamma=\eta\gamma^0\gamma^1...\gamma^{d-1}=-\eta\gamma^1\cdots\gamma^{d-1}\gamma^0.
\ee
We require $\bar{\gamma}^2=1$ such that $(1\pm\bar{\gamma})/2$ are projectors. This
implies for the phase $\eta$
\be\label{2.5A}
\eta^2=(-1)^{\frac{d}{2}-s}.
\ee
With this phase $\bar{\gamma}$ is hermitean. The matrix $\bar{\gamma}$ anticommutes with
all Dirac matrices $\gamma^m$ and therefore indeed commutes with $\Sigma^{mn}$. We
summarize the properties of $\bar{\gamma}$ by
\be\label{2.5B}
\bar{\gamma}^2=1,~\bar{\gamma}^\dagger=\bar{\gamma},~\{\gamma^m,\bar{\gamma}\}=0,~
[\Sigma^{mn},\bar{\gamma}]=0.
\ee
Finally, we fix the phase as
\be\label{2.5C}
\eta=(-i)^{\frac{d}{2}-s}.
\ee
Weyl spinors \footnote{In analogy to the four dimensional notation in Minkowski space one may identify $\psi_+=\psi_L,~\psi_-=\psi_R$.} are defined as
$\psi_{\pm}=\frac{1}{2}(1\pm\bar{\gamma})\psi$.

The spinor kinetic term reads
\be\label{2.9}
S_{kin}=\int d^dx\bar\psi(i\gamma^\mu\partial_\mu)\psi.
\ee
For flat space one has $\gamma^\mu=\delta^\mu_m\gamma^m$, whereas 
for more general geometries the vielbein $e_m{^\mu}$ replaces
$\delta^\mu_m$. The kinetic term is invariant under both Lorentz and internal $U(N)$
transformations. (Gauge invariance and general coordinate
invariance can be implemented as usual by employing
covariant derivatives instead of $\partial_\mu$.)
We emphasize that the kinetic term is given
by eq. (\ref{2.9}) for arbitrary signature and arbitrary dimension.
This fixes\footnote{Different conventions for the kinetic
term can be related to ours by multiplication of $\bar\psi$ with a
phase or with $\bar{\gamma}$.}
the relative phase convention between $\psi$ and
$\bar \psi$.

In even dimensions a Dirac spinor is composed from two Weyl
spinors with opposite ``helicity''
\be\label{2.10}
\psi_\pm=\frac{1}{2}(1\pm\bar\gamma)\psi\quad,\quad
\bar\psi_\pm=\frac{1}{2}\bar\psi(1\mp\bar\gamma).\ee
We have chosen here conventions for $\bar\psi_\pm$ such that
$\bar\psi_+\psi_+=0,\bar\psi\gamma^m\psi=
\bar\psi_+\gamma
^m\psi_++\bar\psi_-\gamma^m\psi_-$ as one is used
to from Minkowski space in four dimensions\footnote{Note
that $\bar\psi_+$ denotes here the ``plus component of
$\bar\psi$'' rather than the  Lorentz
representation which is the complex conjugate
to the Weyl spinor $\psi_+$. The latter would
be $\overline{(\psi_+)}=\frac{1}{2}\bar\psi(1+(-1)^s\bar\gamma)$
\cite{W1}. In this respect our notations differ from ref. \cite{W1}.}. The kinetic term (\ref{2.9}) decomposes then
into independent kinetic terms for the Weyl spinors
$\psi_+$ and $\psi_-$
\be\label{2.11}
S_{kin}=\int d^dx\left\{\bar\psi_+(i\gamma^\mu
\partial_\mu)\psi_++\bar\psi_-(i\gamma^\mu\partial_\mu)
\psi_-\right\}.\ee
For arbitrary signature it is invariant under separate
``chiral rotations'' of $\psi_+$ and $\psi_-$.
Such a chiral rotation acting only on $\psi_+$
\be\label{2.12}
\psi^i_+\longrightarrow U^i_{\ j}\psi^j_+\quad,\quad
\psi^i_-\longrightarrow\psi^i_-\ee
must transform $\bar\psi_\pm$ according to
\be\label{2.13}
\bar\psi^i_+\longrightarrow\bar\psi_+^{\ j}(U^\dagger)_{j}
^{\ i}\quad,\quad
\bar\psi^i_-\longrightarrow\bar\psi^i_-.
\ee
The Weyl spinors $\bar\psi_+$ and $\psi_+$ correspond to inequivalent spinor representations of the Lorentz  group in $d=4$ mod $4$, and to equivalent ones in  $d=2$ mod $4$. 

\section{Analytic continuation}
\setcounter{equation}{0}
\label{Analyticcontinuation}

The Minkowski signature $s=1$ and euclidean signature $s=0$ can be connected by analytic
continuation. Rather than continuing the time coordinate we will use here a formulation
with a vielbein. The analytic continuation multiplies the $0-m$-components of the vielbein
with a phase. (This is, of course, equivalent to the usual continuation of time.) One may compute Greens functions in the background of a Vielbein whose value can be analytically continued from a euclidean to a Minkowski signature of the metric. Since the vielbein enters directly the definition of squared momenta this version of analytic continuation is equivalent to the usual analytic continuation of the momenta. The physical Greens functions in Minkowski space can be computed in this way by analytic continuation of the euclidean Greens functions. 

Let us start with euclidean signature with a free massless Dirac spinor
\begin{equation}\label{X3.1}
S_E=S_{kin}=i\int d^dxe\bar{\psi}\gamma^me_m{^\mu}\partial_\mu\psi
\end{equation}
with vielbein $e_\mu{^m}=\delta^m_\mu$, inverse vielbein $e_m{^\mu}$ obeying
$e_m{^\mu}e_\mu{^n}=\delta^n_m$, and $e=\det(e_\mu{^m})$. We may now consider
$e_\mu{^m}$ as a free variable on which the partition function depends.
In particular, we consider the choice for the vielbein
\begin{equation}\label{X3.2}
e_0{^m}=e^{i\varphi}\delta^m_0~,~e_k{^m}=\delta^m_k,~k=1\dots d-1.
\end{equation}
Correspondingly, one has $e=e^{i\varphi}$ and the inverse vielbein obeys
$e_m{^0}=e^{-i\varphi}\delta^0_m,~e_m{^k}=\delta^k_m$. We do neither transform the coordinates nor the Grassmann variables $\psi,\bar{\psi}$. As usual in general relativity we define
the matrices
\begin{equation}\label{X3.3}
\gamma^\mu=\gamma^m e_m{^\mu},~\{\gamma^\mu,\gamma^\nu\}=e_m{^\mu}e_n{^\nu}
\delta^{mn}=g^{\mu\nu}.
\end{equation}
If we specialize to $\varphi=\pi/2,~e^{i\varphi}=i$, \cite{CWPO} we find $g^{\mu\nu}=\eta^{\mu\nu}_M$
where $\eta^{\mu\nu}_M$ has now the Minkowski signature $s=1$, i.e. $\eta^{00}_M=-1$.
Correspondingly, we may identify the Dirac matrices $\gamma^\mu$ with the matrices
defined by eq. (\ref{2.3}) for $s=1$. This also holds for
$\Sigma^{\mu\nu}=-\frac{1}{4}[\gamma^\mu,\gamma^\nu]$ which now generates the Lorentz group
$SO(1,d-1)$. In short, the theory with $s=1$ (and vielbein $e_\mu{^m}=\delta^m_\mu$) can
be seen equivalently as a theory with euclidean signature $(s=0)$ and complex vielbein
given by eq. (\ref{X3.2} with $e_0\ ^0=i$). This continuation procedure is partly analogous to the continuation of the metric proposed in ref. \cite{MEH1}, but we do not employ the additional rotation between  $\gamma^0$ and $\gamma^5$ proposed in this paper. The latter is not needed if the fundamental Grassmann variables are $\psi$ and $\bar\psi$. 

For the Dirac algebra the only change from euclidean to Minkowski signature is
therefore the relation
\begin{equation}\label{X3.4}
\gamma^0_M\equiv\gamma^{\mu=0}_M=-i\gamma^0_E.
\end{equation}
The matrix $\bar{\gamma}$ remains identical for both euclidean and Minkowski signature
since by virtue of eq. (\ref{2.5C}) one finds for $s=1$ the phase
$\eta_M=i\eta_E$. We may replace in the definition \eqref{2.4}, \eqref{2.5C} the factor $i^s$ by $e$,
\be\label{3.4A}
\bar\gamma=(-i)^{\frac d2}e\prod_\mu\gamma^\mu=(-i)^{\frac d2}\prod_m\gamma^m,
\ee
with properly ordered and symmetrized products. This clearly shows that $\bar\gamma$ is not affected by analytic continuation.

If we formulate $S_{kin}$ in terms of $\gamma^\mu_{(M)}$ the analytic continuation of eq. \eqref{X3.1} becomes in Minkowski space $i S_{kin}$, where the factor $i$ results from the continuation of the determinant of the vielbein $e$. Thus for the partition function the kinetic term contributes a factor $\exp(-S_{kin})$ for euclidean signature, and $\exp  (-iS_{kin})$ for Minkowski signature. For euclidean signature the natural definition of the action is $S^{(s=0)}_E=\int d^4xe_EL_E$ with $e_E=1$ in flat space. For a Minkowski signature the historical convention contains an additional minus sign $S^{(s=1)}_M=\int d^4x{\cal L}_M=-\int d^4xL_E$ (anal. contd), corresponding to ${\cal L}_M\sim$ kinetic energy minus potential energy ($T-V$). ($V$ does not change under analytic continuation). This implies 

\begin{equation}\label{eu1}
S^{(s=1)}_M=iS^{(s=0)}_E \textup{ (anal. contd.)}
\end{equation}
where  in addition to the minus sign a factor $i$ accounts for the fact that the definition of $S_E$ (anal. contd.) includes the analytical continuation of the volume element $e$ which is not included in the definition of $S_M$. (The kinetic term contributes $S_M=-S_{kin}+\dots)$  The transition from $L_E$ to $-{\cal L}_M$ could be realized by a change from euclidean to Minkowski $\gamma^m$-matrices with a simultaneous change of the signature for $\eta_{mn}$. Instead, we keep here $\gamma^m$ and $\eta_{mn}$ fixed and perform the analytic continuation \eqref{X3.2} in the vielbein.\footnote{In flat space the analytic continuation in the vielbein can be replaced by an analytic continuation in the time variable. Our convention corresponds to $t=-i\tau$, with $\tau/t$ the time variable in euclidean/Minkowski space, respectively.} Going from euclidean to Minkowski signature the weight factor in the partition function changes then only by a change of the ``background'' vielbein, with $\exp(iS_M)=\exp(-S^{(s=0)}_E$ (anal. contd)). In particular, if one would construct gravity theories where the vielbein is a complex integration variable one has the same weight factor for all signatures. This also holds for theories where no explicit vielbein or $\eta_{mn}$ appears, like for spinor gravity \cite{SG}, \cite{CWPO}. Note that for all procedures the analytical continuation of $\gamma^\mu=\gamma^me^{\ \mu}_m$ and $g^{\mu\nu}=\frac{1}{2}\{\gamma^\mu,\gamma^\nu\}$ is the same. 

The analytic continuation of the vielbein is convenient for its simplicity. Equivalent versions obtain by applying a generalized Lorentz-transformation on the spinors and the vielbein which leave the action invariant. This does not change the analytic continuation of the metric. There are also alternative formulations of analytic continuation that keep the metric fixed and change the time coordinate instead. Again, this can be accompanied by a generalized Lorentz transformation \cite{MEH}. A Lorentz transformation of the spinors can also be transferred to an equivalent Lorenz transformation of the $\gamma^\mu$-matrices, as in \cite{MEH1}.

\section{Euclidean Majorana spinors}
\label{EuclideanMajoranaspinors}
\setcounter{equation}{0}
In a group theoretical sense Majorana spinors can be defined whenever it is possible to identify the Grassmann variables $\bar\psi_\gamma(x)$ with suitable linear combinations of variables $\psi_\gamma(x)$, such that this identification is consistent with the generalized Lorentz transformations. For Dirac spinors $\psi$ and $\bar\psi$ belong to equivalent representations of the Lorentz group. We can therefore always find a linear combination
\be\label{4.1N}
\psi^c_\gamma(x)=W_{\gamma\delta}\bar\psi_\delta(x) 
\ee
which transforms under Lorentz transformations \eqref{2.2} in the same way as $\psi_\gamma(x)$. (In sect. \ref{Hermiteanspinor} the matrix $W$ is identified with $W_1=(C^T)^{-1}=C^*$.) The identification $\psi^c=\psi$ therefore expresses $\bar\psi_\gamma(x)=W^{-1}_{\gamma\delta}\psi_\delta(x)$ as a linear combination of $\psi_\gamma(x)$ and is compatible with the Lorentz transformations. Thus the identification 
\be\label{4.2N}
\psi^c_\gamma(x)=\psi_\gamma(x)
\ee
defines a Majorana spinor in a group theoretical sense. This is obviously possible for arbitrary dimension and signature. 

For physical Majorana spinors we have to impose an additional requirement, namely that eqs. \eqref{4.1N}, \eqref{4.2N} are compatible with a kinetic term \eqref{2.9}, such that a free propagating fermion can be described by the Grassmann functional integral. This is not the case for arbitrary dimension. Using the Majorana constraint we can write
\be\label{4.3N}
S_{kin}=i\int_x\psi^T(W^T)^{-1}\gamma^\mu\partial_\mu \psi.
\ee
This expression is non-vanishing only if the matrix $(W^T)^{-1}\gamma^\mu$ is symmetric. The Pauli principle is expressed by the anticommuting properties of the Grassmann variables, and one finds for an arbitrary matrix $A$ by partial integration
\be\label{4.4N}
\int_x\psi^T A\partial_\mu \psi=-\int_x(\partial_\mu\psi)^T A^T\psi=\int_x\psi^T A^T\partial_\mu\psi,
\ee
such that an antisymmetric part of $A$ does not contribute. Physical Majorana spinors therefore require the condition 
\be\label{4.5N}
\big((W^T)^{-1}\gamma^\mu\big)^T=(W^T)^{-1}\gamma^\mu
\ee
or 
\be\label{4.6N}
W^T(\gamma^\mu)^T=\gamma^\mu W.
\ee

This condition can be met only if the symmetric product of two identical Majorana spinors contains a Lorentz-vector. This group-theoretical property depends on the dimension, but not on the signature. For odd dimensions, the symmetric product of two (identical) fundamental spinor representations contains a vector for $d=3,9$ mod $8$, but not for $d=5,7$ mod $8$. Physical Majorana spinors can therefore not be implemented for $d=5,7$ mod $8$. For even $d$ the issue is more subtle since the irreducible spinor representations also obey a Weyl constraint. For $d=2$ mod $8$ a vector is contained in the symmetric product of two identical Weyl spinors. For $d=2$ mod $8$ we can therefore implement Majorana-Weyl spinors for arbitrary signature. For $d=4,8$ mod $8$ the symmetric product of two identical Weyl spinors does not contain a vector, such that physical Majorana-Weyl spinors are not possible. The vector is contained in the product of two inequivalent Weyl spinors, as manifest in eq. \eqref{2.11} where $\bar\psi_+$ and $\psi_+$ belong to inequivalent Lorentz representations for  $d=4,8$ mod $8$ . We can, however, still impose a Majorana constraint by identifying $\bar\psi_-$ with $\psi_+$. This allows for the implementation of physical Majorana spinors for $d=4,8$ mod $8$ which are equivalent to Weyl spinors. Finally, for $d=6$ mod $8$ both $\bar\psi_+$ and $\psi_+$ are in equivalent representations of the Lorentz group. A Majorana constraint would therefore have to identify $\bar\psi_+$ with $\psi_+$. However, the symmetric product of two identical Weyl spinors does not contain a vector, such that physical Majorana spinors do not exist for $d=6$ mod $8$. We summarize that physical Majorana spinors exist for $d=2,3,4,8,9$ mod $8$, independently of the signature. These group theoretical properties will be reflected by the explicit construction of Majorana spinors in sect. \ref{Generalized Majorana}. For a complex Grassmann algebra the reduction of the number of independent Grassmann variables by imposing a Majorana constraint does not interfere with analytic continuation. Analytic continuation is possible for Majorana spinors in arbitrary dimensions. 

Symmetry transformations are defined by their action on $\psi$ and $\bar\psi$. A given symmetry is consistent with a Majorana constraint \eqref{4.2N} if $\psi$ and $\psi^c$ transform identically. For example, in $d=4$ mod $4$ dimensions a global unitary transformation of $\psi_+$ must be accompanied by the same transformation of $\bar\psi_-$, since the Majorana constraint identifies $\psi_+$ and $\bar\psi_-$. In these dimensions we can actually define all symmetries for Majorana spinors in terms of the Weyl spinors $\psi_+$ and $\bar\psi_+$. The identification of $\psi_+$ and $\bar\psi_-$, as well as $\bar\psi_+$ and $\psi_-$, by the Majorana constraint {\em defines} in this case the appropriate symmetry transformations of $\psi_-$ and $\bar\psi_-$. This transfers the symmetry transformations from the Weyl basis $(\psi_+,\bar\psi_+)$ to the Majorana basis $(\psi_+,\psi_-)$. The analytic continuation of symmetries for Majorana spinors in Minkowski space to a euclidean signature can be implemented in this way. One first expresses the symmetries in the Weyl basis by their action on $\psi_+,\bar\psi_+$. In this basis the action is analytically continued, and the euclidean symmetries correspond to transformations of $\psi_+$ and $\bar\psi_+$ which leave the action for euclidean signature invariant. These transformations can then be expressed as transformations of the equivalent euclidean Majorana spinors. This construction applies for all symmetries, including supersymmetry. For the definition of the action of supersymmetry transformations on euclidean Majorana spinors it is sufficient to formulate the supersymmetry for euclidean Weyl spinors.

A different issue concerns the question if a quantum theory for Majorana spinors can be formulated in terms of a real Grassmann algebra. In other words, we may ask if it is possible to employ a ``real'' functional integral, where all coefficients in the euclidean action $S_E$ are real. Analytic continuation preserves the symmetry properties of matrices as $(W^T)^{-1}\gamma^\mu$ and does therefore not interfere with the existence of Majorana spinors. However, it changes the properties of such matrices under complex conjugation. If $S_E$ is real for a given dimension and signature, its analytic continuation will no longer share this property. One may therefore expect that the reality properties of $S_E$ can depend on the signature.

An observation that a given $S_E$ involves complex numbers is, however, not sufficient to exclude that a real Grassmann algebra can be formulated for a suitable basis of Grassmann variables. A complex similarity transformation among the Grassmann variables changes the reality properties of $S_E$ and may be used in order to transform $S_E$ into a real object. General obstructions which would forbid the formulation of a real Grassmann algebra are difficult to formulate since many possible complex structures can be formulated for a given Grassmann algebra. (By a ``real Grassmann algebra'' we understand here a Grassmann algebra using only linear combinations of $\psi_\gamma(x)$ as basis elements - or of $\psi_\gamma(x)$ and $\bar\psi_\gamma(x)$ in case of Dirac or Weyl spinors - and involving only real coefficients in a suitable basis. It is trivial to reformulate any complex Grassmann algebra as a real Grassmann algebra with twice the number of elements.)

Consider the presence of an involution, $g\to\bar\theta(g)$, within a complex Grassmann algebra. We suppose the property $\bar\theta(\lambda g)=\lambda^*\bar\theta(g)$. The involution property $\bar\theta^2=1$ implies that the spinors (Grassmann elements linear in the Grassmann variables $\psi_u$) can be divided into real and imaginary spinors
\be\label{4.7N}
\psi_R=\frac12 \big(\psi+\bar\theta(\psi)\big)~,~i\psi_I=\frac{1}{2}
\big(\psi-\bar\theta(\psi)\big).
\ee
The spinors $\psi_R$ and $i\psi_I$ are even and odd under $\bar\theta$, respectively. If the action $S_E$ is invariant under the involution, $\bar\theta(S_E)=S_E$, it can be written in the form 
\be\label{4.8N}
S_E=S_{{\rm even}}+S_{{\rm odd}},
\ee
where $S_{{\rm even}}$ involves only terms with an even number of $i\psi_I$ and real coefficients $a_R$, while $S_{{\rm odd}}$ involves terms with an odd number of $i\psi_I$ and imaginary coefficients $ia_I$. Switching to variables $\psi_R$ and $\psi_I$, $S_{{\rm odd}}$ involves an odd number of $\psi_I$ with real coefficients $\sim a_I$. For the formulation of a real Grassmann algebra it is therefore sufficient to find a suitable involution $\bar\theta$ such that $S_E=\bar\theta(S_E)$. In this case $S_E$ is an element of a real Grassmann algebra. (This construction involves no reordering of Grassmann variables, but it can be realized to involutions which also involve a total reordering of all Grassmann elements as discussed in sects. \ref{Complexconjugation}, \ref{Realfermionicactions}.) Whenever a formulation in terms of a real Grassmann algebra is possible, the partition function \eqref{2.2AA} can be interpreted as a real functional of the real coefficients $a_R,a_I$. Taking derivatives with respect to $a_R,a_I$ yields real values for the associated Greens functions for composite multi-fermion operators.

Every involution $\bar\theta$ defines a complex structure in the space of spinors, which associates  to $\psi$ a complex conjugate spinor $\psi^*=\bar\theta(\psi)$. However, the Lorentz transformations are not compatible with every complex structure. Compatibility requires that $\psi^*$ transforms as
\be\label{4.9N}
\delta\psi^*=-\frac12 \epsilon_{mn}(\Sigma^{mn})^*\psi^*
\ee
or 
\be\label{4.10N}
\bar\theta\Sigma^{mn}\bar\theta=(\Sigma^{mn})^*.
\ee
(A general discussion of complex structures and consistency requirements can be found in appendix B.) If this condition is not obeyed, both the Lorentz transformations and the complex structure $\bar\theta$ remain well defined. Only the appropriate choice of the complex structure changes under Lorentz transformations. We may consider a general action $S$, not necessarily Lorentz invariant, which is real with respect to the involution $\bar\theta,\bar\theta (S)=S$. Applying a Lorentz-transformation yields an action $S'$, and for $\bar\theta$ not obeying eq. \eqref{3.2} one may find $\bar\theta(S')\neq S'$. In this case a different involution $\bar\theta'$ will leave $S'$ invariant, $\bar\theta'(S')=S'$. 

Another independent question asks if the Lorentz group admits a real representation for Majorana spinors. This means that we can find real $2^{\left[\frac d2\right]}\times 2^{\left[\frac d2\right]}$ matrices $\Sigma^{mn}$ which obey the commutation relations for the generators of the (generalized) Lorentz group. Furthermore, we require that a non-vanishing kinetic term can be constructed from Majorana spinors transforming in this representation. This issue is discussed in detail in ref. \cite{W1}. We only remark here that real $2^{\frac d2}\times 2^{\frac d2}$ matrices $\Sigma^{mn}$ exist for all even $d$. This is trivial since the complex $2^{\left(\frac d2-1\right)}\times 2^{\left(\frac d2-1\right)}$ matrices representing the Lorentz generators in the Weyl representation can always be written as real $2^{\frac d2}\times 2^{\frac d2}$ matrices (cf. eq. \eqref{BB8}). However, it is not always possible to construct a kinetic term from a single Weyl spinor. We emphasize that in quantum field theory the existence of Majorana spinors neither requires the formulation in terms of a real Grassmann algebra, nor the existence of real representation of the Lorentz group. 

Before proceeding in the next sections to an explicit discussion of complex structures, Majorana spinors and discrete symmetries for arbitrary dimension and signature, it may be useful to illustrate the general statements of the present section by the example of a Majorana spinor in four dimensions. For Minkowski signature $(s=1)$ we consider the four component spinor $\psi_\gamma(x)$ which transforms under Lorentz transformations \eqref{2.2} with real generators
\be\label{y}
\Sigma^{0k}=-\frac12 T_k~,~\Sigma^{kl}=-\frac12 \epsilon^{klm}\tilde I T_m.
\ee
They involve the real matrices
\be\label{BB2N}
T_1=\left(\begin{array}{cc}
0,&1\\1,&0\end{array}\right)~,~T_2=
\left(\begin{array}{cc}
0,&c\\-c,&0\end{array}\right)~,~T_3=
\left(\begin{array}{cc}
1,&0\\0,&-1\end{array}\right)~,~\tilde I=-
\left(\begin{array}{cc}
c,&0\\0,&c\end{array}\right),
\ee
where $c=i\tau_2$. For a Majorana spinor $\bar\psi$ is related to $\psi$ and the action can be written in terms of $\psi$ alone. The kinetic term is an element of a real Grassmann algebra, 
\be\label{BB3N}
S_E=-iS_M=\int_x\psi^T(\partial_0-T_k\partial_k)\psi.
\ee
(For more details and the Lorentz invariance of $S_E$ see appendix A.) The complex structure is trivial, $\theta_M(\psi)=\psi$, and the Lorentz transformations are compatible with it since we use a real representation of the Lorentz group. 

Analytic continuation to euclidean signature results in
\be\label{BB4N}
S_E=\int_x\psi^T(\partial_0+iT_k\partial_k)\psi.
\ee
The $SO(4)$ rotations of the generalized Lorentz group are represented by generators
\be\label{BB5N}
\Sigma^{0k}=-\frac i2 T_k~,~\Sigma^{kl}=-\frac12\epsilon^{klm}\tilde I T_m,
\ee
which are no longer all real. There is no need to introduce independent spinors $\bar\psi$ - our setting describes euclidean Majorana spinors. For both Minkowski and euclidean signature the chiral transformation of a single Weyl spinor is described for the  equivalent Majorana spinor by the infinitesimal transformation
\be\label{BB6N}
\delta\psi=\alpha\tilde I\psi.
\ee
Since $\tilde I$ commutes with $T_k$ and $\tilde I^T=-\tilde I~,~\tilde I^2=-1$, both actions \eqref{BB3N} and \eqref{BB4N} are invariant. It is rather obvious that there is no particular problem for the analytic continuation from Minkowski to euclidean signature.

With respect to the trivial involution $\theta_M$ the action \eqref{BB4N} is no longer real for euclidean signature. Also the generalized Lorentz transformations are not compatible with $\theta_M$ since $(\Sigma^{0k})^*=-\Sigma^{0k}$. For euclidean signature we define a different involution
\be\label{BB7N}
\theta\psi(t,\vec x)=\psi(-t,\vec x).
\ee
Its action in the Grassmann algebra involves a complex conjugation of all coefficients and a total reordering of all Grassmann variables. The invariance of the action \eqref{BB4N}, $\theta(S_E)=S_E$, reflects Osterwalder-Schrader positivity \cite{OS}. With respect to $\theta$ the even elements are even functions of $t$, and the odd elements are odd in $t$. We will define ``real and imaginary spinors'' as 
\ba\label{BB8N}
\psi_R(t,x)&=&\frac12 \big(\psi(t,x)+\psi(-t,x)\big),\nn\\
i\psi_I(t,x)&=&\frac12\big(\psi(t,x)-\psi(-t,x)\big).
\ea

It is instructive to understand this complex structure in Fourier space by putting (euclidean) time on a torus with circumference $1/T$, employing antiperiodic boundary conditions for $\psi(t,x)$. The $\theta$-even part $\psi_R(t,x)$ can be expanded as
\be\label{BB10N}
\psi_R(t,x)=\sum^\infty_{n=0}\cos\big[(2n+1)\pi T t\big]\psi_{Rn}(\vec x),
\ee
while the odd part reads
\be\label{BB11N}
i\psi_I(t,x)=\sum^\infty_{n=0}\sin\big[(2n+1)\pi T t\big]\psi_{In}(\vec x).
\ee
This describes the setting at nonzero temperature $T$ with Matsubara frequences given by integers $n$. We note that no negative $n$ contribute to the sums - they would not give independent contributions. (We recall that we have only one Grassmann variable $\psi_\gamma(t,\vec x)$ for every $t,\vec x$ and $\gamma$.) Inserting the expansions \eqref{BB10N}, \eqref{BB11N} into the action \eqref{BB4N} yields
\ba\label{BB12N}
S_E&=&\frac{1}{2T}\int\limits_{\vec x} \sum^\infty_{n=0}\Big\{(2n+1)\pi T
\big[\psi^T_{Rn}(\vec x)\psi_{In}(\vec x)-\psi^T_{In}(\vec x)\psi_{Rn}(\vec x)\big]\nn\\
&&+i\big[\psi^T_{Rn}(\vec x)T_k\partial_k\psi_{Rn}(\vec x)+
\psi^T_{In}(\vec x)T_k\partial_k\psi_{In}(\vec x)\big]\Big\}.
\ea
We note that $S_E$ changes sign if $\psi_{In}(x)\to -\psi_{In}(x)$, together with a complex conjugation of all coefficients. (It remains invariant if one reverses, in addition, the order of the Grassmann variables.) We may define new Grassmann variables as $\psi_{Rn}(x)=\sqrt{-i}\varphi_{Rn}(x),\psi_{In}(x)=\sqrt{i}\varphi_{In}(x)$. Under this similarity transformation the first term $\sim T$ in eq. \eqref{BB12N} keeps its form, while for the second term one observes $i\psi_R(T_k\partial_k)\psi_R\to\varphi_R(T_k\partial_k)\varphi_R$ and $i\psi_I(T_k\partial_k)\psi_I\to-\varphi_I(T_k\partial_k)\varphi_I$. In terms of $\varphi_R$ and $\varphi_I$ the euclidean action \eqref{BB12N} only involves real coefficients. We can formulate the euclidean quantum field theory for free massless Majorana spinors in terms of a real Grassmann algebra.

We finally perform a Fourier transform to momentum space,  
\be\label{BB13N}
\psi_{Rn}(\vec x)+i\psi_{In}(\vec x)=\int_{\vec q}
e^{i\vec q\vec x}
\big (\tilde\psi_{Rn}(\vec q)+i\tilde \psi_{In}(\vec q)\big)
\ee
which reads for the variables $\psi_{Rn},\psi_{In}$
\ba\label{BB14N}
\psi_{Rn}(\vec x)&=&\int\limits_{\vec q}\big\{\cos(\vec q\vec x)\tilde \psi_{Rn}(\vec q)-\sin 
(\vec q\vec x)\tilde \psi_{In}(\vec q)\big\}\nn\\
\psi_{In}(\vec x)&=&\int\limits_{\vec q}\big\{\cos(\vec q x)\tilde 
\psi_{In}(\vec q)+\sin(\vec q\vec x)\tilde\psi_{Rn}(\vec q)\big\}.
\ea
The action \eqref{BB12N} becomes, with $\omega_n=(2n+1)\pi T$,
\ba\label{Bb15N}
S_E&=&\frac{1}{2T}\sum^\infty_{n=0}\int\limits_{\vec q}\Big\{(2n+1)\pi T
\big[\tilde\psi^T_{Rn}(\vec q)\tilde \psi_{In}(\vec q)-\tilde\psi^T_{In}(\vec q)\psi_{Rn}(\vec q)\big]\nn\\
&&-i\big[\tilde\psi^T_{Rn}(\vec q)q_kT_k\tilde\psi_{In}(\vec q)
-\tilde \psi^T_{In}(\vec q)q_kT_k\psi_{Rn}(\vec q)\big]\Big\}\nn\\
&=&\frac 1T\sum^\infty_{n=0}\int\limits_{\vec q}\tilde \psi^T_{Rn}(\vec q)(\omega_n-iq_kT_k)
\tilde \psi_{In}(\vec q).
\ea
The action \eqref{Bb15N} is invariant under a transformation $\psi_{Rn}(\vec q)\to\psi_{Rn}(-\vec q),\psi_{In}(\vec q)\to \psi_{In}(-\vec q)$, together with a complex conjugation of all coefficients. This involution permits the formulation of real and imaginary spinors according to eq. \eqref{4.7N}, and again a formulation in terms of a real Grassmann algebra. It is obvious that the question if the Grassmann algebra is real or complex depends on the choice of the Grassmann variables. 

\section{Complex conjugation for Minkowski signature}
\label{Complex conjugation for Minkowski signature}
\setcounter{equation}{0}
In the next four sections we develop the issue of possible complex structures for Dirac spinors in arbitrary dimensions and with arbitrary signature. A more abstract discussion of complex structures can be found in appendix B. 

In Minkowski space we are used to the notion of a complex conjugation
which relates\footnote{We will not always make
a distinction between $\bar\psi$ and $\bar\psi^T$ such
that we may also write $\bar\psi=\epsilon\psi^\dagger D$.
In order to avoid confusion with ref. \cite{W1} we emphasize
that in \cite{W1} $\bar\psi$ denotes the conjugate
of $\psi$ in a group-theoretical sense for which (\ref{2.16})
holds for arbitrary signature. In other words, the same complex conjugation is chosen in ref. \cite{W1} for all signatures, in contrast to the approach of the present paper. In the present paper $\bar\psi$
denotes the spinor associated to $\psi$ by the kinetic term (\ref{2.9}).
The introduction of $\epsilon$
is motivated by standard conventions for charge conjugate spinors
to be explained later. Both $D$ and $\epsilon D$ obey the relation
(\ref{2.7}).} $\psi^*$ and $\bar\psi$
\be\label{2.16}
\bar\psi=\epsilon D^T\psi^*\ , \quad\epsilon^2=1.\ee
Here the matrix $D$ acts only on spinor indices and
obeys.  
\be\label{2.7}
D\Sigma^{mn}D^{-1}=-(\Sigma^{mn})^\dagger~,~ D^\dagger D=1.\ee
For even $d$ these conditions can be fulfilled if $D$ is given either by $D_1$ or $D_2$ which obey
\begin{eqnarray}\label{4.2A}
&&D_1\gamma^mD^{-1}_1=(\gamma^m)^\dagger~,~D_2\gamma^mD^{-1}_2=-(\gamma^m)^\dagger,\nonumber\\
&&D^{\ \dagger}_1=D_1~,~D^\dagger_2=-D_2~,~D_2=i^{s-1}D_1\bar{\gamma}.
\end{eqnarray}
The matrix $D_1$ can be chosen as $D_1=1,\gamma^0\bar\gamma,\gamma^0,\bar\gamma$
for $s=0,1,d-1$ and $d$, respectively, with $\gamma^0=
\gamma^{0\dagger}$ for $s=d-1$ and $\gamma^0=-\gamma^{0\dagger}$ for
$s=1$. (A systematic discussion and details can be found in appendix $C$. There we use a different naming such that for $s=d-1$ the hermitean matrix is $\gamma^{d-1}$.) The kinetic term (\ref{2.9}) is hermitean (cf. sect. \ref{Hermiteanspinor}) for both $D_1$ and $D_2$. Eq. \eqref{2.16} defines the notion of the complex conjugate spinor which has not yet been introduced so far. It is an element of the Grassmann algebra which can be written in terms of $\bar\psi$ as
\be\label{4.3A}
\psi^*=\epsilon(D^T)^{-1}\bar\psi.
\ee
We will see later that other definitions are also possible. 

A linear transformation
\be\label{2.71}
\psi(x)\longrightarrow A\psi(x')\ee
is compatible with this complex conjugation if
\be\label{2.7A}
\psi^*(x)\to A^*\psi^*(x'),
\ee
or if the associated spinor $\bar\psi$ transforms as
\be\label{2.8}
\bar\psi(x)\longrightarrow \bar\psi(x')D^{-1}A^\dagger D.\ee
We conclude that Lorentz transformations are compatible
with the complex conjugation.
Another example is a unitary
matrix $A$ acting only on internal indices $a=1...N$
with $\bar\psi\to \bar\psi A^\dagger$. Then the spinor
$\bar\psi$ transforms as the complex conjugate representation
of $\psi$ with respect to the corresponding symmetry group
($U(N)$ or a subgroup of it) and $\bar\psi\psi$ is
invariant.
We emphasize that transformations $A$ not obeying the
compatibility condition (\ref{2.8}) or even more general
transformations mixing $\psi$ and $\bar\psi$ remain well
defined. They simply do not obey the rule (\ref{2.7A}) and can therefore
not be written as
complex matrix multiplication for complex spinors.
We will see in sect. 6 that there is an option to define a complex
conjugation different from eq. (\ref{2.16}) in case of an euclidean signature.

Using the identity
\be\label{2.14}
D^{-1}\bar\gamma D=(-1)^s\bar\gamma=(-1)^{d-s}\bar\gamma\ee
we see that the chiral transformations (\ref{2.12}) are compatible with eq. 
(\ref{2.8}) only if the number of dimensions with
negative signature is odd.
This indicates that the association (\ref{2.16}) of $\bar\psi$ with
a complex conjugate spinor $\psi^*$ is  
not compatible with chiral rotations for 
a euclidean signature. With eq. \eqref{2.16}, $\psi^*$ does not rotate with the opposite phase as $\psi$. In contrast, for a Minkowski signature
the usual definition of the complex conjugate spinor $\psi^*$
according to eq. (\ref{2.16}) is compatible with the chiral rotations.
The incompatibility of the identification
(\ref{2.16}) with the chiral structure of the theory is well known
and translates a simple property of the spinor representations
of the generalized Lorentz group $SO(s,d-s)$ with respect to
complex conjugation: In four dimensions and for Minkowski
signature ($s=1$ or 3) the two inequivalent
irreducible spinor representations $2_L$ and $2_R$ are complex
conjugate to each other. For an euclidean signature $(s=0$ or $4)$
the corresponding spinor representations of $SO(4)$ are (pseudo)real.

In four dimensions, a spinor bilinear transforming as a four vector $(2_L,2_R)$
cannot be formed from an irreducible two-component spinor and 
its complex conjugate if the signature is euclidean and complex conjugation is defined by eq. \eqref{4.3A}.  Similar features and complications generalize to all even dimensions.
This observation has often led to the opinion that the spinor
degrees must be doubled when the signature is changed from
Minkowski to euclidean. We emphasize, however, that the number
of independent spinors $\psi^{\ a}_\gamma(q),\bar\psi^{\ a}_\gamma
(q)$ is exactly the same for {\it all} signatures. Only the
standard complex structure in the 
Grassmann algebra which defines $\psi^*$ in the usual way 
(\ref{2.16}) is not
compatible with chiral rotations for a euclidean signature.

The complex and hermitean conjugates of an element $g$ of the Grassmann
algebra\footnote{The index $u,v$ combines here internal indices
and space coordinates.}
\be\label{3.1}
g=a^{\gamma_1...\gamma_p\ \delta_1...\delta_q}_{u_1...u_p\ v_1
...v_q}\psi^{u_1}_{\gamma_1}\psi^{u_2}_{\gamma_2}...\psi^{u_p}_{\gamma_p}
\bar\psi^{v_1}_{\delta_1}
...\bar\psi^{v_q}_{\delta q}
\ee
are defined by
\ba\label{3.2}
&&g^*=(a^{\gamma_1...\gamma_p\delta_1...\delta_q}_{u_1...u_p v_1...v_q})^*
(\psi^{u_1}_{\gamma_1})^*...(\psi^{u_p}_{\gamma_p})^*(\bar\psi^{v_1}_{\delta
_1})^*...(\bar\psi^{v_q}_{\delta_q})^*,\nonumber\\
&&
g^\dagger=\left(a^{\gamma_1...\gamma_p\ \delta_1...\delta_q}_{u_1...u_p\ v_1
...v_q}\right)^*\left(\bar\psi^{v_q}_{\delta_q}\right)^*
...\left(\bar\psi^{v_1}_{\delta
_1}\right)^*\left(\psi^{u_p}_{\gamma_p}\right)^*...
\left(\psi^{u_1}_{\gamma_1}\right)^*.\ea
Here we note that $g^\dagger$ involves a transposition which amounts to a total reordering of all Grassmann variables. The kinetic term 
(\ref{2.9}) is therefore invariant under hermitean conjugation. More
generally, hermiticity of the action $S_M$ in Minkowski space is believed
to be crucial for the consistency of a field theory since it is 
closely related to unitarity. 

More generally, a complex conjugation in the space of spinors (linear combinations of $\psi$ and $\bar\psi$) can be defined as a linear map $\bar\theta$ in the space of spinors, which is an involution, $\bar\theta^2=1$. In particular, this involution maps every Grassmann variable $\psi^u_\gamma,\bar\psi^u_\gamma$ into a linear combination of Grassmann variables, $\psi^u_\gamma\to\bar\theta(\psi^u_\gamma)=(\psi^*)^u_\gamma~,~\bar\psi^u_\gamma\to \bar\theta(\bar\psi^u_\gamma)=
(\bar\psi^*)^u_\gamma$. The involution property implies that the only possible eigenvalues of $\bar\theta$ are $\pm 1$, and we can divide the spinors into even and odd elements with respect to $\bar\theta$. In our case the numbers of even and odd elements are equal. We can identify the even elements with the real parts of complex spinors, and the odd elements with the imaginary parts.

With Minkowski signature this involution is realized by the identification (\ref{2.16}), i.e. by the mapping $\psi\to \theta_M\psi,\bar\psi\to\theta_M\bar\psi$,
\be\label{3.AA}
\psi^*=\theta_M\psi
=\epsilon D^*\bar\psi\ ,\quad \bar\psi^*=\theta_M\bar\psi=\epsilon D^\dagger\psi.
\ee
The map $\theta_M$ involves also a complex conjugation of the coefficients of the Grassmann elements linear in $\psi,\bar\psi$. One easily verifies that $\theta_M$ is an evolution (using $D^\dagger D=1)$,
\be\label{5.4B}
\theta_M(\lambda\psi)=\lambda^*\theta_M\psi~,~\theta_M(\lambda\bar\psi)
=\lambda^*\theta_M\bar\psi~,~\theta^2_M=1.
\ee
Even spinors $\psi_R$ and odd spinors $\psi_I$ are given by
\be\label{5.4A}
\psi_R=\frac12 (\psi +\epsilon D^*\bar\psi)~,~i\psi_I=
\frac12(\psi-\epsilon D^*\bar\psi).
\ee
We can use $\psi^u_\gamma$ and $(\psi^*)^u_\gamma$ instead of $\psi^u_\gamma$ and $\bar\psi^u_\gamma$ as basis elements of the Grassmann algebra. They are related by the similarity transformation $\psi^*=\epsilon D^*\bar\psi$. Alternatively, we could also use $(\psi_R)^u_\gamma$ and $(\psi_I)^u_\gamma$ as new Grassmann variables and interpret eq. \eqref{5.4A} as a change of basis of the Grassmann algebra. The expression of $\psi$ in terms of $\psi_R$ and $\psi_I$ takes the standard form 
\be\label{3.3}
\psi^u_\gamma=(\psi_R)^u_\gamma+i(\psi_I)^u_\gamma.
\ee
Correspondingly,  $\psi^*$ is defined in the usual way as $\psi^*=\psi_R-i\psi_I$. 

An arbitrary element of the Grassmann algebra can be represented in different ways. Instead of eq. \eqref{3.1}  one may use
\be\label{5.4}
g=b^{\gamma_1\dots \gamma_p\delta_1\dots \delta_q}_{u_1\dots u_p v_1\dots v_q}
\psi^{u_1}_{\gamma_1}\dots \psi^{u_p}_{\gamma_p}
(\psi^*)^{v_1}_{\delta_1}\dots(\psi^*)^{v_q}_{\delta_q}
\ee
or
\be\label{5.4D}
g=c^{\gamma_1\dots\gamma_p\delta_1\dots\delta_q}_{u_1\dots u_pv_1\dots v_q}
(\psi_R)^{u_1}_{\gamma_1}\dots(\psi_R)^{u_p}_{\gamma_p}
(\psi_I)^{v_1}_{\delta_1}\dots
(\psi_I)^{v_q}_{\delta_q}.
\ee
Identities of the type
\ba\label{5.4E}
\psi^u_\gamma(\psi^*)^v_\delta&=&-(\psi^*)^v_\delta\psi^u_\gamma\nn\\
&=&(\psi_R)^u_\gamma(\psi_R)^v_\delta+(\psi_I)^u_\gamma (\psi_I)^v_\delta-
i(\psi_R)^u_\gamma(\psi_I)^v_\delta+
i(\psi_I)^u_\gamma(\psi_R)^v_\delta
\ea
permit to change between the expansions \eqref{5.4} and \eqref{5.4D}, and $\psi^*$ is related to $\bar\psi$ by eq. \eqref{5.4B}. In order to define a complex structure for a Grassmann algebra one has to extend the involution $\bar\theta$ from the space of spinors to the space of all Grassmann elements. This defines the complex conjugate Grassmann element $g^*=\bar\theta(g)$. 

The map $g\to g^*$ can also be interpreted as a map of the coefficients $a\to\bar\theta(a),b\to\bar\theta(b)$ or $c\to\bar\theta(c)$ for fixed basis elements by expanding $g^*$ in a given basis. This map has to obey $\bar\theta^2=1$ and should be compatible with the multiplication by complex numbers (in particular $i$) $\bar\theta(\lambda a)=\lambda^*\bar\theta(a)$. There are obviously many different possibilities to define a complex structure. One possibility based on $\theta_M$ uses $g^*$ as defined in eq. \eqref{3.2}. An alternative extension of $\theta_M$ to the Grassmann algebra involves in addition a total reordering of all Grassmann variables, $\theta_M(g)=g^\dagger$ according to eq. \eqref{3.2}. The notion of real and imaginary elements of the Grassmann algebra depends on the choice of the involution. As another example for a possible involution one may take a simple complex conjugation of all coefficients in the basis \eqref{3.1}, $\theta'(a)=a^*$. This would constitute a perfectly valid alternative candidate for a complex conjugation. In this case all Grassmann variables $\psi$ and $\bar\psi$ would be invariant with respect to $\theta'$. 

\section{Modulo two periodicity in the signature}
\label{Modulo two periodicity in the signature}
\setcounter{equation}{0}
At this place it is time to ask what should be the euclidean
correspondence of the operations ``complex conjugation'' or 
``hermitean conjugation'' well known for spinors in Minkowski
space. The principle replacing hermiticity for a euclidean signature is
Osterwalder-Schrader positivity \cite{OS}. This requires invariance
of the action with respect to a particular type of reflection
of one coordinate $\theta(x_0,x_1...x_{d-1})=(-x_0,x_1...x_{d-1}).$
This transformation transforms $\psi$ into $\bar\psi$ and vice versa
\ba\label{3.4}
&&\theta(\psi(x))=H^*\bar\psi(\theta x),\nonumber\\
&&\theta(\bar\psi(x))=H^{-1}\psi(\theta x).\ea
For an element of the Grassmann algebra,
it involves in addition a complex conjugation of all coefficients
$a^{\gamma...}_{u...}$  as well as a total reordering
of all Grassmann variables similar to $g^\dagger$ in eq. (\ref{3.2}). One has, in particular,
\be\label{3.5}
\theta(\bar\psi(x)\psi(x))=\bar\psi(\theta x)H^\dagger H^{-1}
\psi(\theta x)\ee
and, more generally, $\theta (g)$ is given by 
replacing in the formula (\ref{3.2}) for $g^\dagger$ the
factors $\psi^*$ and $\bar\psi^*$ by $\theta(\psi)$
and $\theta(\bar\psi)$. Similar to the conjugation defined by $\theta_M$ the 
transformation $\theta$ is an involution,\footnote{For a short discussion of
the logical possibility $\theta^2=-1$ see ref. \cite{W3}.}
\be\label{3.6} \theta^2=1.\ee
Here we remind that the action in the Grassmann algebra implies
$\theta(H^*\bar\psi(\theta(x)))=H\theta(\bar\psi(\theta(x))=\psi(x)$. We observe the close analogy between $\theta$ in
euclidean space and  hermitean conjugation in
Minkowski space. 

The operation $\theta$ is defined for
arbitrary signature. For Minkowski signature it plays the role of a time reversal operation. 
Similarly, the Minkowskian hermitean conjugation
(\ref{2.16}) can be extended to arbitrary signature by introducing  the
transformation $\theta_M$
\ba\label{3.7}
\theta_M(\psi(x))=\epsilon D^*\bar\psi(x)~,~
\theta_M(\bar\psi(x))=\epsilon D^{-1}\psi(x),
\ea
with a suitable matrix $D$. Again, the action of $\theta_M$ on the element of the Grassmann algebra involves a complex conjugation of all coefficients and a total reordering of all Grassmann variables, cf. eq. \eqref{3.2}. This formulation allows the definition of the operation $\theta_M$
for arbitrary signature without that $\psi^*$ is necessarily defined in this way.
One simply has to replace in eq. (\ref{3.2}) $g^\dagger$ by $\theta_M(g)$ as well as $\psi^*$ and $\bar\psi^*$
by $\theta_M(\psi)$ and $\theta_M(\bar\psi)$. Invariance of the kinetic
term (\ref{2.9}) under the transformation $\theta_M$
follows from the property (\ref{2.7}) for the matrix $D$.
The involution property can be written in the form
\be\label{3.8A}
\theta^2_M=1.\ee

The invariance of the kinetic term  with respect to $\theta$
necessitates the following property of $H$
\ba\label{3.8}
&&H^\dagger(\gamma^0)^\dagger H^{-1}=-\gamma^0\nonumber\\
&&H^\dagger(\gamma^i)^\dagger H^{-1}=\gamma^i\ea
which is equivalent to
\be\label{3.9}
H^\dagger=-\gamma^0H\gamma^0=\gamma^iH\gamma^i.\ee
We show below that an appropriate matrix $H$ exists for arbitrary
$d$ and $s$. In consequence we always choose $H$ obeying (\ref{3.9})
such that for Dirac spinors
the kinetic term is invariant under both transformations $\theta_M$ and
$\theta$.

We next consider the properties of the matrix $H$ in some more 
detail. Since $\theta$ corresponds to a time reflection combined with
a complex conjugation of the Lorentz representation we demand
that $\theta\psi$ and $\theta\bar\psi$ transform under Lorentz
transformations according to
\ba\label{3.10}
&&\delta(\theta\psi)=\frac{1}{2}\epsilon_{mn}
\tilde\Sigma^{Tmn}\theta\psi\nonumber\\
&&\delta(\theta\bar\psi)=-\frac{1}{2}\epsilon_{mn}
\tilde\Sigma^{mn}\theta\bar\psi\ea
with, for $i,j\not=0$,
\be\label{3.11}
\tilde\Sigma^{0i}=-\Sigma^{0i}\quad,\quad
\tilde\Sigma^{ij}=\Sigma^{ij}.\ee
This implies the properties
\ba\label{3.12}
&&H\Sigma^{mn}H^{-1}=-\tilde\Sigma^{\dagger mn}\nonumber\\
&&H^{-1}\Sigma^{\dagger mn}H=-\tilde\Sigma^{mn}.\ea
The consistency of these relations is guaranteed\footnote{The
fact that $\zeta$ must be real can be derived by combining
(\ref{3.12}) with (\ref{3.8}).} for 
\be\label{3.13}
H^\dagger=\zeta H,\quad \zeta^2=1.\ee
From (\ref{3.8}) we see that $H$ either commutes or anticommutes with
$\gamma^m$ according to the signature and the value of $\zeta$, using $(\gamma^m)^\dagger=\eta_{mm}\gamma^m$ and
\ba\label{3.14}
&&H(\gamma^0)^\dagger=-\zeta\gamma^0H,\nonumber\\
&&H(\gamma^i)^\dagger=\zeta\gamma_iH.\ea
Therefore $H^\dagger H=\zeta H^2$ commutes with all $\gamma^m$ matrices
and we can use an appropriate scaling of $H$ such that
\be\label{3.15}
H^\dagger H=1.\ee

Let us first consider an even number of dimensions. The matrices
$[(\gamma^0)^\dagger,-(\gamma^i)^\dagger]$ and $[-(\gamma^0)^\dagger,
(\gamma^i)^\dagger]$ obey the same Clifford algebra
as $[\gamma^0,\gamma^i]$
and there is only one irreducible representation of the
Clifford algebra by complex $2^{d/2}\times 2^{d/2}$ matrices up to 
equivalence transformations. There must therefore exist matrices
$H_1,H_2$ with
\be\label{3.16}
H_1\gamma^0H^{-1}_1=-(\gamma^0)^\dagger,\ H_1\gamma^iH^{-1}_1=
(\gamma^i)^\dagger,\ee
\be\label{3.17}
H_2\gamma^0H^{-1}_2=(\gamma^0)^\dagger,\ H_2\gamma^iH^{-1}_2=
-(\gamma^i)^\dagger.\ee
By an appropriate scaling of $H_i$ one always has
\be\label{3.18}
H_1^\dagger H_1=H_2^\dagger H_2=1.\ee
Both $H_1$ and $H_2$ obey eq. (\ref{3.12}). We note that
(\ref{3.16}) and (\ref{3.17}) fix $H_1$ uniquely up to an overall
phase and similar for $H_2$. We determine this phase by the
conditions
\be\label{3.19}
H_1^\dagger=H_1,\quad H_2^\dagger=-H_2,
\ee
such that both $H_1$ and $H_2$ fulfill eq. (\ref{3.8}).

The commutation properties of $H_i$ with $\bar\gamma$ depend
on the number of dimensions with negative signature
\be\label{3.20}
H_i\bar\gamma=(-1)^{s+1}\bar\gamma H_i,
\ee
and we note that the same relation for $H$ can also be extracted
from (\ref{3.12}). In the following we will identify $H$ either
with $H_1$ or $H_2$, the two matrices being related by
\be\label{3.21}
H_2=i^sH_1\bar\gamma.\ee
A few special cases are of interest. For
euclidean signature $s=0$, $H_2$ commutes
with $\gamma^0$ and anticommutes with $\gamma^i$
and one finds\footnote{For $H=H_2$ this
convention agrees with ref. \cite{W3}.}
$H_2=-i\gamma^0$. For the opposite
signature $s=d$ one has $H_1=- i\gamma^0$. For a Minkowski type
signature $s=d-1$ we see that $H_2$ commutes with all matrices
$\gamma^m$ and obtain $H_1=\bar{\gamma}~,~  H_2=i^s$, whereas
for $s=1$ the corresponding choices are $H_1=1~,~H_2= i\bar{\gamma}$.

For odd dimensions $d$ the Clifford algebra can be constructed from the
even dimensional Clifford algebra in one dimension less by adding
the $\bar\gamma$ matrix of the preceding even dimensional algebra
\be\label{3.22}
\gamma^{d-1}=\xi\bar\gamma.\ee
Here $\xi$ must obey $\xi=\pm 1$ if the additional dimension
has positive signature $(\gamma^{d-1}=(\gamma^{d-1})^\dagger)$
whereas $\xi=\pm i$ otherwise. Corresponding to the signature of the
additional dimension only one of the matrices $H_1$ {\it or} $H_2$ exists, since the
commutation properties of $H_i$ with $\gamma^{d-1}$ are fixed by
the relation (\ref{3.20}). One obtains (with $s$ the
number of negative signature dimensions in the odd dimensional
theory)
\be\label{3.23}
H=\left\lbrace\begin{array}{llll}
H_1&\ {\rm for}&\ s&\ {\rm odd}\\
H_2&\ {\rm for}&\ s&\ {\rm even}\end{array}\right..\ee
These relations are one aspect of the modulo two
periodicity in $s$ mentioned in the introduction.

For the discrete transformations $\theta$ the crucial relation for the modulo two periodicity in $s$
is equation (\ref{3.20}). For even $d$ this is translated
directly into the transformation properties of Weyl spinors
\ba\label{3.24}
&&\psi_+\stackrel{\theta}{\leftrightarrow}\bar\psi_+, \quad
\psi_-\stackrel{\theta}{\leftrightarrow} \bar\psi_-\quad
{\rm for}\ s\ {\rm even}\nonumber\\
&&\nonumber\\
&&
\psi_+\stackrel{\theta}{\leftrightarrow}\bar\psi_-,\quad
\psi_-\stackrel{\theta}{\leftrightarrow} \bar\psi_+,\quad
{\rm for}\ s\ {\rm odd}.
\ea
The mappings (\ref{3.24}) are to be compared with the
action of  $\theta_M$ for which the
crucial relation for the modulo two periodicity in $s$
is given by eq. (\ref{2.14})
\ba\label{3.25}
&&\psi_+\stackrel{\theta_M}{\leftrightarrow}\bar\psi_+,\quad
\psi_-\stackrel{\theta_M}{\leftrightarrow} \bar\psi_-\quad{\rm for}\ s\ {\rm odd}\nonumber\\
&&\nonumber\\
&&\psi_+\stackrel{\theta_M}{\leftrightarrow}\bar\psi_-,
\quad \psi_-\stackrel{\theta_M}{\leftrightarrow}
\bar\psi_+ \quad{\rm for}\ s\ {\rm even}.
\ea
For given even $d$ we see how the change of the number of negative
signature dimensions by one unit switches the role of $\theta$
and $\theta_M$.   In our context this is the most important feature of the
modulo two periodicity in the signature. 

In consequence, the
simultaneous change of the signature of all dimensions does
not affect the role of $\theta$ or $\theta_M$. On the other
hand, changing only the signature of the 0-direction (time-like dimension)
amounts to an exchange of the role of $\theta$ and
$\theta_M$. This also holds if we keep positive signature of the 0-direction
and switch to negative signature for the remaining $s=d-1$ dimensions.
In particular, we may compare the euclidean signature
$s=0$ and the Minkowski type signature $s=d-1$ for
$d$ even. In both cases we can choose $(\gamma^0)^\dagger=\gamma^0$
and represent $\bar\gamma$ and $\gamma^0$ in
a Weyl basis
\be\label{3.26}
\psi={\psi_+ \choose \psi_-},\quad \bar\gamma=
\left(\begin{array}{llc}
1&,&0\\0&,&-1
\end{array}\right)~,~
\gamma^0=
\left(\begin{array}{lll}
0&,&1\\1&,&0\end{array}\right).
\ee
For the euclidean signature one has $H_2=-i\gamma^0,D_1=1$ whereas
for the Minkowski signature $s=d-1$ one may choose $D_1=\gamma^0,\ H_2=i^s$,
illustrating the close correspondence of eqs. (\ref{3.4}) and
(\ref{3.7}).

We may also discuss the close relation between $\theta$ and $\theta_M$ for neighboring signature $s$ under the aspect of analytic continuation. Indeed, the euclidean matrices $H_i$ for $s=0$ and the Minkowski-matrices $\epsilon D_i$ for $s=1$ are identical
\begin{eqnarray}\label{eu2}
H^{(s=0)}_1&=&\epsilon D^{(s=1)}_1=-i\gamma^0_E\bar{\gamma}=\gamma^0_M\bar{\gamma},\nonumber\\
H^{(s=0)}_2&=&\epsilon D^{(s=1)}_2=-i\gamma^0_E=\gamma^0_M.
\end{eqnarray}
In this sense, the hermitean conjugation $\theta_M$ for $s=1$ is the analytic continuation of the $\theta$-transformation for $s=0$: the matrices appearing in eqs. (\ref{3.4}) and
(\ref{3.7}) are identical. The only modification is the additional flip of the zero momentum component $q^0$ which is related to the additional minus sign under complex conjugation of the phase in the analytic continuation of the vielbein component $e^{\ m}_0$, as discussed in sect. 3.

We will mainly use a definition of $\theta$ and $\theta_M$ based on $H_2=-i\gamma^0_E~,~D_2=\gamma^0_M$ for $s=0,1$. The ``dynamical part'' of the action, which is the part of $S_{kin}$ containing a time derivative, takes then in Minkowski space the standard form
\be\label{5.26a}
S_{E,dyn}=-iS_{M,dyn}=\int_x\psi^\dagger\partial_t\psi.
\ee
If we define for euclidean signature $\psi^\dagger=\bar\psi H^\dagger_2=i\bar\psi\gamma^0_E$ the relation \eqref{5.26a} remains valid. The dynamical part remains invariant under analytic continuation since it contains both a time integration $dt$ and a derivative $\partial_t$. In our formulation the factor $-i$ from the analytic continuation of $e_m \ ^0$ is canceled by a factor $i$ from $e$.

\section{Generalized complex conjugation with euclidean signature}
\label{Complexconjugation}
\setcounter{equation}{0}
Having established the close correspondence of $\theta$ for
euclidean signature with $\theta_M$ for Minkowski signature
suggests that $\theta$ can be used to define a new operation
of generalized complex conjugation within Grassmann algebras with euclidean
signature. Our guiding principle is the observation that $\theta_M$
defines a complex conjugation in the basis $(\psi,\bar\psi)$.
If a similar complex conjugation based
on $\theta$ is defined in the basis $(\psi,\bar\psi)$ for
euclidean signature, this operation will induce similar mappings
between Weyl spinors as $\theta_M$ for a Minkowski signature
(cf. (\ref{3.24}), (\ref{3.25})). With respect to the complex
conjugation $\theta$ the Majorana spinors with euclidean signature should
behave similarly to Majorana spinors with Minkowski signature and conjugation $\theta_M$.
Also the positivity properties related to hermiticity in Minkowski
signature should carry over to euclidean signature.

In a Fourier representation
\be\label{4.13}
\psi(x)=\sum_q\exp(iq_\mu x^\mu)\psi(q)\quad,\quad
\bar\psi(x)=\sum_q\exp(-iq_\mu x^\mu)\bar\psi(q)\ee
we define for euclidean Dirac spinors the generalized complex conjugation
\ba\label{4.22}
\psi^{**}(q)&=&H^*\bar\psi(\theta q),\nonumber\\
\bar\psi^{**}(q)&=&H^{-1}\psi(\theta q).
\ea
Here we use the symbol $\psi^{**}$ for convenience of the reader in order to recall that $q^0$ is reversed. From the formal point of view $\psi^{**}(q)$ is the complex conjugate of $\psi(q)$. With our conventions $H=-i\gamma^0$ for the signature
$s=0$ this relation is the same as in Minkowski space
$s=1$(\ref{3.AA})
up to a $q_0$ reflection, $\theta(q_0,\vec q)=(-q_0,\vec q)$,
\ba\label{4.23}
\psi^{**}(q)&=&i(\gamma^0)^*\bar\psi(\theta q),\nonumber\\
\bar\psi^{**}(q)&=&i\gamma^0\psi(\theta q),\ea
or, in the usual ``row representation'' of $\bar\psi\quad (\bar
\psi^T\to\bar\psi)$
\be\label{4.24}
\bar\psi(q)=-i\psi^\dagger(\theta q)\gamma^0=\psi^\dagger(\theta)q)\gamma^0_M.
\ee
We can again define real and imaginary parts of $\psi$ by the appropriate
linear combinations
\ba\label{4.21}
\psi(q)&=&\psi_R(q)+i\psi_I(q),\nonumber\\
\psi^{**}(q)&=&\psi_R(q)-i\psi_I(q).\ea
This demonstrates again the complete analogy of spinor
degrees of freedom in an Minkowski or euclidean formulation.
Instead of $\psi$ and $\bar\psi$ we can also take
$\psi_R$ and $\psi_I$ or $\psi$ and $\psi^{**}$ as
independent fermionic degrees of freedom. The new complex
structure does not necessitate a ``doubling'' of euclidean spinor
degrees of freedom.

The compatibility of a (momentum-independent) linear
transformation $\psi\to A\psi$ with the complex conjugation requires again that
$\psi^{**}$ transforms according to $\psi^{**}\to A^*\psi^{**}$. This
replaces eq. (\ref{2.8}) by the requirement
\be\label{4.25A}
\bar\psi\to \bar\psi H^{-1}A^\dagger H.\ee
Despite the close similarity of the new complex conjugation
(\ref{4.22})
with the formulation for
Minkowski signature we recall that the transformation 
properties of spinors are originally formulated in terms of
$\psi$ and $\bar\psi$. Not all transformations
need to be consistent with
the complex structure. (This is similar to the example of $2M$ real 
scalar fields which can be combined into $M$ complex fields.
Among the possible $O(2M)$ symmetry transformations only the
subgroup $U(M)$ is compatible with the complex structure.)
Compatibility with the complex structure holds trivially
for vector-like global unitary transformations where $U$ does not
act on spinor indices. As a consequence of the modulo two 
periodicity discussed in the last section the chiral transformations
are compatible with the complex structure only if we choose
$\theta$ for euclidean signature and $\theta_M$ for Minkowski 
signature. 

The Lorentz transformations, in
contrast, are only compatible with the complex structure $\theta_M$
and {\it not} with $\theta$ (cf. (\ref{3.12}). For euclidean 
signature and complex conjugation $\theta$ the Lorentz transformations
mix real and imaginary parts of $\psi$ in a way which cannot be
expressed in terms of multiplication with a complex matrix. These
properties, together with the reflection of one coordinate,
constitute the main qualitative difference between possible
complex structures for Minkowski or euclidean signature. Whereas
for Minkowski signature the complex conjugation
$\theta_M$ is compatible
with both chiral and Lorentz transformations, the same cannot be
realized for euclidean signature. For euclidean Dirac spinors
we can, in principle, define the two different complex structures
$\theta$ and $\theta_M$. One is compatible with chiral transformations
($\theta$), the other with Lorentz transformations ($\theta_M$), but
none with both.

In practice, it is most convenient for euclidean spinors to define
all transformations for $\psi$ and $\bar\psi$. We will use the complex
conjugation $\theta$ since only this is compatible with the chiral
structure and related to the necessary positivity properties of the
fermionic action. This choice also guarantees a close analogy 
between euclidean and Minkowski signature. We only have to remember
that the Lorentz transformations of $\psi$ and $\bar\psi$
cannot be represented in the basis $(\psi_R,\psi_I)$ by a multiplication
of a complex spinor $\psi$ with a 
complex matrix. One rather needs the transformations
of both $\psi$ and $\bar\psi$ and can then infer the appropriate 
transformations of $\psi_R$ and $\psi_I$ using
$\psi_R(q)=\frac{1}{2}(\psi(q)+i\gamma^0\bar\psi(\theta q)),
\psi_I(q)=-\frac{i}{2}(\psi_R(q)-i\gamma^0\bar\psi(\theta q))$.

\section{``Real'' fermionic actions}
\label{Realfermionicactions}
\setcounter{equation}{0}

A fermionic action $S_\psi$ which is invariant under either one of the
transformations $\theta_M$ or $\theta$ will be called real. This can
be generalized to other involutions which involve a complex conjugation.
We will collectively denote such transformations by $\bar\theta$.
In this section we establish for a euclidean signature that the fermionic functional integration (\ref{2.2AA})
with a real fermionic action gives a real result,
provided $S_\psi$ contains only terms with an even number of
Grassmann variables. The partition function $Z$ is then real. The reality properties for Minkowski signature follow by analytic continuation. 

The mere existence of the 
involution $\bar\theta$ which maps $\psi$ onto
$\bar\psi$ requires that the fields $\psi$ and
$\bar \psi$ come in associated pairs $(\psi_u,\bar\theta\psi_u)=
(\psi_u,E_{uv}\bar\psi_v)$. (Note that $\bar\theta^2=1$ forbids
$\bar\theta\psi_u=0$.) Using $\det E=1$ one can always write the 
functional measure in the form used in (\ref{2.1a}) and establish
its invariance under $\bar\theta$
%\newpage
\ba\label{5.1}
&&\int {\cal D}\psi {\cal D}\bar\psi\equiv\int\prod_{u'}(d\psi_{u'}
d\bar\psi_{u'})=\int\prod_{u'}(d\psi_{u'}d(\bar\theta\psi_{u'}))
\nonumber\\
&&=\int\prod_{u'}(d(\bar\theta\bar\psi)_{u'}
d(\bar\theta\psi)_{u'})
=\int\prod_{u'}((d\bar\theta\psi)_{u'}d(\bar\theta\bar\psi)_{u'}).
\ea
We assume here and in the following that the number of degrees
of freedom $\psi_u$ is even. Then $\det H=\pm 1$ or $\det \epsilon D
=\pm1$ implies $\det E=1$.

\medskip\noindent
{\bf Real euclidean action}

First we consider a euclidean quadratic fermionic action
\be\label{5.2}
S_E=\bar\psi_uA_{uv}\psi_v.\ee
Invariance under $\bar\theta$ requires $\bar{\theta}(A)=A$ according to
\be\label{5.3}
\bar\theta(S_E)=(\bar\theta\psi)_vA^*_{uv}(\bar\theta\bar\psi)_u
=\bar\psi_u(\bar\theta(A))_{uv}\psi_v=S_E.\ee
For the transformations $\theta_M$ and $\theta$ one has
\ba\label{5.4a}
\theta_M(A)&=&D^\dagger A^\dagger D^{-1},\nonumber\\
\theta(A)&=&SH^\dagger A^\dagger H^{-1}S,\ea
where $S$ operates a time reflection, i.e. $S(\psi(q))=
\psi(\theta(q)),\ S^2=1$. In consequence, the functional integration
(\ref{2.1a}) yields a real result if $S_E$ is $\bar\theta$
invariant
\be\label{5.5}
\det (\bar\theta(A))=\det A^\dagger=(\det A)^*=\det A.\ee
(We employ here the fact that for any variable $\psi(q)$ there
 is a variable~ $\psi(\theta(q))$ ~such that the number of
variables ~$\psi_u$ ~is even and therefore~ $\det(H^\dagger H^{-1})=1,~ \det D^\dagger D^{-1}=1$. The partition function 
$Z=\det(-A)=\det A$ (\ref{2.2AA}) is therefore real.)

For a real euclidean action $S_E$ the reality of $Z$ can be generalized for fermionic
actions containing arbitrary polynomials of $\psi,\bar{\psi}$. We expand $S_E$ in powers of $\psi$ and
$\bar\psi$ and denote by $b_i^{(n)}$ the coefficients 
of such an expansion, with $n$ the number of Grassmann variables
for a given term. (On the quadratic level the
$b_i^{(2)}$ correspond to $A_{uv}$ and to coefficients of bilinears
of the type $\psi\psi$ and $\bar\psi\bar\psi$.) We consider $Z$ as a functional of $b_i^{(n)}$ and first establish the relation
\begin{equation}\label{7.6a}
\big(Z[b^{(n)}_i]\big)^*=Z[\bar{\theta}(b^{(n)}_i)]
\end{equation}
with $\bar{\theta}(b^{(n)}_i)$ defined after eq. (\ref{5.6}). This relation follows essentially from the $\bar{\theta}$-invariance of the functional measure and does not yet assume the reality of $S$. 

In order to
simplify the discussion we consider here a finite number of degrees
of freedom $\psi_u,\bar\psi_u$
-- the limit to infinitely many degrees of freedom can be taken
at the end. By definition of the Grassmann integration the ``functional''
integration yields a polynomial in the (complex) quantities
$b_i^{(n)}$ with real coefficients
\be\label{5.5a}
Z[b^{(n)}_i]=\int\prod_{u'}(d\psi_{u'}d\bar\psi_{u'})e^{-S_E[\psi,\bar\psi;
b_i^{(n)}]}=P(b_i^{(n)}).\ee
Using the fact that $\bar\theta$ involves a complex conjugation
of the $b_i$ we can write
\be\label{5.6}
\bar\theta(S_E[\psi,\bar\psi;\ b_i^{(n)}])=S_E[\chi,\bar\chi;
\eta^{(n)}(b_i^{(n)})^*]=S_E[\psi,\bar\psi;\ \bar{\theta}(b^{(n)}_i)],
\ee
with $\chi_u=(\bar\theta\psi)_u,\bar\chi_u=(\bar\theta\bar\psi)_u$.
The factor $\eta^{(n)}$ results from the total reordering of
Grassmann variables, $\eta^{(n)}=+1$ for $n=0,1\ mod\ 4,,
\ \eta^{(n)}=-1$ for $n=2,3\ mod\ 4$. The last equation defines $\bar{\theta}(b^{(n)}_i)$ in close analogy to eq. (\ref{5.4}), i.e. by appropriate multiplication of $(b^{(n)})^*_i$ with matrices $D$ or $HS$. 

We now perform a variable
transformation $\psi\to\chi,\bar\psi\to\bar\chi$
in the functional integral and exploit the fact that the Jacobian
is unity (cf. eq. (\ref{5.1})). One therefore finds
\be\label{5.7}
P\big(\bar{\theta}(b_i^{(n)}\big)=P(\eta^{(n)}(b^{(n)}_i)^*).\ee
If the total number of $\psi_u$ is $2m$ $(u=1...2m)$, only those
products of $b_i^{(n)}$ contribute in $P$
for which the sum of all $n$ equals $4m$. (This follows from
a Taylor expansion of the exponential and (\ref{2.1}).) For
$S_E$ containing only terms with an even number of Grassmann
variables one always needs an even number of $b^{(n)}$ with
$n=2\ mod\ 4$.
One infers
\be\label{5.8}
P(\eta^{(n)}b^{(n)}_i)=P(b^{(n)}_i)~,~
P\big(\bar{\theta}(b_i^{(n)})\big)=P((b_i^{(n)})^*)=(P(b_i^{(n)}))^*,
\ee
where the last identity exploits that the coefficients
in the polynomial $P(b_i^{(n)})$ are real. If these properties are preserved in the limit of infinitely many degrees of freedom,  $Z[b^{(n)}_i]$ must be an even functional of the $b^{(n=2mod 4)}_i$. Eq. (\ref{5.8}) then establishes the relation (\ref{7.6a}), with $\big(Z[b^{(n)}_i]\big)^*=Z^*\big[(b^{(n)}_i)^*\big]=Z\big[(b^{(n)}_i)^*\big].$

In absence of other fields a real fermionic action requires $\bar{\theta}(b^{(n)}_i)=b^{(n)}_i$. The relation (\ref{7.6a}) then directly implies that the functional integral over a
real $S_E$ with only even numbers of $\psi$ or $\bar\psi$ is
real 
\begin{equation}\label{XAB1}
Z=\int {\cal D}\psi {\cal D}\bar{\psi} \exp (-S_E)=Z^*.
\end{equation}
Typical fermionic actions only involve even numbers of
fermions -- for example, this is a necessary requirement
for Lorentz-invariance. Osterwalder-Schrader positivity for
euclidean fermions implies then immediately the reality\footnote{We observe that functional integration over fermionic actions with even and odd numbers of
fermions would not give a real result even in case of a real
action. For a given $m$ one may replace, for example, three
terms with two fermions $((\eta^{(2)})^3=-1)$ by two terms with three
fermions $((\eta^{(3)})^2=1)$ such that $P$ would necessarily
have an imaginary part. Our discussion does not cover
the case that the $b_i^{(n)}$ are Grassmann variables themselves
as, for example, linear source terms. A generalization to this case is straightforward.} of the
fermionic functional integral. We note that we have not assumed
continuity of space-time such that this observation applies
directly to lattice theories. 

It is straightforward to show that the expectation values of all $\bar{\theta}$-even operators are real
whereas the ones for $\bar{\theta}$-odd operators are purely imaginary. (We
assume that the operators involve an even number of Grassman variables.) Let us
denote the even and odd operators by $O^k_R$ and $O^l_I,~ \bar{\theta}(O_R)=O_R,~
\bar{\theta}(O_i)=-O_I$. One obtains (with real $f_k,f_l)$
\begin{eqnarray}\label{XAB2}
\langle f_kO^k_R+if_lO^l_I\rangle
&=&Z^{-1}\int{\cal D}\psi {\cal D}\bar{\psi}(f_kO^k_R+if_lO^l_I)\exp(-S_E)\nonumber\\
&=&\langle f_kO^k_R+if_lO^l_I\rangle^*
=f_k\kl O^k_R\kr -if_l\kl O^{l}_I\kr^*.
\end{eqnarray}
(We may consider $-ln(f_kO^k_R+if_lO^l_I)$ as an additional part of a real fermionic
action.) By suitable linear combinations one finds for an arbitrary operator
\begin{equation}\label{XAB3}
\langle\bar{\theta}(O)\rangle=\langle O \rangle^*.
\end{equation}

\medskip\noindent
{\bf Fermions and bosons}

We can extend our discussion to the case where the $b^{(n)}_i$ depend on additional bosonic fields $\phi$. If the bosonic fields also transform under $\bar{\theta}$ we have to replace in eq. (\ref{5.6}) 
$(b^{(n)}_i)^*$ by $\bar{b}^{(n)}_i=b^{(n)*}_i\big(\bar{\theta}(\phi)\big)$ and correspondingly generalize the meaning of $\bar{\theta}(b^{(n)}_i)$. Repeating the discussion above generalizes eq. \eqref{7.6a}, 
\begin{equation}\label{7.10A}
Z\Big[\bar{\theta}\big(b^{(n)}_i(\phi)\big)\Big]
=Z\Big[\big(b^{(n)}_i\big)^*\big(\bar{\theta}(\phi)\big)\Big]
=Z^*\Big[\big(b^{(n)}_i\big)^*\big(\bar{\theta}(\phi)\big)\Big].
\end{equation}
Typically, for the euclidean reflection $\theta$ the transformation of a scalar field with positive $\theta$-parity also involves a reflection of the time coordinate $\theta\big(\phi(x)\big)=\phi^*(\theta x)$. (In this case a Yukawa coupling $b^{(2)}(\phi)=h\phi(x)$ implies $\big(b^{(2)}\big)^*\big(\bar{\theta}(\phi)\big)\equiv h^*\phi^*\big(\theta(x)\big).$) A real action obeys again $\bar{\theta}\big(b^{(n)}_i(\phi)\big)=b^{(n)}_i(\phi)$ and therefore
\be\label{7.10B}
Z^*\big[\bar{\theta}(\phi)\big]=Z[\phi].
\ee
In other words, if we expand $Z[\phi]$ in terms of polynomials of $\phi$ and $\phi^*$ with complex coefficients $c$, the replacement $\phi\to\bar\theta(\phi),\phi^*\to\bar\theta(\phi^*),c\to c^*$ leaves $Z[\phi]$ invariant. For a conjugation $\theta$ with $\theta\big(\phi(x)\big)=\phi^*(\theta x)$ we can write eq. \eqref{7.10B} as 
\be\label{8.15A1}
\Big(Z\big[\phi(x)\big]\Big)^*=
Z^*\big[\phi^*(x)\big]=Z\big[\phi(\theta x)\big].
\ee
Eq. \eqref{8.15A1} can be used for establishing properties of the effective bosonic action $S_{eff}[\phi]$ which obtains by integrating out the fermionic fields
\be\label{5.18}
Z[\phi]=\exp(-S_{eff}[\phi])=\int D\psi D\bar\psi\exp(-S[\phi,\psi,\bar\psi]).
\ee
In general, the fermionic functional integral with a real action can lead to an
imaginary part of $Z$ if $\bar\theta\big(\phi^*(x)\big)\neq\phi(x)$,
\ba\label{5.20}
Im\ \exp(-S_{eff}[\phi])&=&-\frac{i}{2}\big\{\exp\big(-S_{eff}[\phi]\big)-\exp\big(-S^*_{eff}[\phi^*]\big)\big\}\nonumber\\
&=&-\frac{i}{2}\big \{\exp(-S_{eff}
[\phi])-\exp(-S_{eff}[(\bar\theta(\phi^*))^*]\big\}.\ea
As an example we may consider euclidean signature with a real scalar field obeying $\theta\big(\phi(x)\big)=\phi(\theta x)$. For a real effective action, $S^*_{eff}[\bar\theta(\phi)])=S_{eff}[\phi]$, a possible local term in the effective action with an odd number of time derivatives must have an imaginary coefficient. An example for such a local term is ($\alpha$ real).
\be\label{8.15A}
S_{eff}=i\alpha \int_x\epsilon^{\mu\nu\rho\sigma}\partial_\mu\phi\partial_\nu\phi\partial_\rho\phi\partial_\sigma\phi +\dots
\ee
We may also evaluate $Z[\phi]$ for a given time odd scalar field configuration $\phi(\theta x)=-\phi(x)$, with 
\be\label{8.15B}
Im Z[\phi]=-\frac{i}{2}\big(Z[\phi]-Z[-\phi]\big). 
\ee
The imaginary part of $Z[\phi]$ does not vanish if $S_{eff}[\phi]$ contains terms that are odd in $\phi$.  

\medskip\noindent
{\bf Real expectation values}

Often one is interested in the expectation value of some operator
\be\label{5.21}
<O[\phi,\psi,\bar\psi]>=\int D\phi D\psi D\bar\psi O[\phi,\psi,\bar
\psi]\ \exp(-S[\phi,\psi,\bar\psi]).
\ee
Let us assume that the functional measure $\int D\phi$ is invariant
under $\bar\theta$ in the sense that it can be written in the
form $\prod\int d\phi d(\bar\theta(\phi))$. For real $S$ one concludes
\be\label{5.22}
<\bar\theta(O[\phi,\psi,\bar\psi])>~=~<O[\phi,\psi,
\bar\psi]>,
\ee
such that the expectation values of all $\bar\theta$-odd operators
vanish.  We can therefore restrict the discussion to $\bar\theta$-even operators. 
For the case $\bar\theta(\phi)=\phi^*$ the imaginary
part of the fermionic integration vanishes and the remaining
bosonic integral is real, implying $<O>=<O>^*$. This situation
is typically realized for $\bar\theta\equiv \theta_M$. On the
other hand, for an euclidean signature the action may only be invariant
under $\theta$ and not under $\theta_M$. Then the time reflection
implies $\bar\theta(\phi)\not=\phi^*$. In this case we further assume
that the bosonic functional measure can be written as a product $\prod
\int d\phi d(\bar\theta(\phi))^*$. We distinguish between real operators
obeying
\be\label{5.23} O_R[\phi,\psi,\bar\psi]=\bar\theta O_R[\phi^*,
\psi,\bar\psi]\ee
and imaginary operators for which (\ref{5.23}) involves a minus
sign. From $\bar\theta(\phi^*)=(\bar\theta(\phi))^*$ and \eqref{7.10B} 
one concludes
\be\label{5.24}
<\bar\theta O[\phi^*,\psi,\bar\psi]>=<O[\phi,
\psi,\bar\psi]>^*.
\ee
Real operators have therefore always real expectation values, whereas
the expectation values of imaginary operators are purely imaginary.
For real operators the imaginary parts (\ref{5.20}) of the 
contributions to the functional integral from $\phi$ and
$(\bar\theta(\phi))^*$ exactly cancel. They can therefore simply
be omitted and one can replace $\exp(-S_{eff})$ by its real part.

These properties have implications for euclidean lattice
simulations of gauge theories with fermions. For a computer
simulation the fermionic functional integral has to be performed
analytically and the result is in general not real. As long as
only the expectation values of real operators are computed, one
can simply omit the imaginary parts of the fermionic integration
and work with a real effective action defined by
\be\label{5.25}
\exp(-\bar S_{eff}[\phi])=Re\ \exp(-S_{eff}[\phi])=\frac{1}{2}
\{\exp(-S_{eff}[\phi])+\exp(-S_{eff}[(\bar\theta(\phi))^*])\}.
\ee

\vspace{-0.3cm}
\noindent
The result remains exact. More precisely, the l.h.s. of eq. \eqref{5.25} is a real, but not necessarily positive weight factor. The formal expression $\exp(-S_{eff})$ in eq. \eqref{5.18} stands for $Z[\phi]$. It is possible that the real part of $Z$ becomes negative. Examples for $\bar\theta$-even and real operators are $\phi(x)+\phi^*(\theta x),\phi^*(\theta x)\phi(x)$, or all hermitean fermion bilinears (see next section) which do not involve $\phi$.

\medskip\noindent
{\bf Bosonization}

Finally we notice that the positivity properties should be 
respected if one (partially) bosoni\-zes a fermionic theory. Bosonization 
typically replaces a fermion bilinear by a bosonic field. Let
us consider an euclidean signature with fermion bilinears
represented by Lorentz scalars $\phi$. If one wants to represent
the action of $\theta$ on $\phi$ by standard complex
conjugation $\theta(\phi(x))=\phi^*(\theta x)$, one has to make
sure that $\phi(x)+\phi^*(\theta x)$ corresponds to a real 
fermionic bilinear operator. For $d$ even and $a,b$ 
internal indices for different fermion species, the real scalar
bilinears are
\be\label{5.26}
\bar\psi_-^a(x)\psi_+^b(x)-\bar\psi_+^b(\theta x)\psi_-
^a(\theta x)\quad,\quad i(\bar\psi_-^a(x)\psi^b_+(x)+\bar\psi_+^b(\theta x)
\psi_-^a(\theta x)).
\ee
We emphasize that $\bar\psi\psi$ is an imaginary operator
whereas $\bar\psi\bar\gamma\psi$ is real. Operators of the type
$\bar\psi^a\psi^b, \bar\psi^a
\bar\gamma\psi^b$ have no definite reality property for
$a\not=b$. Corresponding to (\ref{5.26}) one
may bosonize the bilinear $\bar\psi_-^a(x)\psi_+^b(x)$
by a complex scalar field $\phi^{ab}(x)$. The complex
conjugated field $(\phi^{ab}(x))^*$ corresponds then
to $-\bar\psi_+^b(x)\psi_-^a(x)$, whereas 
$\bar\psi_+^a(x)\psi_-^b(x)$ is replaced by
$-(\phi^\dagger)^{ab}(x)$. The bilinears 
$i\bar\psi^a(x)\psi^b(x)$ and $\bar\psi^a(x)\bar
\gamma\psi^b(x)$ transmute into hermitean scalar fields.

\medskip\noindent
{\bf Reality for Minkowski signature}

Formally, a real quadratic Minkowski action $S_M$ is also sufficient to guarantee a
real partition function $Z=\int{\cal D}\psi\exp(iS_M)$. For $S_M=\bar{\psi}_u A^M_{uv}\psi_v$ we now have the relation
\begin{equation}\label{5.5A}
Z=\det(iA^M)=\det\big(i\bar{\theta}(A^M)\big)=
\det\big(i(A^M)^\dagger\big)=
\big(\det(-iA^M)\big)^*=\big(\det(iA^M)\big)^*.
\end{equation}
For the last relation we use once more the property that the number of spinors $\psi_u$
is even. Again, $Z^*=Z$. Furthermore, if the number of spinors $\psi_u$ is $4~mod~4$ one may use $\det(-iA^M)=\det(A^M)$. For an interacting theory this formal reality of $Z$ does not remain valid, however, in the presence of the poles characteristic for the momentum integrals with Minkowski signature. In this case it is destroyed by the non-hermitean regulator terms that have to be added to $S_M$ in order to make the momentum integrals well defined. As a result, the momentum integrals appearing in the computation of $\ln Z$ produce a factor $i$ such that the fluctuation contribution to $-i\ln Z$ is real, just as the classical part, $-i\ln Z_{cl}=S_M[\psi=0].$ 

As an example, we investigate a Yukawa coupling between spinors and a complex scalar field $\phi$ for $d=4$
\be\label{h14a}
-S_M=\int_xi\bar{\psi}\gamma^\mu\partial_\mu\psi+h(\bar{\psi}_R\phi\psi_L-\bar{\psi}_L\phi^*\psi_R).
\ee
With real $h$ and $\theta_M(\phi)=\phi^*$ the action is real with respect to $\theta_M,\theta_M(S_M)=S^\dagger_M=S_M$. Performing the Gaussian integral for the fermions yields formally for constant $\phi$, with $\qslash=\gamma^\mu q_\mu$,
\be\label{h14b}
\ln Z[\phi]=tr \int \frac{d^4q}{(2\pi)^4}\ln\left(-\qslash +h\phi\frac{1+\bar{\gamma}}{2}-h\phi^*\frac{1-\bar{\gamma}}{2}\right).
\ee
We are interested in the $\phi$ dependence of $\ln Z ~(\Omega_4$: four volume, $\rho=\phi^*\phi)$
\be\label{h14c}
\frac{1}{\Omega_4}\frac{\partial^2\ln Z}{\partial\phi\partial\phi^*}=2h^2\int_q\frac{q^2}{(q^2+h^2\rho)^2}=\tilde{\mu}^2(\rho).
\ee
In order to avoid discussing the complications of the UV-regularization we consider here the UV-finite derivative
\be\label{h14d}
\frac{\partial^2\tilde{\mu}^2}{\partial\rho^2}=12h^6\int_q\frac{q^2}{(q^2+h^2\rho)^4}=\frac{i}{48\pi^2h^2\rho}.
\ee
Even though the $q-$ integral is formally real the momentum integration is well defined only by adding to $q^2$ a small imaginary part, $q^2\rightarrow q^2-i\epsilon, \epsilon>0$. The poles at $q_0=\pm\sqrt{\vec{q}^2+h^2\rho-i\epsilon}$ are now away from the real axis and the integration yields an imaginary result. This demonstrates that the formally real expression $\ln\det A_M$ turns actually out to be imaginary.

More generally, we expect that functional integration over fermions with a real Minkowski-action $S_M$ yields a real effective action\footnote{Strictly speaking, this holds provided we choose a regularization which is consistent with this property.} 
\be\label{h14e}
S_{M,eff}[\phi]=-i\ln Z[\phi].
\ee
and therefore purely imaginary $\ln Z$. Indeed, if the Minkowski-action $S_M$ obtains from the euclidean action $S_E$ by analytic continuation $(-iS_M=S_E$ (anal. contd.)) we may exploit analyticity also for the effective action. As demonstrated above, a $\theta$-invariant $S_E$ yields a real effective action $S_{eff}$. By analytic continuation one infers that $S_M$ is $\theta_M$-invariant. Also the analytic continuation of $S_{eff}$ yields $-iS_{M,eff}$. In turn, $S_{M,eff}$ should now be real with respect to $\theta_M$ (hermitean conjugation).

\section{Hermitean spinor bilinears}
\label{Hermiteanspinor}
\setcounter{equation}{0}
Let us now concentrate on hermitean spinor bilinears. In particular, from hermitean bilinears containing both $\psi$ and $\bar{\psi}$ one can easily construct real fermionic
actions obeying $\bar{\theta}(A)=A$ in eq. (\ref{5.2}). For $\theta_M$ and $D=D_1$ the action of $\theta_M$ reorders the $\gamma$-matrices according to
\be\label{5.9}
\theta_M(\bar\psi(q)\gamma^m\gamma^n...\gamma^p\psi(q))
=\bar\psi(q)\gamma^p...\gamma^n\gamma^m\psi(q).\ee
For even dimensions one also may use
\be\label{5.10}
\theta_M(\bar\psi(q)\gamma^m\gamma^n...\gamma^p
\bar\gamma\psi(q))=(-1)^{s+Q}\bar\psi(q)\gamma^p...\gamma^n\gamma^m
\bar\gamma\psi(q),
\ee
with $Q$ the total number of matrices $\gamma^m$. For $D=D_1$ the invariants
with respect to $\theta_M$ can then be constructed by multiplying
the following terms with real coefficients\footnote{The kinetic
term reads $-\int\frac{d^dq}{(2\pi)^d}\bar\psi(q)q_\mu\gamma^\mu
\psi(q)$.}
\be\label{5.11}
\bar\psi\psi\ , \ i^s\bar\psi\bar\gamma\psi\ ,\ \bar\psi\gamma^m
\psi\ ,\ i^{s-1}\bar\psi\gamma^m\bar\gamma\psi\ ,
\ i\bar\psi\Sigma^{mn}\psi~,~i^{s-1}\bar{\psi}\Sigma^{mn}\bar{\gamma}\psi.\ee
If, instead, we use $D=D_2$ the hermitean ($\theta_M$-invariant) bilinears are 
\begin{equation}\label{7.13A}
i\bar{\psi}\psi~,~i^{s-1}\bar{\psi}\bar{\gamma}\psi~,~\bar{\psi}\gamma^m\psi~,
~i^{s-1}\bar{\psi}\gamma^m\bar{\gamma}\psi~,~\bar{\psi}\Sigma^{mn}\psi~,
~i^s\bar{\psi}\Sigma^{mn}\bar{\gamma}\psi.
\end{equation}
In consequence, the kinetic term (\ref{2.9}) is invariant under $\theta_M$ for both $D_1$ and $D_2$. 

The corresponding bilinear transformation rules for $\theta$ are
given by
\ba\label{5.12}
&&\theta(\bar\psi(q)\gamma^m\gamma^n...\gamma^p\psi(q))
=\kappa\bar\psi(\theta q)\tilde\gamma^p...\tilde\gamma^n\tilde
\gamma^m\psi(\theta q),\nonumber\\
&&\theta(\bar\psi(q)\gamma^m\gamma^n...\gamma^p\bar\gamma\psi(q))
=(-1)^{s+1+Q}\kappa\bar\psi(\theta q)\tilde \gamma^p...\tilde\gamma^n
\tilde\gamma^m\bar\gamma\psi(\theta q),\ea
with
\be\label{5.13}
\tilde\gamma^0=-\gamma^0,\ \tilde\gamma^i=\gamma^i\quad {\rm for}
\quad i\not=0\ee
and 
\be\label{5.14}
\kappa=
\begin{cases}
1&{\rm for} H=H_1\\
(-1)^{Q+1}&{\rm for} H=H_2
\end{cases}.
\ee

For $H=H_2$ one finds among the
possible $\theta$-invariants with real coefficients $c_j$
(with $\gamma^i, i\not=0$)
\ba\label{5.15}
&&S_E=\int\frac{d^dq}{(2\pi)^d}\bar\psi(q)(c_jO_j)\psi(q)\nonumber\\
&&O_j=i\ ,\ i^s\bar\gamma\ ,\ i\gamma^0\
,\ \gamma^i\ ,\ q_\mu\gamma^\mu\ ,\ i^{s+1}\gamma^0
\bar\gamma\ ,\ i^s\gamma^i\bar\gamma\ ,...\ea
As it should be the list (\ref{7.13A}) of $\theta_M$-invariants for $s=1~,~D=D_2$ corresponds precisely\footnote{This correspondence holds modulo a factor factor $i$ for each $\gamma^0$ factor in the $\theta$-invariants} to the $\theta$-invariants (\ref{5.15}) for $s=0~,~H=H_2$, reflecting analytic continuation. For irreducible Weyl, Majorana or Majorana-Weyl spinors some of the bilinears vanish identically. This is discussed in detail in ref. \cite{W1}, where tables of allowed and forbidden bilinears, in particular mass terms, can be found. For euclidean spinors with Osterwalder-Schrader positivity a fermion mass term in our convention $(H=H_2)$ is either $im
\bar\psi\psi$ or $m\bar\psi\bar\gamma\psi$. 

It is worth
mentioning that the euclidean fermion number operator is given by
\be\label{5.16}
N_\psi=i\int\frac{d^{d-1}q}{(2\pi)^{d-1}}\bar\psi(q)\gamma^0
\psi(q)=\int\frac{d^{d-1}q}{(2\pi)^{d-1}}\psi^\dagger (q)\psi(q)\ee
(cf. eq. (\ref{4.24})). Invariance of $N_\psi$ with respect
to $\theta$ guarantees that the fermionic functional integral
results in a real determinant even in presence of a nonvanishing
chemical potential associated to $N_\psi$. Indeed, the analytic continuation of a chemical potential from Minkowski to euclidean signature is straightforward in our formalism. For a Minkowski signature $(s=1)$ the chemical potential contributes to the Minkowski action $S_M$ a piece
\be\label{9.10}
S^{(\mu)}_M=\mu\int d^dx\psi^\dagger\psi.
\ee
We use a complex structure with $\epsilon D_2=\gamma^0_M$ as in eq. \eqref{eu2}. In the vielbein formalism the corresponding piece in the action $S_E$ reads then
\be\label{9.11}
S^{(\mu)}_E=\mu\int d^d xe\bar\psi\gamma^0_M\psi,
\ee
with $e=i$ for Minkowski signature. 

Analytic continuation replaces in eq. \eqref{9.11} $e=1$, according to the euclidean signature $s=0$. (Note that no inverse vielbein appears in $S^{(\mu)}$, in contrast to the term containing a time derivative. The analytic continuation of the combination $-i\partial_t-\mu$ for $s=1$ reads therefore $\partial_t-\mu$ for $s=0$, as appropriate for the Matsubara formalism where $t$ is usually denoted by $\tau$.) Using further $\gamma^0_M=-i\gamma^0_E$ results for $s=0$ in the euclidean action 
\be\label{9.12}
S^{(\mu)}_E=-i\mu\int d^dx\bar\psi\gamma^0_E\psi.
\ee
According to eq. \eqref{5.15} this piece is real with respect to the involution $\theta$ for $H=H_2=\gamma^0_M=-i\gamma^0_E$. We conclude that the reality of a euclidean action for euclidean signature is not changed if a piece containing the chemical potential is added. This statement refers to the complex conjugation based on $\theta$.

The action \eqref{9.12} is antihermitean with respect to the transformation $\theta_M$, both for $D_1$ and $D_2$, as can be seen from eqs. \eqref{5.11}, \eqref{7.13A}. In presence of a nonvanishing chemical potential the action does therefore not remain real with respect to $\theta_M$. As we have discussed in sect. \ref{Realfermionicactions}, reality with respect to $\theta_M$ would imply a real effective action after integrating out the fermions. Using reality with respect to $\theta$ implies for the effective action only eq. \eqref{8.15A1} or
\be\label{9.13}
\Big(Z\big[\phi_i(x)\big]\Big)^*=Z\big[\phi_i(\theta x)\big].
\ee
Here $\phi_i(x)$ stands for bosonic fields obeying $\theta\big(\phi(x)\big)=\phi^*\big(\theta x\big)$, while generalizations to more complex transformations of bosonic fields are straightforward. As we have discussed in sect. \ref{Realfermionicactions}, however, the imaginary part of $Z[\phi]$ can be omitted for the computation of expectation values of real operators. Real operators can be used, for example, for an exploration of the phase diagram of QCD. Even in presence of a chemical potential their expectation values can be computed using real (but not necessarily positive) weight factors
\be\label{9.14}
Z_R\big[\phi_i(x)\big]=\frac12 \Big(Z\big[\phi_i(x)\big]+Z\big[\phi_i(\theta x)\big]\Big).
\ee

\section{Generalized charge conjugation}
\label{Generalizedcharge}
\setcounter{equation}{0}

The properties of the discrete symmetries charge conjugation, parity
reflection and time reversal can depend on the dimension and the 
signature. Formulated in terms of the spinors $\psi$ and $\bar\psi$ the charge conjugation is found to be independent of the signature. Only if we use a complex structure the expression of charge conjugation in terms of $\psi$ and $\psi^*$ will also depend on the generalized signature. We call a generalized euclidean signature (E) the case
where $\bar\theta=\theta$ is used to define a complex conjugation. For $d$
even $(E)$ corresponds to $s$ even. Conversely, for a generalized
Minkowski signature $(M)$ one uses $\bar\theta=\theta_M$ for the
definition of the complex conjugation -- for $d$ even this corresponds to $s$
odd. 

\bigskip\noindent
{\bf Charge conjugation and Lorentz symmetry}

\medskip
Since $\psi_u$ and $\bar\psi_v$ are independent Grassmann variables the spinors $\psi(x)$ and $\bar\psi(x)$ correspond in a formal group theoretical sense to two distinct Dirac spinors. If we define a combined spinor $\hat \psi(x)=\big(\psi(x),\bar\psi(x)\big)$ this will always be a reducible representation of the Lorentz group, with $2^{\left[\frac d2\right]+1}$ components. We therefore expect that in the group theoretical sense there always exists a generalized Majorana-type constraint which can reduce $\hat\psi$ to a single Dirac spinor, thus eliminating half of the degrees of freedom. This extends to Weyl spinors if we define $\hat\psi_+=(\psi_+,\bar\psi_\pm)$ with a choice of $\bar\psi_\pm$ such that $\hat\psi_+$ contains two equivalent Weyl spinors. We can define generalized Majorana- and Majorana-Weyl spinors in this way in a group theoretical sense. 

To what extent these representations of the Lorentz group can be used for a description of physical particles will depend on the possible existence of a kinetic term in the action. This will not be realized for all generalized Majorana and Majorana-Weyl spinors. We will see in sect. \ref{Generalized Majorana} that for euclidean signature $(E)$ physical Majorana spinors exist for $d=2,3,4,8,9$ mod $8$, and Majorana-Weyl spinors for $d=2$ mod $8$. These are precisely the dimensions for which Majorana and Majorana-Weyl spinors exist for Minkowski signature \cite{W1}. They differ from the dimensions for which euclidean Majorana and Majorana-Weyl spinors exist in an algebraic sense as ``real representations'' \cite{W1}. We also emphasize that the compatibility of the Majorana constraint with complex conjugation depends on the choice of the involution $\bar\theta$, typically $\theta_M$ for $(M)$ and $\theta$ for $(E)$.

A generalized charge conjugation ${\cal C}_W$ is a map from $\psi$
to $\bar\psi$ which commutes with the
Lorentz transformations. Expressed in the basis $\psi,\bar{\psi}$ it is purely a map in the space of spinors. In contrast to the transformations $\theta$ or $\bar{\theta}$ it does not involve an additional complex conjugation of the coefficients in the Grassmann algebra. Thus ${\cal C}_W$ constitutes a symmetry if the action is invariant. We emphasize that the charge conjugation makes no use of the complex structure. This observation will be crucial for the definition of Majorana spinors in the next section. Majorana spinors will be associated to eigenstates of ${\cal C}_W$. 

It is convenient to represent ${\cal C}_W$ by a matrix acting on $\hat\psi(x)$
\be\label{6.1}
{\cal C}_W\hat\psi(x)=
{\cal C}_W
\left(\begin{array}{c}
\psi(x) \\ \bar\psi(x)
\end{array}\right)
=\left(\begin{array}{ccc}
0&,& W_1 \\ W_2&,& 0
\end{array}\right)
\left(\begin{array}{c}
\psi(x) \\ \bar\psi ( x)
\end{array}\right)
=\left(\begin{array}{c}
\psi^c(x)\\ \bar\psi^c(x)
\end{array}\right).
\ee
Corresponding to the convention (\ref{4.13}), ${\cal C}_W$
involves an additional reflection in momentum space, i.e. $\psi(q)\to
W_1\bar\psi(- q)$.  In momentum space the combined spinor is therefore defined as $\hat\psi(q)=\big(\psi(q),\bar\psi(-q)\big)$. Commutation with the Lorentz transformations,
\be\label{6.2}
[{\cal C}_W,\ \hat\Sigma]=0\qquad,\qquad
\hat\Sigma=-\frac{1}{2}\epsilon_{mn}
\left(\begin{array}{ccc}
\Sigma^{mn}&,&0\\0&,&-(\Sigma^{mn})^T
\end{array}\right),
\ee
requires
\be\label{6.3}
W_1^{-1}\Sigma^{mn}W_1=W_2\Sigma^{mn}W_2^{-1}=-(\Sigma^{mn})^T.
\ee
This guarantees that $\psi^c$ and $\psi$ transform identically under Lorentz transformations, i.e.
\be\label{9.3A}
\delta\psi^c=-\frac12\epsilon^{mn}\Sigma_{mn}\psi^c.
\ee
At this stage it is always possible to choose ${\cal C}_W$ as an involution, in which case 
\be\label{9.3B}
{\cal C}^2_W=1~,~W_2=W^{-1}_1.
\ee
We will below also consider more general possibilities for ${\cal C}_W$.

\bigskip\noindent
{\bf Invariance of kinetic term}

\medskip
Invariance of the fermion kinetic term $-\Sigma_q	\bar
\psi(q)\gamma^\mu q_\mu\psi(q)$ necessitates
\be\label{6.4}
W_2^T\gamma^mW_1=(\gamma^m)^T.\ee
We emphasize that a generalized charge conjugation obeying eq. \eqref{6.3} can be defined
without the condition (\ref{6.4}). The latter should be viewed
as a condition for a dynamical theory of fermions of the standard
type.
We concentrate here on the case ${\cal C}^2_W=\epsilon_W=\pm1$ where
\be\label{6.x/8.4a}
W_1=(C^T)^{-1},\quad W_2=\epsilon_W C^T,
\ee
and
\be\label{6.x}
C^T\Sigma^{mn}(C^T)^{-1}=-(\Sigma^{mn})^T,
\ee
such that eq. (\ref{6.4}) results in the condition
\be\label{6.5}
C\gamma^m(C^T)^{-1}=\epsilon_W(\gamma^m)^T.
\ee
(We omit the possible alternative for even dimensions
where ${\cal C}^2_W=\pm\bar\gamma$.) We will see in sect. \ref{Generalized Majorana} that eq. \eqref{6.5} restricts the choice of $\epsilon_W$, such that eqs. \eqref{9.3B} and \eqref{6.5} are not compatible for all dimensions. In turn, this will restrict the dimensions for which physical Majorana spinors can be defined. 

A simultaneous solution of the conditions (\ref{6.3})
and (\ref{6.4}) obtains for
\ba\label{6.6}
C^T&=&\delta C,\quad \delta^2=1,\nonumber\\
C\gamma^mC^{-1}&=&\epsilon_W\delta(\gamma^m)^T,\nonumber\\
C\Sigma^{mn}C^{-1}&=&-(\Sigma^{mn})^T.
\ea
We use the labels \cite{W1} $C=C_1,\ \delta=\delta_1$ for
$\epsilon_W\delta=-1$ and $C=C_2,\ \delta=\delta_2$ for $\epsilon_W\delta=1 $, with $C_2=C_1\bar\gamma$ for $d$ even. The properties of the matrices $C_1$ and $C_2$ are summarized in appendix C. Consistency of the relation $\big((\gamma^\mu)^T\big)^\dagger=\big((\gamma^\mu)^\dagger\big)^T$ requires
\be\label{9.10A}
CDC^\dagger=\eta_DD^T,
\ee
with $\eta_D$ some phase. The values $\delta_{1,2}$ characterize the symmetry properties of $C$ and depend on the number of dimensions
\cite{5}, \cite{6}, \cite{W1}
\ba\label{6.8}
&&\delta_1=\left\{\begin{array}{ccc}
1&{\rm for}& d=6,7,8\ mod\ 8\\
-1& {\rm for}& d=2,3,4\ mod\ 8\end{array}\right.\nonumber\\
&&\delta_2=\left\{\begin{array}{ccc}
1&{\rm for}& d=2,8,9\ mod\ 8\\
-1& {\rm for}& d=4,5,6\ mod\ 8.\end{array}\right.
\ea
The matrix $C_1$ does not exist for $d=5$  mod $4$, and $C_2$ does not exist for $d=3$ mod $4$.
In even dimensions one finds from eq. (\ref{6.x}) or \eqref{6.6}
\be\label{6.10}
\bar\gamma^T=(-1)^{\frac{d}{2}}C^T\bar\gamma (C^T)^{-1}=(-1)^{\frac d2}C\bar\gamma C^{-1}.
\ee

For even $d$ there is an alternative  possibility to fulfill eq. (\ref{6.x}),
\be\label{6.7}
iC^T\bar\gamma\gamma^m(C^T)^{-1}=
\epsilon_W\delta(\gamma^m)^T,
\ee
and eq. (\ref{6.5}) follows from
\be\label{6.7a}
C=i\delta C^T\bar\gamma\quad, \quad \delta^2=1.
\ee
In this case one may use $C=\tilde C_1,\ \delta=\tilde \delta_1$ for
$\epsilon_W\delta=-1$ and $C=\tilde C_2,\ \delta=\tilde \delta_2$
for $\epsilon_W\delta=1$. For $d$ even the existence of the matrices $C_1,C_2,\tilde C_1,\tilde C_2$ obeying eqs. \eqref{6.6}, \eqref{6.7} is guaranteed by the fact that $(\gamma^m)^T,-(\gamma^m)^T~,~\gamma^m$ and $i\bar\gamma\gamma^m$ all obey the same defining relation for the Clifford algebra.

\bigskip\noindent
{\bf Charge conjugation for Weyl spinors}

\medskip
In order to understand the action of ${\cal C}_W$ on Weyl spinors 
we first introduce a matrix 
\be\label{6.11AA}
\hat\Gamma=\left(\begin{array}{ccc}
\bar\gamma&,&0\\
0&,&(-1)^{d/2}\bar\gamma^T\end{array}
\right),
\ee
which represents the action of $\bar\gamma$ on the eight-component spinor $\hat\psi=(\psi,\bar\psi)$. The eigenvectors to the eigenvalues $\pm1$ correspond
to the two inequivalent spinor representations. By the convention
(\ref{2.10})  we always have for block-diagonal $\bar\gamma$
\be\label{6.A}
\psi={\psi_+ \choose \psi_-}\quad,\quad\bar\psi={\bar\psi_- \choose
\bar\psi_+},
\ee
and we recall that $\bar\psi_-$ is in a representation
equivalent to $\psi_+$ for $d=4\ mod\ 4$, whereas $\bar\psi_+$
and $\psi_+$ are equivalent for $d=2\ mod\ 4$. From eq. (\ref{6.10})
one concludes 
\be\label{6.B}
[{\cal C}_W,\ \hat\Gamma]=0.
\ee
The generalized charge conjugation maps equivalent spinor 
representations in $\psi$ and $\bar\psi$ into each other.

The matrix $\hat\Gamma$ is, however, not compatible with
every complex structure. We demonstrate this for the 
complex structure $\theta_M$ and introduce 
\be\label{6.C}
\hat{\bar\gamma}=
\left(\begin{array}{ccc}
\bar\gamma&,&0\\0&,&D^T\bar\gamma D^*
\end{array}\right)=
\left(\begin{array}{ccc}
\bar\gamma&,&0\\0&,&-\bar\gamma^T
\end{array}\right).
\ee
For $d=2\ mod\ 4$ one observes compatibility with the complex
structure $\hat\Gamma=\hat{\bar\gamma}$, whereas for $d=4\ mod\ 4$ the matrices $\hat\Gamma$ and $\hat{\bar\gamma}$ differ and the mapping $\hat\psi\to \hat\Gamma\hat\psi$ cannot be expressed
by the rules $\psi\to\bar\gamma\psi,\ \bar\psi=\epsilon
D^T\psi^*\to\epsilon D^T(\bar\gamma\psi)^*$ which
rather correspond to $\hat\psi\to\hat{\bar\gamma}\hat\psi$.
For the transformation $\hat\psi\to\hat{\bar\gamma}\hat\psi$ one
finds 
\ba\label{6.11}
[{\cal C}_W,\hat{\bar\gamma}]=0&{\rm for}& d=2\ mod\ 4\nonumber\\
\{{\cal C}_W,\hat{\bar\gamma}\}=0&{\rm for}& d=4\ mod\ 4.
\ea
For the second solution (\ref{6.7}) in even dimensions
one finds
\be\label{7.4}
\bar\gamma^T=-C\bar\gamma C^{-1}=-C^T\bar\gamma(C^T)^{-1}
=(-1)^{d/2}C^T\bar\gamma(C^T)^{-1}.
\ee
One concludes that this solution is possible only for
$d=2\ mod\ 4$ where $[{\cal C}_W,\hat{\bar\gamma}]=0$. We will not consider this possibility further.

We emphasize that for the generalized
charge conjugation the properties (\ref{6.10})-(\ref{6.11})
depend only on the dimension and not on the signature (in contrast to 
the algebraic charge conjugation ${\cal C}$ discussed in \cite{W1}).
This is related to the fact that
the definition of ${\cal C}_W$ is based on mappings between $\psi$ and
$\bar\psi$ and therefore involves the properties of spinor
representations under transposition\footnote{The algebraic
charge conjugation ${\cal C}$ \cite{W1} is based on
a mapping $\psi\to\psi^*$ and therefore involves the properties of
representations under complex conjugation which depend on the signature.},
which are independent of the signature. This setting
guarantees a close analogy between Minkowski and
euclidean signature. In terms of $\psi$ and $\bar\psi$ we will use the same definition of charge conjugation both for Minkowski and euclidean signature. We also note that all relations of this section remain valid if $C$ is multiplied by an arbitrary phase, $C\to \eta_C C~,~|\eta_C|=1$.

\section{Generalized Majorana spinors}
\setcounter{equation}{0}
\label{Generalized Majorana}

Generalized Majorana spinors correspond to an identification of
$\psi(x)$ and $\pm{\cal C}_W(\psi(x))$. Therefore,
$\frac{1}{2}(1\pm{\cal C}_W)$ must be projectors, which requires
${\cal C}_W^2=1$. Then, for 
\be\label{7.1}
{\cal C}_W^2=1\quad,\quad \epsilon_W=1,
\ee
generalized Majorana spinors are defined as
\be\label{7.2}
\hat\psi_{M\pm}=\frac{1}{2}(1\pm {\cal C}_W)\hat \psi,\quad \hat\psi=
{\psi \choose \bar\psi}.
\ee
Furthermore,  since ${\cal C}_W$ commutes
with $\hat\Gamma$, one can have
generalized Majorana-Weyl spinors
which are both eigenstates of ${\cal C}_W$ and $\hat\Gamma$. We 
observe that the commutation of ${\cal C}_W$ with
the Lorentz transformations (\ref{6.x}) gives no restriction on
$\epsilon_W$. In a group-theoretical sense we can therefore
define generalized Majorana-Weyl spinors in all even dimensions.
They correspond to the irreducible spinor representations of the 
Lorentz group.

\bigskip\noindent
{\bf Physical Majorana spinors}

\medskip
The requirement of invariance of the spinor kinetic term
changes the picture. The value of $\epsilon_W$ depends
on $\delta$ and therefore only on the dimension and not
on the signature. The second eq. \eqref{6.6} and the definitions of $C_1,C_2$ imply
\ba\label{7.3}
\epsilon_W=-\delta_1 \text{ for }C=C_1,\nn\\
\epsilon_W=\delta_2 \text{ for }C=C_2.
\ea
In the following we will fix $\epsilon_W$ by eq. \eqref{7.3}. The sign $\epsilon$ in the definition of $\theta_M$ in eq. \eqref{2.16} will be identified with $\epsilon_W$ and we often drop the index $W$ from now on. Comparing eq. \eqref{7.3} and eq. (\ref{6.8}) we infer that 
an invariant kinetic term for Majorana spinors is not
possible for all dimensions. Independently of the signature ``physical'' generalized
Majorana spinors are possible only for  $d=2,3,4,8,9\ mod\ 8$.
Physical generalized Majorana-Weyl spinors exist for $d=2\ mod\ 8$. These physical Majorana spinors are similar to the ones used by H.~Nicolai in the context of euclidean supersymmetric theories \cite{NIC}.
 
For the solution (\ref{6.6}), $d=2,3,4,8,9\ mod\ 8,\ $ and
Minkowski signature the generalized Majorana spinors 
correspond to the standard formulation of Majorana
spinors as real representations of the Lorentz group \cite{W1}.
This is not the case anymore for euclidean signature. Indeed,
for $d=4$ the euclidean rotation group $SO(4)$ has no real
representations for Majorana spinors. A definition of Majorana spinors as real 
representations would then imply that four euclidean dimensions do not admit
Majorana spinors \cite{W1}, in contrast to the setting of this paper. Using the
definitions of sect. \ref{Generalizedcharge} the charge conjugate spinor obeys
\ba\label{7.6}
&&\psi^c(x)={\cal C}_W(\psi(x))=
(C^T)^{-1}\bar\psi(x),
\nonumber\\
&&\bar\psi^c(x)={\cal C}_W(\bar\psi(x))=\epsilon C^T\psi(x),
\ea
with $\epsilon$ given by eq. \eqref{7.3}. This definition will be used for all signatures. Majorana spinors identify $\psi^c$ and $\psi$. This is possible for $\epsilon=1$ which can be realized for $d=2,3,4,8,9$ mod $8$.

\bigskip\noindent
{\bf Compatibility with complex structure}

\medskip
We next investigate if the definition \eqref{7.6} is compatible with a generalized notion of complex conjugation which corresponds to the involution $\bar\theta$. In other words, we want to express the charge conjugate spinor $\psi^c$ in terms of the complex conjugate spinor $\psi^{**}$ as
\be\label{11.4A}
\psi^c(x)=B^{-1}_W\psi^{**}(\bar\theta x),
\ee
with $\bar\theta x=x\ (M)$ or $\bar\theta x=\theta x\  (E)$, depending on the generalized signature. (For the complex conjugation based on $\theta_M$ the generalized complex conjugate spinor $\psi^{**}$ corresponds to the usual notation of complex conjugation for fixed coordinates, that we denote by $\psi^*$.) Inserting the relations between $\psi^*$ and $\bar\psi$ \eqref{2.16} or \eqref{4.13}, the comparison of eqs. \eqref{7.6} and \eqref{11.4A} yields
\ba\label{7.7}
&&B_W=B=\epsilon(D^T)^{-1}C^T=\epsilon\delta D^*C,\quad(M),\nonumber\\
&&B_W=(H^T)^{-1}C^T=\delta H^*C. \hspace{1.7cm}(E).
\ea
The choice $H^*(s=0)=\epsilon D^*(s=1)$, which guarantees consistency with analytic continuation, implies that we actually use the same matrix $B_W$ for $s=0$ and $s=1$. 

For a Minkowski type signature $(M)$ the matrix $B$ can be identified with the matrix $B_1$ or $B_2$ discussed in appendix C. From   $(\gamma^\mu)^*=\big((\gamma^\mu)^\dagger)\big)^T$ one infers the relations
\ba\label{7.5AA}
&&(\Sigma^{mn})^*=B\Sigma^{mn}B^{-1},\nonumber\\
&&(\gamma^m)^*=\left \{\begin{array}{rll}
-\sigma B\gamma^mB^{-1}&{\rm for}&C=C_1\\
\sigma B\gamma^mB^{-1}&{\rm for}&C=C_2\end{array}\right.,
\ea
with $\sigma=1$ for $D=D_1$ and $\sigma=-1$ for $D=D_2$. Thus the matrix $B$ coincides (up to a possible phase) with the matrix $B_1$ in appendix $C$ if we choose for $C$ and $D$ the pairs $(C_1,D_1)$ or $(C_2,D_2)$. For the other possible pairs $(C_1,D_2)$ or $(C_2,D_1)$ it coincides with $B_2$. Both $B$ and $C$ are unitary, 
\be\label{XX}
B^\dagger B=1~,~C^\dagger C=1.
\ee

The involutive property $\big((\gamma^\mu)^*\big)^*=\gamma^\mu$ implies
\be\label{7.5AB}
B^*B=\epsilon_B=\pm 1.
\ee
The consistency of the complex structure with the Majorana constraint $\psi^c=\psi$ requires $\psi^{**}=B\psi$ and therefore $(\psi^{**})^{**}=B^*\psi^{**}=B^{*}B\psi=\psi$. This is possible only for $\epsilon_B=1$. The values of $\epsilon_B$ are discussed in appendix C. They depend on $d$ and $s$. In particular, for a Minkowski signature $s=1$ and a suitable choice of $B$ and $C$ one finds according to eq. \eqref{C.1} 
\be\label{11.7A}
\epsilon_B=\epsilon_W=\epsilon,
\ee
such that the charge conjugation is compatible with the complex structure. For this purpose we select $B_1$ and $C_2$ for $d=5,8,9$ mod $8$, while for $d=3,4,7$ mod $8$ we take $B_2$ and $C_1$. For $d=2,6$ mod $8$ the choice is arbitrary. The Majorana spinors for $d=2,3,4,8,9$ mod $8$ obey then 
\be\label{11.7B}
\psi^c=B^{-1}\psi^*=\psi.
\ee
This coincides with the definition in \cite{W1}. 

For euclidean signature $(E)$ the situation is different since the complex conjugation involves now also a time reflection (and replaces $\epsilon D^*\to H^*$ for $B_W$). In terms of complex conjugate spinors $\psi^{**}$ the
euclidean generalized Majorana spinor $\psi_M$ is now a non-local superposition
\be\label{7.8}
\psi_{M+}(x)=\frac{1}{2}\{\psi(x)+C^*H^T\psi^{**}
(\theta x)\}.
\ee
This should not obscure the property that in terms of
the basic spinor fields $\psi(x)$ and $\bar\psi(x)$ the 
generalized euclidean Majorana spinor is a perfectly local
object
\be\label{7.9}
\psi_{M\pm}(x)=\frac{1}{2}\{\psi(x)\pm(C^T)^{-1}\bar\psi(x)\}.
\ee
We will not discuss here in detail the compatibility of the charge conjugation with the complex structure based on $\theta$. For the most interesting case $s=0$ it is guaranteed by analytic continuation from $s=1$. 

\bigskip\noindent
{\bf Majorana spinors and group theory}

\medskip
The action of the generalized charge conjugation is
perhaps most apparent in a basis $\big((C^T)^{-1}=C^*\big)$
\be\label{7.10}
\hat\psi_s=
\left(\begin{array}{ccc}
1&,&0\\0&,&(C^T)^{-1}\bar\psi
\end{array}\right)
\hat\psi=\left(\begin{array}{c}
\psi\\(C^T)^{-1}\bar\psi 
\end{array}\right),
\ee
where the Lorentz transformations are represented by
a diagonal matrix $(\Sigma=-\frac{1}{2}\epsilon_{mn}\Sigma^{mn}$)
\be\label{7.11}
\hat\Sigma_s=
\left(\begin{array}{ccc}
1&,&0\\0&,&(C^T)^{-1}\end{array}\right)
\hat\Sigma
\left(\begin{array}{ccc}
1&,&0\\0&,&C^T\end{array}\right)=
\left(\begin{array}{ccc}
\Sigma&,&0\\0&,&\Sigma\end{array}\right).
\ee
It simply reads
\be\label{7.12}
{\cal C}_W(\hat\psi_s(x))=
\left(\begin{array}{ccc}
0&,&1\\\epsilon_W&,&0\end{array}\right)
\hat\psi_s(x),
\ee
such that for $\epsilon_W=1$ the upper and lower components of $\hat\psi_s$ 
are exchanged.
For $d$ even $\hat{\bar\gamma}$ and $\hat\Gamma$ take in this basis the form
\ba\label{7.13}
\hat{\bar\gamma}_s&=&
\left(\begin{array}{ccc}
1&,&0\\0&,&(C^T)^{-1}\end{array}\right)
\left(\begin{array}{ccc}
\bar\gamma&,&0\\0&,&-\bar\gamma^T\end{array}\right)
\left(\begin{array}{ccc}
1&,&0\\0&,&C^T\end{array}\right)
=\left(\begin{array}{ccc}
\bar\gamma &,& 0 \\0&,&\tau_s\bar\gamma
\end{array}\right)\quad,\quad 
\nn\\
\hat\Gamma_s&=&
\left(\begin{array}{ccc}
\bar\gamma&,&0\\0&,&\bar\gamma
\end{array}\right),
\ea
with $\tau_s=(-1)^{\frac{d}{2}-1}$. 
Since $\psi$ and $\bar\psi$ always contain equivalent
Lorentz representations, $\hat\psi_s$ is a reducible representation
which is decomposed into irreducible representations by suitable
projectors $\frac{1}{2}(1\pm {\cal C}_W)$ (and $\frac{1}{2}
(1\pm\hat\Gamma)$ for $d$ even). From a group-theoretical point of
view it is obvious that one can always define the transformation
(\ref{7.12}) with $\epsilon_W=1$ and perform the decomposition.

From a group theoretical point of view one still has to answer the question if a non-vanishing
kinetic term exists for only one irreducible representation
of the Lorentz group. This requires that a Lorentz vector is
contained in the symmetric product of two identical irreducible
spinor representations, and is the case for $d=2,3,9\ mod\ 8$.
For these dimensions invariant actions involving a single
irreducible generalized Majorana spinor $(d=3,9\ mod\ 8$) or an 
irreducible generalized Majorana-Weyl spinor $(d=2\ mod\ 8)$ can be 
formulated. For $d=4\ mod\ 8$ the kinetic term involves two
inequivalent spinor representations of the Lorentz group.
One can still use generalized Majorana-Weyl spinors, but
not a single one. The minimal setting for an invariant kinetic
term consists of two inequivalent generalized Majorana-Weyl
spinors, which can be combined into one Weyl spinor
or one generalized Majorana spinor.
Similarly, for $d=5,6,7\ mod\ 8$ the vector is in the
antisymmetric product of two equivalent irreducible spinors. A kinetic
term therefore requires at least two generalized Majorana spinors,
corresponding to a Dirac spinor for $d=5,7\ mod\ 8$ or a Weyl 
spinor for $d=6\ mod\ 8$. The restrictions from eq. \eqref{6.4} for the possible values of $\epsilon_W$ have a simple group theoretical origin. 

In conclusion, the algebraic notion of Majorana
spinors is based on the transformation
properties of Lorentz representations under complex conjugation
according to a classification into real, pseudo-real or
complex representations. This classification depends on the
signature \cite{W1}. The generalized  Majorana spinors are related
to the decomposition of $\psi+\bar\psi$ into irreducible 
Lorentz representations. This is independent of the signature,
and only distinguishes between $d$ even or odd.
The existence of physical generalized Majorana spinors is 
related to the decomposition of a symmetric product of two
irreducible representations. This depends on the dimension,
but not on the signature. The notion of physical Majorana spinors is therefore compatible with analytic continuation in arbitrary dimension $d$. For Minkowski signature, the 
algebraic notion of Majorana representations coincides with the physical
Majorana spinors. For euclidean signature, 
algebraic and physical Majorana spinors differ: the physical
Majorana spinors do not coincide with real representations
of $SO(d)$. They correspond to a single spinor representation, where the spinors $\bar\psi_\gamma$ are given by appropriate linear combinations of $\psi_\gamma$, rather than being independent variables. 

\section{Parity and time reversal}
\setcounter{equation}{0}
\label{Parityandtimereversal}
The Lorentz transformations contain the reflections of an even number of 
coordinates with equal signature. Possible additional discrete
symmetries involve a reflection of an odd number of coordinates. We
first consider parity in even dimensions with euclidean $(s=0)$
or Minkowski $(s=1)$ signature. It involves the
reflection of $d-1$ coordinates $x^i, \ i=1...d-1$, whereas
$x^0$ remains unchanged (with $\eta_{00}=-1$ for Minkowski
signature). We work here in a fixed coordinate system, where
the reflections act only on the spinor fields. 
On Dirac spinors a parity transformation can be defined as
\ba\label{8.1}
&&P_1(\psi(q))=\eta_P\gamma^0\psi(P(q)),\nonumber\\
&&P_1(\bar\psi(q))=\bar\eta_P(\gamma^0)^T\bar\psi(P(q)),
\ea
with $P(q^0)=q^0,\ P(q^i)=-q^i$. Invariance of the kinetic
term requires 
\be\label{8.2}\eta_P\bar\eta_P\eta_{00}=1.\ee
For $s=0,1$ this 
implies $\bar\eta_P=(-1)^s/\eta_P$. 

Using the connection (\ref{2.16}) between $\bar{\psi}$ and $\psi^*$ one finds for a Minkowski signature $(s=1)$
\be\label{10.AA}
P_1\psi^*(x)=\bar{\eta}_PD^*(\gamma^0)D^T\psi^*
\big(P(x)\big)=\sigma\bar{\eta}_P(\gamma^0)^*\psi^*\big(P(x)\big),
\ee
with $\sigma=+1$ for $D=D_1$ and $\sigma=-1$ for $D=D_2$.  Compatibility of $P_1$ with the complex structure requires
\be\label{10.AB}
P_1\psi^*(x)=\big(P_1\psi(x)\big)^*=\eta^*_P(\gamma^0)^*\psi^*\big(P(x)\big),
\ee
and therefore
\be\label{10.AC}
\bar{\eta}_P=\sigma\eta^*_P.
\ee
Combining with eq. (\ref{8.2}) this yields
\be\label{10.AD}
\eta^*_P\eta_P=\sigma\eta_{00}.
\ee
For $s=1$ the compatibility of the parity transformation (\ref{8.1}) with the complex structure requires the choice $D=D_2$ such that $\sigma=-1,\sigma\eta_{00}=1$. For a euclidean signature $(s=0)$ we use the same parity transformation as for $s=1$ in order to remain compatible with analytic continuation, i.e. 
$\eta^{(E)}_P\gamma^0_E=\eta^{(M)}_P\gamma^0_M$ or 
$\eta^{(E)}_P=-i\eta^{(M)}_P,\bar{\eta}^{(E)}_P=-i\bar{\eta}^{(M)}_P$.

Let us now concentrate on $D=D_2, \eta^*_P\eta_P=1,\bar{\eta}_P=(-1)^s\eta^*_P$. We still remain with the freedom of a phase in $\eta_P$. In order to fix $\eta_P$, we 
first impose $P^2_1\psi=\pm\psi$ or $\eta^2_P=\bar\eta_P^2=\pm1$. Next
we require that the parity reflection commutes with the generalized
charge conjugation ${\cal C}_W$,
\be\label{8.3}
[{\cal C}_W,P_1]=0.\ee
This allows us to define the action of parity also for generalized
Majorana spinors. Using (\ref{6.6}) this yields
\be\label{8.4}
\eta^2_P=\epsilon\delta\eta_{00}.\ee
In particular, for $d=4,\ s=1,\ C=C_1$ (or $\epsilon\delta=-1)$
one finds $\eta_P=-\bar\eta_P=\pm1,\ P^2_1\psi=-\psi$. For euclidean
signature $C=C_1$ implies an imaginary phase $\eta_P=-\bar\eta_P
=\pm i,\ P^2_1\psi=-\psi$. We choose $\eta^{(M)}_P=1,\bar{\eta}^{(M)}_P=-1,\eta^{(E)}_P=-i,\bar{\eta}^{(E)}_P=i$. The transformation of the fermion bilinears involving $\bar{\psi}$ and $\psi$ is independent of the phase $\eta_P$ and independent of the signature:
\ba\label{8.5}
P_1:&&\bar\psi_+i\gamma^\mu\partial_\mu\psi_+\leftrightarrow
\bar\psi_-i\gamma^\mu\partial_\mu\psi_-\quad,\quad
\bar\psi^a_+\psi^b_-\leftrightarrow \bar\psi_-
^a\psi_+^b,\nonumber\\
&&\bar\psi\psi\rightarrow\bar\psi\psi\quad,\quad\bar\psi\bar\gamma\psi
\rightarrow-\bar\psi\bar\gamma\psi.\ea

In even dimensions there is an alternative version of the parity transformation involving $\bar{\gamma}$
\ba\label{10.AE}
P_2\big(\psi(q)\big)&=&\eta_P\gamma^0\bar{\gamma}\psi\big(P(q)\big),\nonumber\\
P_2\big(\bar{\psi}(q)\big)&=&-\bar{\eta}_P\gamma^{0T}\bar{\gamma}^T\bar{\psi}
\big(P(q)\big).
\ea
The transformations $P_2$ and $P_1$ are related by a chiral transformation $P_2=P_1R_-$ where $R_-$ changes the sign of $\psi_-$ and $\bar{\psi}_-$ while $\psi_+,\bar{\psi}_+$ are invariant. The relations (\ref{8.2}) and (\ref{10.AC}) also hold for $P_2$. We conclude that for $s=1$ only the complex structure built on the choice $D=D_2$ is compatible with the parity transformation. Using the same phases $\eta_p,\bar\eta_p$ for $P_1$ and $P_2$ one finds $P^2_2=-P^2_1$, such that for $d=4,s=1,C=C_1$ one obtains $P^2_2=1$. 
With respect to $P_2$ the transformation properties of spinor bilinears are
\ba\label{??}
P_2:&&\bar{\psi}_+i\gamma^\mu\partial_\mu\psi_+\leftrightarrow \bar{\psi}_-i\gamma^\mu
\partial_\mu\psi_-~,~
\bar{\psi}^a_+\psi^b_-\leftrightarrow-\bar{\psi}^a_-\psi^b_+\nonumber\\
&&\bar{\psi}\psi\rightarrow-\bar{\psi}\psi~,~\bar{\psi}\bar{\gamma}\psi\rightarrow
\bar{\psi}\bar{\gamma}\psi.
\ea   
Since for $s=1,D=D_2$ the bilinear $\bar{\psi}\bar{\gamma}\psi$ is real (cf. eq. (\ref{7.13A})) it seems natural to use the parity transformation $P_2$. In this case a mass term $m\bar{\psi}\bar{\gamma}\psi$ with real $m$ conserves parity.

For an odd number of dimensions the parity reflection of $d-1$ coordinates
is contained in the continuous Lorentz transformations and needs not to be discussed separately. There still exist non-trivial reflections of an odd number of space coordinates. 

Generalized time reversal $T_W$ is a reflection of the
remaining $0$-component of the coordinates. This 
transformation also maps $\psi$ into $\bar\psi$ and vice versa. In
even dimensions one has\footnote{Note the additional minus sign
in momentum space connected to the definition (\ref{4.13}).}
\ba\label{8.6}
T_W(\psi(q))&=&\eta_T\gamma^0\bar\gamma(C^T)^{-1}\bar\psi(P(q)),
\nonumber\\
T_W(\bar\psi(q))&=&\bar\eta_T(\gamma^0)^T\bar\gamma^TC^T\psi(P(q)).
\ea
Invariance of the kinetic term requires
\be\label{8.7}
\bar\eta_T\eta_T=\epsilon\eta_{00}.
\ee
We may choose the phases such that $[{\cal C}_W, T_W]=0$,
implying
\ba\label{8.8}
\bar\eta_T&=&(-1)^{d/2}\delta\eta_T,\nonumber\\
\eta^2_T&=&(-1)^{\frac{d}{2}}\epsilon\delta\eta_{00}=(-1)^{\frac{d}{2}}
\eta^2_P.
\ea
For the convention (\ref{8.4}) the ``time
reflection'' $T_W$ anticommutes then with the parity
reflection, $\{T_W,P\}=0$, independent of the choice of the phases
$\eta_T$ and $\bar\eta_T$. The definition \eqref{8.6} for time reversal is not unique. One may use modified reflections, similar to the possibilities $P_1$ and $P_2$ for parity

One observes that the combination $T_W{\cal C}_W$ is a mapping
$\psi\to\psi$,
\ba\label{8.9}
T_W{\cal C}_W(\psi(q))&=&\epsilon\eta_T\gamma^0\bar\gamma\psi(-P(q)),
\nonumber\\
T_W{\cal C}_W(\bar\psi(q))&=&\bar\eta_T(\gamma^0)^T\bar\gamma^T
\bar\psi(-P(q)).
\ea
In euclidean space this combination acts similarly to the
reversal of any other ``spacelike'' coordinate. Finally, one 
obtains for the generalized CPT-transformation for even dimensions $(P=P_1)$
\ba\label{8.10}
PT_W{\cal C}_W(\psi(q))&=&\epsilon\eta_P\eta_T\eta_{00}\bar\gamma\psi(-q),
\nonumber\\
PT_W{\cal C}_W(\bar\psi(q))&=&\bar\eta_P\bar\eta_T
\eta_{00}\bar\gamma^T\bar\psi(-q).
\ea

For euclidean signature a combination of Lorentz rotations
with angle $\pi$ in the (01)(23)... planes results in
\be\label{8.11}
\psi(q)\to (i)^{\frac{d}{2}}\bar\gamma\psi(-q)\quad,\quad
\bar\psi(q)\to(-i)^{\frac{d}{2}}\bar\gamma^T\bar\psi(-q).
\ee
One concludes that for even $d$ and a euclidean signature
the combined transformation $PT_W{\cal C}_W$ is a pure
$SO(d)$ transformation. Invariance of the action under CPT
is therefore a simple consequence of the Lorentz symmetry!
On the other hand, for odd dimensions the reflection $PT_WC_W$ as
well as $T_WC_W$ are not contained in the continuous Lorentz
transformations.

\section{Continuous internal symmetries}
\setcounter{equation}{0}
\label{Continuousinternalsymmetries}
We discuss here general continuous global symmetries in even dimensions
which commute with the Lorentz transformations. In the basis (10.10)
$\hat\psi_s=(\psi^a, (C^T)^{-1}\bar\psi^a)$,
with $a=1...N$ internal indices, they are represented by regular
complex matrices
\be\label{9.1}
{\cal A}(\hat\psi_s)=\hat A_s\hat\psi_s
\quad,\quad [\hat A_s,\hat\Sigma_s]=0.
\ee
This implies that $\hat A_s$ also commutes with $\hat\Gamma_s$ and therefore does not mix inequivalent Lorentz representations. We choose
a convention with $\bar\gamma=\bar\gamma^T=diag(1,-1)$ and define for $d=4$ mod $4$
\be\label{9.2}
\hat\psi_+(q)={\psi_+(q) \choose (C^T)^{-1}\bar\psi_-(-q)},
\quad \hat{\bar\psi}_+={(C^T)^{-1}\bar\psi_+(q) \choose
\rho\psi_-(-q)},
\ee
with $\rho=\pm1$ or $\pm i\bar\gamma$. Taking $\rho=-\delta_1$ or
$\rho=\delta_2$ for the solution (\ref{6.6}) with $C=C_1$ or
$C_2$, respectively, and $\rho=-i\tilde\delta_1\bar\gamma$
or $\rho=i\tilde\delta_2\bar\gamma$ for the solution (\ref{6.7a})
with $\tilde C_1$ or $\tilde C_2$, the kinetic term reads simply
\be\label{9.3}
S_{kin}=-\int\frac{d^dq}{(2\pi)^d}(\hat{\bar\psi}_+(q))^TC\gamma^\mu q_\mu\hat\psi_+(q).
\ee

In this basis $[\hat A_s,\hat\Sigma_s]=0$ implies
\be\label{9.4}
{\cal A}(\hat\psi_+)=A\hat\psi_+,\quad {\cal A}
(\hat{\bar\psi}_+)=(\tilde A^{-1})^T\hat{\bar\psi}_+,
\ee
with $A$ and $\tilde A$ complex $2N\times 2N$-matrices
not acting on the spinor indices of $\psi_+$ and
$(C^T)^{-1}\bar\psi_-$. Invariance of the kinetic term
requires $A=\tilde A$ and the existence of the inverse $A^{-1}$.
We conclude that for even dimensional Dirac spinors the most
general continuous invariance group of the kinetic term is
$SL(2N,C\!\!\!\!C)$. These transformations also leave a standard functional
measure invariant $(\det \hat A_s=1)$. The chiral transformations $U
(N)\times U(N)$ are a subgroup of $SL(2N,C\!\!\!\!C)$ with
\be\label{9.5}
A=
\left(\begin{array}{ccc}
U_+&,&0 \\ 0&,& U_-
\end{array}\right)
\quad, \quad U^\dagger_+U_+=U_-^\dagger U_-=1.
\ee
As we have discussed before, this subgroup is compatible with the complex
structure, whereas the transformations (\ref{9.4}), in general, are not.
Nevertheless, they are genuine symmetries of the kinetic term.

One is often interested in situations where part of the
continuous symmetries are gauged. One wants to know which additional
global symmetries are consistent with the gauge symmetries. We discuss
this topic here in a QCD-like setting with $N_F$ flavors where
$N=3N_F$, and the internal index is replaced 
by a double index $(a,i), \ a=1...N_F$. The gauge group $SU(3)_c$ acts
(infinitesimally) on the color indices $i=1...3$,
\be\label{9.6}
\delta_G\hat\psi_+^{ia}=\frac{i}{2}\theta_z(\hat\lambda_z)^{ij}
\hat\psi_+^{ja}\quad,\quad \hat\lambda_z=
\left(\begin{array}{ccc}
\lambda_z&,& 0 \\0&,&\lambda^*_z\end{array}\right),
\ee
with $\lambda_z$ the Gell-Mann matrices, $Tr(\lambda_y\lambda_z)
=2\delta_{yz},\ Tr\lambda_z=0$. 

Gauge invariance of the spinor kinetic term induces a ``quark-gluon
coupling'' (with $g$ the gauge coupling)
\be\label{9.7}
{\cal L}_g=\frac{1}{2}g\hat{\bar\psi}_+^T(x)C\gamma^\mu A^z_\mu
(x)\hat\lambda_z\ \hat\psi_+(x).
\ee
Global transformations leaving (\ref{9.7}) invariant obey (with
invariant\footnote{More generally, the global transformations could
be accompanied by
a global $SU(3)_c$ transformation of $A^z_\mu$.}  $A^z_\mu$)
\be\label{9.8}
A^{-1}\hat\lambda_z A=\hat\lambda_z\quad,\quad A=
\left(\begin{array}{ccc}
A_+&,&0 \\0&,&A^*_-\end{array}\right),
\ee
where $A_\pm$ act only on the flavor indices. Since the matrices
$A_\pm$ are arbitrary regular complex matrices in flavor space,
the global symmetries form the group $SL(N_F,C\!\!\!\!C)\times SL(N_F,
C\!\!\!\!C)$. This group 
is only a subgroup of the invariance group of the
kinetic term. It is, however, still larger than the usually considered
global flavor symmetry $U(N_F)\times U(N_F)$.

It is instructive to investigate the action of (\ref{9.8})
on the Lorentz scalars,
\be\label{9.8A}
{\cal L}_s=\bar\psi_-G^w_+\lambda_w\psi_++
\bar\psi_+G^w_-\lambda_w\psi_-,
\ee
where $w$ runs from 0 to 9 with
$\lambda_0=\sqrt{2/3}$ and $G^w_\pm$ arbitrary flavor matrices.
This  follows from the transformation $\psi_+\to A_+\psi_+,\ \psi_-
\to (A^\dagger_-)^{-1}\psi_-,\ \bar\psi_+\to(A^T_+)^{-1}\bar\psi_+,\bar
\psi_-\to A^*_-\bar\psi_-$. One finds
\be\label{9.8B} G^w_+\to A^\dagger_-
G^w_+A_+,\ G^w_-\to A^{-1}_+G^w_-(A^\dagger_-)^{-1}.
\ee
A flavor-singlet
mass term is invariant for $A^\dagger_-A_+=1$. For unitary $A_\pm
=U_\pm$ this reduces the symmetry to the vector-like flavor group
$U(N_F)$ with
$U_+=U_-$. For our more general transformation, a singlet
mass term reduces $SL(N_F,C\!\!\!\!C)\times SL(N_F,C\!\!\!\!C)$
to the ``diagonal'' subgroup $SL(N_F,C\!\!\!\!C)$
with $A^\dagger_-=A^{-1}_+$.

The fact that $SL(N_F,C\!\!\!\!C)\times SL(N_F,C\!\!\!\!C)$
transformations leave the covariant kinetic term and therefore
the classical QCD-action invariant does not yet guarantee that
they are symmetries of the full quantum theory. For this property
one also has to require invariance of the functional measure.
For a local gauge theory this is not trivial. The measure
has to be consistent with gauge symmetry which is not the case
for the ``naive'' $SL(N_F,C\!\!\!\!C)\times SL(N_F,C\!\!\!\!C)$
invariant measure. In fact, there is an incompatibility between
 part of the $SL(N_F,C\!\!\!\!C)\times SL(N_F,C\!\!\!\!C)$
transformations and gauge transformations, leading to an anomaly
similar to the usual axial anomaly. (The axial $U(1)$ transformations
are actually part of $SL(N_F,C\!\!\!\!C)\times SL(N_F,C\!\!\!\!C)$.)
This anomaly can be most easily seen by an investigation of
the anomalous 't Hooft multi-fermion interaction \cite{H}.
The bilinear $\tilde\varphi^{(1)}_{ab}=\bar\psi_{+b}\psi_{-a}$
transforms as $\tilde\varphi^{(1)}\to(A^\dagger_-)^{-1}
\tilde\varphi^{(1)}A^{-1}_+$ and $\det\ \tilde\varphi^{(1)}$
is invariant only for $\det A_+(\det A_-)^*=1$. From $\tilde\varphi_{ab}^{(2)}=-\bar\psi_{-b}
\psi_{+a},\ \tilde\varphi^{(2)}\to A_+\tilde\varphi^{(2)}
A^\dagger_-$ one sees that this is also the condition for the
invariance of $\det\ \tilde\varphi^{(2)}$. It is straightforward
to verify that the 't Hooft interaction is invariant
precisely under the transformations obeying 
$\det A_+(\det A_-)^*=1$.

Consider QCD with two massless quarks, $d=4,N_F=2$. The $SL(2,{\mathbbm C})$-transformations leaving the kinetic term and a singlet chiral condensate $\bar\psi\bar\gamma\psi$ invariant read
\ba\label{12.11}
\psi'_L&=&A\psi_L~,~\bar\psi'_L=\bar\psi_LA^{-1},\nn\\
\psi'_R&=&A\psi_R~,~\bar\psi'_R=\bar\psi_R A^{-1}.
\ea
The compact subgroup consists of the unitary transformations $SU(2)\times U(1)$. A typical transformation that is not contained in the unitary transformations is the scaling by a real factor $\lambda$
\be\label{12.12}
\psi'=\lambda\psi~,~\bar\psi'=\lambda^{-1}\bar\psi.
\ee
It is allowed since $\psi_u$ and $\bar\psi_v$ are independent Grassmann variables. The functional measure is clearly invariant under this transformation. On the other hand, the scaling \eqref{12.12} is not compatible with the complex structure. A priori the compatibility with the complex structure is not a restriction that has to be imposed. Symmetries not compatible with a complex structure are well known from other examples, as for a scalar field theory for a complex doublet which has the symmetry $SO(4)$ and not only $SU(2)\times U(1)$ - only the latter being compatible with the complex structure. 

The question arises what happens if the extended symmetries are spontaneously broken. The QCD-action for our case is invariant under
\ba\label{12.13}
\psi'_L&=&A\psi_L~,~\bar\psi'_L=\bar\psi_L A^{-1},\nn\\
\psi'_R&=&B\psi_R~,~\bar\psi'_R=\bar\psi_R B^{-1},
\ea
while the regular complex matrices $A,B$ obeying the constraint
\be\label{12.14}
\det (AB^*)=1
\ee
also leave the t'Hooft interaction invariant $(A=A_+,B=A_-)$. The matrix $A$ can be represented as 
\be\label{12.15}
A=\exp(i\alpha_w\tau_w)~,~\alpha_w\in {\mathbbm C},
\ee
with $w=0,1,2,3,\tau_0=1,\tau_k$ the Pauli matrices, and a similar representation $B=\exp (i\beta_w\tau_w)$. The difference to the unitary transformations arises from the complex parameters $\alpha_w,\beta_w$ instead of real ones. For infinitesimal transformations the condition \eqref{12.14} reads
\be\label{12.16}
\alpha_0=\beta^*_0.
\ee
The real part of $\alpha_0$ corresponds to the $U(1)$ transformations representing fermion number conservation, while $Im(\alpha_0)$ describes the scaling
\ba\label{12.17}
\psi'_L&=&\lambda\psi_L~,~\bar\psi'_L=\lambda^{-1}\bar\psi_L,\nn\\
\psi'_R&=&\lambda^{-1}\psi_R~,~\bar\psi'_R=\lambda\bar\psi_R.
\ea
The scaling \eqref{12.12} is anomalous.

We may try to represent the color singlet fermion bilinears with fermion number zero by complex scalar fields
\be\label{12.18}
\varphi^{ab}=\bar\psi_R^{a,i}\psi^{b,i}_L~,~
\tilde \varphi^{ab}=-\bar\psi^{a,i}_L\psi^{b,i}_R.
\ee
If we want to guarantee that the real part of $\varphi$ has a real expectation value, and the same for $i Im\varphi$, consistency requires the identification $\tilde\varphi=\varphi^\dagger$ (cf. sect. \ref{Realfermionicactions}). On the level of bosonic fields only those transformations that are consistent with $\tilde\varphi=\varphi^\dagger$ can be consistently realized. From eq. \eqref{12.13} one infers
\ba\label{12.19}
\varphi'&=&(B^T)^{-1}\varphi A^T~,~\tilde\varphi'=(A^T)^{-1}\tilde\varphi B^T,\nn\\
(\varphi^\dagger)'&=&A^*\varphi^\dagger(B^*)^{-1}.
\ea
Consistency of $\tilde\varphi=\varphi^\dagger$ requires $A^\dagger=A^{-1},B^\dagger=B^{-1}$. Only this subgroup of unitary transformations is relevant for the discussion of Goldstone bosons and we remain with the usual scenario. 

\section{Conclusions}
\label{Conclusions}
We have discussed a consistent mapping between Minkowski signature and euclidean signature for all types of spinors, in particular Majorana spinors and Majorana-Weyl spinors. This holds for arbitrary dimension $d$. In particular, no discontinuous ``doubling of degrees of freedom'' is needed when implementing the four-dimensional Majorana spinors for Minkowski signature in the corresponding euclidean quantum field theory. Analytic continuation is always possible.

Care has to be taken, however, for the proper choice of the complex structure. The hermitean conjugation for Minkowski signature has to be replaced by an involution $\theta$ for euclidean signature, which also involves the reflection of the time coordinate. We find a modulo two periodicity in signature for the choice of the complex structure.

We define generalized Majorana spinors based on on a general mapping between the independent Grassmann variables $\psi$ and $\bar\psi$. For Minkowski signature they coincide with the algebraic notion of Majorana spinors. For euclidean signature, however, the generalized Majorana spinors differ from the real spinor representations of the rotation group. The physical euclidean Majorana spinors are based on the notion of generalized Majorana spinors. They exist in precisely the same dimensions as for Minkowski signature. We also use this setting for a consistent implementation of the discrete symmetries parity, time reversal and charge conjugation for euclidean spinors and we discuss the issue of a ``real action'' and real expectation values. Taking things together, we realize a consistent description of spinors for arbitrary dimension and signature of the metric. 

The use of a consistent complex structure for euclidean signature based on $\theta$ defines new notions of ``reality'' and ``hermiticity. For euclidean signature the reality of the action is closely related to Osterwalder-Schrader positivity. For a real action the expectation values of hermitean fermion bilinears are real. Also the Grassmann functional integration over fermions yields a real result. This extends to the presence of bosonic ``background fields'' $\phi$ in the Grassmann functional as, for example, gauge fields. For a real fermion action the result of ``integrating out'' the fermions leads to an effective weight factor (effective action) for the bosons which is, in turn, real with respect to $\theta$. This means that the effective weight factor $Z[\phi]$ is invariant under a generalized complex conjugation which includes the complex conjugation of the bosons as well as a reversal of the time coordinate, $\phi(x)\to\theta\big(\phi(x)\big)=\phi^*(\theta x)$. A real bosonic action (with respect to $\theta$) is real in the usual sense (where all coordinates are kept fixed) if it contains only terms with an even number of time derivatives. Terms with an odd number of time derivatives are purely imaginary if all coordinates are kept fixed. (The ``usual'' complex conjugation maps $\phi(x)\to \phi^*(x)$.)

The presence of different complex structures has interesting consequences for many issues. As one example, we mention a nonvanishing chemical potential in a euclidean quantum field theory with fermions, as relevant for the phase diagram of QCD at nonzero baryon density. In the presence of a chemical potential the euclidean action is not invariant with respect to the usual hermitean conjugation $\theta_M$. After integrating out the fermions the effective weight factor $Z[\phi]$ is then not guaranteed to be a real quantity. However, the action remains real with respect to a generalized complex conjugation based on $\theta$. We have defined the notion of ``real operators'' as, for example, all hermitean fermion bilinears not involving  $\phi$. The expectation values of all real operators can be computed by using only the real part $Z_R[\phi]$ of the fermionic integral. The effective weight factor for real operators is therefore real even in the presence of a chemical potential, permitting perhaps a numerical evaluation.

In summary, the notions of ``real'' and ``imaginary'' depend on the choice of the complex structure. Grassmann integrals for fermions may admit many different complex structures. 

\section*{Appendix A: Spinors in four dimensions}
\renewcommand{\theequation}{A.\arabic{equation}}
\setcounter{equation}{0}

{\bf Clifford algebra}

\medskip
We collect here the most important relations for a particular
representation of the Clifford algebra in four dimensions.
For euclidean signature $(s=0)$ the hermitean matrices $\gamma^m$ are
\ba\label{A.1}
\gamma^0_E&=&
\left(\begin{array}{ccc}
0&,&1\\1&,&0\end{array}\right)
\quad,\quad \gamma^i=
\left(\begin{array}{ccc}
0&,&-i\tau_i\\i\tau_i&,&0
\end{array}\right),\nn\\
\bar\gamma&=&-\gamma^0_E\gamma^1\gamma^2\gamma^3=
\left(\begin{array}{ccc}
1&,& 0 \\0&,&-1
\end{array}\right).
\ea
Here $i=1,2,3$ and $\tau_i$ are the Pauli matrices. 
The hermitean generators of the euclidean Lorentz ($SO(4)$-rotation)
group read
\be\label{A.A1}
\sigma^{mn}=\frac{i}{2}[\gamma^m,\gamma^n]=-2i\Sigma^{mn}
=
\left(\begin{array}{ccc}
\sigma^{mn}_+&,&0\\0&,&\sigma^{mn}_-
\end{array}\right)
\ee
where
\ba\label{A.A2}
&&\sigma^{ij}_+=\sigma^{ij}_-=-\epsilon^{ijk}\tau_k,\nonumber\\
&&\sigma^{0k}_{E,+}=-\tau_k\quad,\quad \sigma^{0k}_{E,-}=\tau_k.
\ea

Corresponding representations for a Minkowski signature are easily obtained from
(\ref{A.1}) by appropriate multiplications with a factor $(-i)$. For
$s=1$ one multiplies $\gamma^0$ by a factor $-i$
whereas for $s=d-1$ the matrices $\gamma^i$ are
multiplied by $-i$. The matrix $\bar{\gamma}=diag(1,1,-1,-1)$ is the same for all signatures $s$. The phase factor in eq. (\ref{2.4}) is $\eta=-1$ for $s=0$ and $\eta=-i$ for $s=1$. In particular, for a Minkowski signature $s=1$ one has 
\begin{equation}\label{eu3}
\gamma^0_M=
\left(\begin{array}{ccc}
0&,&-i\\-i&,&0\end{array}\right)
\quad,\quad
{\sigma}^{0k}_{M,+}=i\tau_k~,~{\sigma}^{0k}_{M,-}=-i\tau_k\quad,\quad \bar{\gamma}=-i\gamma^0_M\gamma^1\gamma^2\gamma^3.
\end{equation}

The matrix $\bar{\gamma}$ is usually called $\gamma^5$ and the Weyl spinors are associated to left handed and right handed fermions
\ba\label{A.AA}
&&\psi_L=\psi_+=\frac{1+\bar{\gamma}}{2}\psi\quad,
\quad\psi_R=\psi_-=\frac{1-\bar{\gamma}}{2}\psi,\nonumber\\
&&\bar{\psi}_L=\bar{\psi}_+=\frac{1-\bar{\gamma}}{2}\bar{\psi}\quad,\quad
\bar{\psi}_R=
\bar{\psi}_-=\frac{1+\bar{\gamma}}{2}\bar{\psi}.
\ea

\bigskip\noindent
{\bf Action}

We use the same kinetic term
\begin{equation}\label{Z1}
L_{kin}=i\bar{\psi}\gamma^\mu\partial_\mu\psi
\end{equation}
for all signatures $s$. The analytic continuation discussed in sect. 3 implies for Minkowski and euclidean signature the convention $e^{-S_E}=e^{iS_M}$. In our convention one has for euclidean signature $S_E=\int d^4xL_{kin}$ and for Minkowski signature $S_M=-\int d^4xL_{kin}$. For a Minkowski signature $s=1$ we use the complex structure
\begin{equation}\label{Z2}
\bar{\psi}=\psi^\dagger\gamma^0_M.
\end{equation}
By virtue of the relation $\gamma^{\mu\dagger}\gamma^{0\dagger}=\gamma^0\gamma^\mu$ we observe that $L_{kin}$ is hermitean, $L^\dagger_{kin}=L_{kin}$. In the notation of sect. \ref{Complex conjugation for Minkowski signature} one has $\epsilon=1~,~D=D_2=\gamma^0_M$. For euclidean signature $s=0$ and the choice $H=H_2=-i\gamma^0_E=\gamma^0_M$ the kinetic term is invariant under the $\theta$-reflection. 

In our conventions the involutions related to the complex structures therefore read
\begin{eqnarray}\label{Z3}
s=0:&&\theta\psi(q)=R\bar{\psi}(\tilde{q})~,~\theta\bar{\psi}(q)=R\psi(\tilde{q}),\nonumber\\
s=1:&&\theta_M\psi(q)=R\bar{\psi}(q)~,~\theta_M\bar{\psi}(q)=R\psi(q),
\end{eqnarray}
with $\tilde q^0=-q^0$ and
\begin{equation}\label{Z3a}
R=H^{*(s=0)}=H^{-1(s=0)}=D^{*(s=1)}=D^{-1(s=1)}=
\left(\begin{array}{ccc}
0&,&i\\ i&,& 0
\end{array}\right),
\end{equation}
or, for $s=1$,
\begin{equation}\label{Z4}
\psi^*(x)=
\left(\begin{array}{ccc}
0&,&i\\i&,&0\end{array}\right)
\bar{\psi}(x).
\end{equation}

For our convention $D=D_2$ for $s=1$ the list of hermitean bilinears $\bar{\psi}O_j\psi$ reads
\begin{equation}\label{Z5}
O_j=i~,~\bar{\gamma}~,~\gamma^m~,~\gamma^m\bar{\gamma}~,
~i\sigma^{mn}~,~\sigma^{mn}\bar{\gamma}.
\end{equation}
For $s=0~,~H=H_2$, this list also corresponds to terms that are invariant under $\theta$ if an even number of indices $m$ equals zero. On the other hand, these bilinears are odd under $\theta$ for an odd number of zero-indices. In particular, a mass term takes the perhaps somewhat unusual form
\begin{equation}\label{Z6}
L_m=m\bar{\psi}_R\psi_L-m^*\bar{\psi}_L\psi_R,
\end{equation}
such that for real $m$ the Dirac operator becomes\footnote{This form is familiar from dimensional reduction of higher dimensional theories \cite{CWDR}.}
\begin{equation}\label{Z7}
{\cal D}=i\gamma^\mu\partial_\mu+m\bar{\gamma}.
\end{equation}
This convention simplifies many computations since the squared Dirac operator in momentum space takes the simple form
\begin{equation}\label{Z8}
{\cal D}^2=q^\mu q_\mu+m^2.
\end{equation}
Such a simple form may constitute an advantage for euclidean lattice gauge theory, since for real $m$ the piece $m^2$ constitutes a gap in the spectrum of ${\cal D}^2$ and therefore acts as an effective infrared cutoff for the fermion fluctuations.

The convention with ${\cal D}=i\gamma^\mu\partial_\mu+m$ would be obtained if we use $D_1,H_1$ instead of $D_2,H_2$ (for $s=1,0)$, with $\bar{\psi}=\psi^\dagger\gamma^0\bar{\gamma}$ instead of eq. (\ref{Z2}). For fixed $\psi^\dagger$ the two conventions would be related by a simple redefinition of $\bar{\psi}~,~\bar{\psi}'=\bar{\psi}\bar{\gamma}$. However, the partition function is defined as a functional integral over $\psi$ and $\bar{\psi}$ as independent variables. The two different choices for $D$ should be interpreted as different choices of the complex structure. In consequence, they are related by a transformation that is not compatible with a given complex structure.

We note that a chiral rotation,
\begin{equation}\label{eu6}
\psi\rightarrow e^{i\alpha\bar{\gamma}}\psi~,~\bar{\psi}\rightarrow\bar{\psi}e^{i\alpha\bar{\gamma}},
\end{equation}
induces a phase change of $m$ in either convention but does not switch from one convention to the other. We could use this chiral phase in order to bring the mass term $m\bar{\gamma}$ in eq. (\ref{Z7}) into the perhaps somewhat more conventional form $im$. (The latter corresponds to purely imaginary $m$ in eq. (\ref{Z6}).) The form \eqref{Z7}, \eqref{Z8} seems, however, most convenient.

For states with a nonzero particle number one adds to the action a piece containing the chemical potential $\mu$. For a Minkowski signature $(s=1)$ it reads
\ba\label{A.15A}
S^{(\mu)}_E&=&i\mu\int d^4x\bar\psi\gamma^0_M\psi=-i\mu\int d^4x\psi^\dagger\psi \nn\\
&=&\mu\int d^4x(\bar\psi_L\psi_L+\bar\psi_R\psi_R).
\ea
The associated Minkowski action 
\be\label{A.15B}
S^{(\mu)}_M=iS^{(\mu)}_E=\mu\int d^4x\psi^\dagger \psi
\ee
is hermitean, while $S^{(\mu)}_E$ is antihermitean. For a euclidean signature $(s=0)$ one finds the hermitean euclidean action
\ba\label{A.15C}
S^{(\mu)}_E&=&-i\mu\int d^4x\bar\psi\gamma^0_E\psi\nn\\
&=&-i\mu\int d^4x(\bar\psi_L\psi_L+\bar\psi_R\psi_R).
\ea
A real euclidean action (with respect to the conjugation $\theta$) remains real if the piece \eqref{A.15C} is added. The determinant associated to the Gaussian integration over fermions remains real. 

With time on a torus with circumference $1/T$ the setting with $s=0$ describes thermal field theory, with partition function 
\be\label{A.15D}
Z=\int {\cal D}\psi\exp(-S_E)
\ee
related to thermodynamic Gibbs free energy $J$ and pressure $p$ by the relation 
\be\label{A.15E}
p=-V^{-1}_3J=V^{-1}_3T\ln Z,
\ee
with $V_3$ the volume. The derivative of $p$ with respect to $\mu$ yields the particle number density
\be\label{A.15F}
n=\frac{\partial p}{\partial\mu}_{|_T}=\kl \bar\psi i\gamma^0_E\psi\kr,
\ee
where we have assumed a homogeneous ground state. The reality of the action guarantees that $p$ and $n$ are real. (From eq. \eqref{Z5} one infers that the bilinear $i\bar\psi\gamma^0_E\psi$ is hermitean.) The non-relativistic approximation for particles with a given charge (no antiparticles) leads to a two-component complex spinor $\varphi$, with action containing the kinetic term \eqref{Z1}, mass term \eqref{Z6} and chemical potential \eqref{A.15C} reduced to $(s=0)$
\be\label{A.15G}
S=\int d^4x\varphi^\dagger\big(\partial_t-\mu-\Delta/(2m)\big)\varphi.
\ee

\bigskip\noindent
{\bf Discrete symmetries}

\medskip
The charge conjugate spinor is defined as
\be\label{A.3}
\psi^c(q)=-C^{-1}\bar\psi(q),\quad \bar\psi^c(q)=-C\psi(q).
\ee
It involves the charge conjugation matrix $(C\equiv C_1$) which is given by
\be\label{eu4}
C=
\left(\begin{array}{ccc}
\tau_2&,&0\\0&,&-\tau_2\end{array}\right),
\ee
and obeys ($\delta=-1$)
\ba\label{eu5}
&&(\gamma^m)^T=-C\gamma^mC^{-1}\quad,\quad(\sigma^{mn})^T=-C\sigma^{mn}C^{-1}~,~
[C,\bar\gamma]=0\quad,\nonumber\\
&&C=-C^*=-C^T=C^\dagger=C^{-1}\quad,\nonumber\\
&&CC^T=-C^2=CC^*=-1.
\ea
It is independent of the signature $s$. 

In presence of internal symmetries the generalized 
charge conjugation ${\cal C}_W$ may also
act on internal spinor indices by a matrix $G$
\ba\label{A.4}
&&{\cal C}_W(\psi)=\psi^c=-GC\bar\psi\quad,\quad
{\cal C}_W(\bar\psi)=\bar\psi^c=-G^*C\psi\nonumber\\
&&G^\dagger G=1\quad,\quad G^*G=\epsilon_G\quad, \quad 
{\cal C}_W^2=\epsilon_G.
\ea
With 
\be\label{A.5}
G=
\left(\begin{array}{ccc}
g_+&,&0\\0&,&g_-\end{array}\right)
\ee
and $g_\pm$ unit matrices in (two-component) spinor space, acting only on internal indices,  one obtains
\ba\label{A.6}
\begin{array}{lll}
\psi^c_+=-g_+\tau_2\bar\psi_-&,& \psi^c_-=g_-\tau_2\bar\psi_+ \\
\bar\psi^c_+=g^*_-\tau_2\psi_-&,& \bar\psi^c_-=-g^*_+\tau_2\psi_+.
\end{array}
\ea
The charge conjugation 
\be\label{A.20a}
{\cal C}_W(\psi_\pm)=\psi^c_\pm=\frac{1\pm\bar{\gamma}}{2}\psi^c
\ee
does not change the Lorentz representation of the spinors, corresponding to mappings $\psi_+\rightarrow\bar{\psi}_-$ etc. Expressed in terms of $\psi$ and $\bar{\psi}$ it acts in spinor space without an additional complex conjugation of the coefficients of the Grassmann algebra (in contrast to $\theta,\bar{\theta}$).

For $g_+=g_-$ the bilinears transform as 
\ba\label{A.7}
{\cal C}_W(\bar\psi_+^a\psi^b_-)&=&\bar\psi_+^b\psi_-^a
\quad,\quad\qquad  {\cal C}_W(\bar
\psi_-^a\psi^b_+)\ \ =\ \bar\psi^b_-\psi^a_+,\nonumber\\
{\cal C}_W(\bar\psi^a_+\gamma^\mu\psi^b_+)&=&-\bar\psi^b_-\gamma^
\mu\psi^a_-\quad,\quad
{\cal C}_W(\bar\psi_-^a\gamma^\mu\psi^b_-)=-\bar\psi^b_+\gamma^\mu
\psi^a_+,\nonumber\\
{\cal C}_W(\bar{\psi}^a_+\sigma^{\mu\nu}\psi^b_-)&=&-\bar{\psi}^b_+\sigma^{\mu\nu}\psi^a_-\quad,\quad
{\cal C}_W(\bar{\psi}^a_-\sigma^{\mu\nu}\psi^b_+)=-\bar{\psi}^b_-\sigma^{\mu\nu}\psi^a_+.
\ea
Here $\bar{\psi}_+\psi_-$ stands for $\bar{\psi}^T_+\psi_-$ and corresponds to a contraction over spinor indices and internal indices on which $G$ acts. The additional ``flavor indices'' $a,b$ denote different species of spinors. (The matrix $G$ acts as a matrix in flavor space.) In particular, a gauge covariant kinetic term for a vector-like gauge symmetry,
\be\label{A.B1}
S_E=i\int d^4x\{\bar{\psi}_+\gamma^\mu(\partial_\mu-igA_\mu)\psi_+
+\bar{\psi}_-\gamma^\mu(\partial_\mu-igA_\mu)\psi_-\},
\ee
is invariant under ${\cal C}_W$ if the gauge field has negative $C$-parity, ${\cal C}_W\big(A_\mu(x)\big)=-A_\mu(x)$. On the other hand, a gauge field with purely axial coupling $(\sim\gamma^\mu\bar{\gamma})$ has positive $C$ parity.

For $G=1$ one has $\bar\psi_R=-\tau_2\psi^c_L~,~\psi_R=\tau_2\bar\psi^c_L$, such that the number density
\be\label{A.24A1}
n(x)=\bar\psi(x)i\gamma^0_E\psi(x)=i(\bar\psi_L\psi_L-\bar\psi^c_L\psi^c_L)
\ee
counts particles and antiparticles with opposite sign. The conserved charge associated to the chemical potential $\mu$ in eqs. \eqref{A.15B} or \eqref{A.15C} can be associated with electric charge. The bilinear $n(x)$ vanishes for Majorana spinors where $\psi^c=\pm \psi$. 

The parity transformation $(P=P_2)$
\be\label{A.B2}
{\cal P}\big(\psi(x^0,\vec{x})\big)= P\psi(x^0,-\vec{x})\quad,\quad
{\cal P}\big(\bar{\psi}(x^0,\vec{x})\big)=-\bar{\psi}(x_0,-\vec{x})P,
\ee
is represented by the matrix
\be\label{A.B3}
P=\gamma^0_M\bar{\gamma}=-i\gamma^0_E\bar{\gamma}=
\left(\begin{array}{ccc}
0&,&i\\-i&,&0
\end{array}\right)
\ee
both for euclidean $(s=0)$ and Minkowski $(s=1)$ signature. The fermion bilinears transform as
\ba\label{A.B4}
\begin{array}{llll}
{\cal P}:\quad&\bar{\psi}_-\psi_+\rightarrow-\bar{\psi}_+\psi_-&,&
\quad\bar{\psi}_+\psi_-\rightarrow-\bar{\psi}_-\psi_+\quad,\\
&\bar{\psi}_+\gamma^\mu\psi_+
\rightarrow-\bar{\psi}_-\tilde{\gamma}^\mu\psi_-&,&
\quad\bar{\psi}_-\gamma^\mu\psi_-\rightarrow-\bar{\psi}_+\tilde{\gamma}^\mu\psi_+,\\
&\bar{\psi}_-\sigma^{\mu\nu}\psi_+\rightarrow -\bar{\psi}_+\tilde{\sigma}^{\mu\nu}\psi_-&,&
\quad\bar{\psi}_+\sigma^{\mu\nu}\psi_-\rightarrow-\bar{\psi}_-
\tilde{\sigma}^{\mu\nu}\psi_+,\\
\end{array}
\ea
with $\tilde{\gamma}^0=-\gamma^0\quad,\quad\tilde{\gamma}^i=\gamma^i\quad,\quad\tilde{\sigma}^{0i}=-\sigma^{0i}\quad,\quad\tilde{\sigma}^{ij}-\sigma^{ij}$. We observe $P^2=1~,~{\cal P}^2=1$. For the combined CP-transformation this implies
\ba\label{A.B5}
\begin{array}{llll}
{\cal C}_W{\cal P}:&\bar{\psi}^a_-\psi^b_+\rightarrow-\bar{\psi}^b_+\psi^a_-&,\quad&
\bar{\psi}^a_+\psi^b_-\rightarrow-\bar{\psi}^b_-\psi_+^a\quad,\quad\\
&\bar{\psi}^a_+\gamma^\mu\psi^b_+\rightarrow\bar{\psi}^b_+\tilde{\gamma}^\mu\psi^a_+
&,\quad&
\bar{\psi}^a_-\gamma^\mu\psi^b_-\rightarrow\bar{\psi}^b_-\tilde{\gamma}^\mu\psi^a_-\quad,\\
&\bar{\psi}^a_-\sigma^{\mu\nu}\psi^b_+\rightarrow\bar{\psi}^b_+\tilde{\sigma}^{\mu\nu}\psi^a_-
&,\quad&
\bar{\psi}^a_+\sigma^{\mu\nu}\psi^b_-\rightarrow\bar{\psi}^b_-\tilde{\sigma}^{\mu\nu}\psi^a_+,
\end{array}
\ea
and we note $({\cal C}_W{\cal P})^2\psi(x)=-\psi(x)$.

In our conventions a general mass term consistent with hermiticity $(s=1)$ or Osterwalder-Schrader positivity $(s=0)$ reads
\be\label{A32}
L_m=\bar{\psi}_a M^{(H)}_{ab}\bar{\gamma}\psi_b+\bar{\psi}_a
M^{(A)}_{ab}\psi_b
\ee
with $M^{(H)}$ and $M^{(A)}$ the hermitean and antihermitean parts of an arbitrary mass matrix $M_{ab}=M^{(H)}_{ab}+M^{(A)}_{ab}~,~
(M^{(H)})^\dagger=M^{(H)}~,~ (M^{(A)})^\dagger=-M^{(A)}$. It is invariant under the charge conjugation ${\cal C}_W$ for $M=M^T$. Invariance with respect to parity requires a hermitean mass matrix $M=M^\dagger$ whereas CP-invariance follows for a real $M=M^*$. It is sufficient that these properties hold for an appropriate basis in spinor-space which may be realized by chiral transformations. In particular, if $M$ can be transformed to a real diagonal form the possible violations of $C,P$ and CP occur in other sectors of the theory and are not related to the fermion masses.

\bigskip\noindent
{\bf Majorana spinors and complex structure}

\medskip
Majorana spinors can be defined by imposing the condition 
\be\label{A.29}
\psi^c_{M+}=\psi_{M+}.
\ee
For both euclidean signature $(s=0)$ and Minkowski signature $(s=1)$ this amounts to the same relation between $\bar\psi$ and $\psi$,
\be\label{A.29B}
\bar\psi=-C\psi,
\ee
where we recall that we use the same matrix $C$ for both signatures. It is obvious that this definition of a Majorana spinor is possible both for euclidean and Minkowski signature.

The Majorana constraint \eqref{A.29B} acts on the Weyl spinors $\psi_\pm$ \eqref{A.AA} as 
\be\label{A.29C}
\bar\psi_-=-\tau_2\psi_+~,~\bar\psi_+=\tau_2\psi_-.
\ee
We may therefore express $\psi_-$ and $\bar\psi_-$ in terms of $\psi_+$ and $\bar\psi_+$ and keep only $\psi_+$ and $\bar\psi_+$ as independent Grassmann variables. This shows that a single Weyl spinor, as described by $\psi_+,\bar\psi_+$, is equivalent to a Majorana spinor for both $s=1$ and $s=0$. We can formulate the action and all symmetry transformations only in terms of the Weyl spinor $\psi_+,\bar\psi_+$. The relation \eqref{A.29C} translates the symmetry transformations to the Majorana spinor. This applies for all symmetries, including supersymmetry, and for both signatures. In particular, if a supersymmetry is formulated for a Weyl spinor in euclidean space, e.g. for the chiral multiplet, this is translated by eq. \eqref{A.29C} in a straightforward way to our version of euclidean Majorana spinors. 

For the notion of complex conjugation in Minkowski space we employ the matrix $B=B_2=\epsilon D^*_2 C^T_1=-D^*_2C_1$, given by
\be\label{A.24A}
B=-(\gamma^0_M)^*C=
\left(\begin{array}{rl}
 0~,&i\tau_2\\-i\tau_2~,&0
\end{array}\right),
\ee
such that
\be\label{A.24B}
\psi^*=B\psi^c~,~\psi^*_{M+}=B\psi_{M+}.
\ee
The matrix $B$ obeys the relations
\be\label{A.24C}
B=B^*=B^T=B^\dagger=B^{-1}
\ee
and 
\be\label{A.24D}
B\gamma^\mu_MB^{-1}=(\gamma^\mu_M)^*.
\ee

For euclidean signature the complex conjugation uses the matrix $B=H^*_2C^T_1=-H^*_2C_1$. Since we employ $H_2=\gamma^0_M$ the matrix $B$ is the same as for Minkowski signature, i.e. given by eq. \eqref{A.24A}. The only difference arises from the fact that the euclidean complex conjugation based on $\theta$ involves also a reversal of time or $q_0$, cf. eq. \eqref{Z3}. We also have to replace eq. \eqref{A.24D} by 
\be\label{A.24E}
B \gamma^\mu_E B^{-1}=(\tilde\gamma^\mu_E)^*,
\ee
where $\tilde\gamma^k_E=\gamma^k_E~,~\tilde\gamma^0_E=-\gamma^0_E$. If we define the complex structure in terms of $\psi$ and $\bar\psi$, it is the same for the signatures $(E)$ and $(M)$, except for $x\to \theta x$, 
\be\label{A.24F}
\psi^*=-BC\bar\psi=(\gamma^0_M)^*\bar\psi.
\ee

\bigskip\noindent
{\bf Majorana representation}

\medskip
The representation of the Clifford algebra \eqref{A.1}, \eqref{eu3} is particularly convenient for a description of Weyl spinors. In this representation $\bar\gamma$ is block diagonal, such that the upper two and lower two components of $\psi$ correspond to the two inequivalent complex two-component representations of the Lorentz symmetry. They describe Weyl spinors with different handedness. The representation \eqref{A.1}, \eqref{eu3} of the Dirac matrices may therefore be called the ``Weyl representation''. In four dimensions Weyl and Majorana spinors are equivalent for Minkowski signature \cite{W1}. This equivalence extends to euclidean signature for the physical Majorana spinors used in this paper. 

Majorana spinors are most conveniently discussed in a ``Majorana representation'' of the Dirac matrices given by
\be\label{A.30}
\gamma^0_{(M)M}=\left(\begin{array}{ccc}
0&,&\tau_1\\-\tau_1&,&0\end{array}\right)~,~
\gamma^0_{(M)E}=i\left(\begin{array}{ccc}
0&,&\tau_1\\-\tau_1&,&0\end{array}\right),
\ee
and (for both euclidean and Minkowski signature)
\be\label{A.31}
\gamma^1_{(M)}=\left(\begin{array}{ccc}
-\tau_1&,&0\\0&,&\tau_1\end{array}\right)~,~
\gamma^2_{(M)}=\left(\begin{array}{ccc}
-\tau_3&,&0\\0&,&-\tau_3\end{array}\right)~,~
\gamma^3_{(M)}=\left(\begin{array}{ccc}
0&,&\tau_1\\ \tau_1&,&0\end{array}\right).
\ee
In this representation $\bar\gamma$ remains block diagonal
\be\label{A.32}
\bar\gamma_{(M)}=-\left(\begin{array}{ccc}
\tau_2&,&0\\0&,&\tau_2\end{array}\right).
\ee
The matrices $\gamma^\mu_{(M)}$ obtain from the Weyl representation $\gamma^\mu_{(W)}$ specified by eqs. \eqref{A.1}, \eqref{eu3} by a similarity transformation
\be\label{A.34}
\gamma^\mu_{(M)}=A\gamma^\mu_{(W)}A^{-1}~,~A^\dagger A=1,
\ee
where 
\be\label{A.35}
A=\frac{1}{\sqrt{2}}
\left(\begin{array}{ccccccc}
1&,&0&,&0&,&1\\-i&,&0&,&0&,&i\\
0&,&1&,&-1&,&0\\
0&,&-i&,&-i&,&0
\end{array}\right).
\ee
We also will employ the real symmetric matrices $T_k=\gamma^0_{(M)M}\gamma^k_{(M)}$ given by
\be\label{A.33}
T_1=\left(\begin{array}{ccc}
0&,&1\\1&,&0\end{array}\right)~,~T_2=
\left(\begin{array}{ccc}
0&,&c\\-c&,&0\end{array}\right)~,~
T_3=\left(\begin{array}{ccc}
1&,&0\\0&,&-1\end{array}\right),
\ee
with 
\be\label{A.33A}
c=i\tau_2=\left(\begin{array}{ccc}
0&,&1\\-1&,&0\end{array}\right).
\ee

For Minkowski signature the matrices $\gamma^\mu_{(M)}$ are all real. The similarity transformation of the charge conjugation matrix $C_{(W)}$ (given by eq. \eqref{eu4}) reads in the Majorana representation 
\be\label{A.36}
\tilde C_{(M)}=AC_{(W)}A^{-1}=
\left(\begin{array}{rr}
0,&c\\-c,&0\end{array}\right).
\ee
However, this is {\em not} the charge conjugation matrix that should be used for the definition of Majorana spinors in the Majorana representation. In every representation the definition of the charge conjugation matrix $C$ is related to the properties of the Dirac matrices under transposition as given by eq. \eqref{6.6}. These properties depend on the particular representation. The $C$-matrices defined by eq. \eqref{6.6} for different representations are not related by a similarity transformation of the type \eqref{A.36}. One rather has to use
\be\label{A.37}
C_{(M)}=A^*C_{(W)}A^{-1}=\gamma^0_{(M)M}=-i\gamma^0_{(M)E},
\ee
which does not coincide with $\tilde C_{(M)}$ since $A$ is not a real matrix. In particular, one observes $C^2_{(M)}=-1$, while $\tilde C^2_{(M)}=1$. 

Majorana spinors are eigenstates of the generalized charge conjugation ${\cal C}_W$
\be\label{A.38}
{\psi\choose\bar\psi}_{M\pm}=\frac12(1\pm{\cal C}_W)
{\psi\choose\bar\psi}=\frac12
\left(\begin{array}{c}
(1\pm W_1)\bar\psi\\(1\pm W_2)\psi\end{array}\right).
\ee
With $W_1=-C^{-1},W_2=-C$ one finds
\be\label{A.39}
{\psi\choose\bar\psi}_{M\pm}=\pm {\cal C}_W
{\psi\choose\bar\psi}_{M\pm}=\mp
{C^{-1}\bar\psi_{M\pm}\choose C\psi_{M\pm}},
\ee
such that the upper and lower components obey 
\be\label{A.40}
\psi_{M\pm}=\mp C^{-1}\bar\psi_{M\pm}~,~\bar\psi_{M\pm}=\mp C\psi_{M\pm}.
\ee
For the Majorana representation \eqref{A.37} this yields
\be\label{A.41}
\bar\psi_{M\pm}=\mp\gamma^0_{(M)M}\psi_{M\pm}.
\ee
In this representation the number density $n(x)$ reads for $s=0$
\be\label{A.59A}
n(x)=\bar \psi(x)i\gamma^0_E\psi(x)=(\psi^c)^T(x)\psi(x)
\ee
and vanishes for Majorana spinors due to the Pauli principle.

With respect to the complex structure $\theta_M$ we find for Minkowski signature
\be\label{A.42}
\bar\psi_{M+}=(\gamma^0_{(M)M})^T\psi^*_{M+}=-\gamma^0_{(M)M}\psi^*_{M+}=-\gamma^0_{(M)M}\psi_{M+},
\ee
such that $\psi_{M+}$ is simply the real part of $\psi$, obeying
\be\label{A.43}
\psi^*_{M+}=\psi_{M+}.
\ee
Correspondingly, $\psi_{M-}$ is the imaginary part with $\psi^*_{M-}=-\psi_M$. Indeed, the matrix $B$ reads in the Majorana basis 
\be\label{A.44}
B_{(M)}=A^*B_{(W)}A^{-1},
\ee
with $B_{(W)}$ given by eq. \eqref{A.24A}. This yields
\be\label{A.45}
B_{(M)}=-(\gamma^0_{(M)M})^*C_{(M)}=1,
\ee
such that for Minkowski signature the complex conjugate spinor reads
\be\label{A.46}
\psi^*=B_{(M)}\psi^c=\psi^c,
\ee
implying for a Majorana spinor $\psi^c_{M+}=\psi_{M+}$ that $\psi$ is real.

\bigskip\noindent
{\bf Real Grassmann algebra for Majorana spinors in Minkowski space and analytic continuation}

\medskip
In the Majorana basis the Majorana spinors are particularly simple: we can simply consider one real four-component spinor $\psi_{M+}$. For Minkowski signature the kinetic part of the action is then an element of a real Grassmann algebra $(\psi=\psi_{M+})$
\ba\label{A.47}
S_E&=&-iS_M=-\int d^4 x\bar\psi\gamma^\mu_{(M)M}\partial_\mu\psi\nn\\
&=&\int d^4 x\psi^T(\partial_0-\sum_kT_k\partial_k)\psi,
\ea
with $T_k$ the real matrices \eqref{A.33}. For a single Majorana spinor no complex quantities appear. While $S_M$ is hermitean or invariant under the involution $\theta_M$, the euclidean action changes sign, $\theta_MS_E=-S_E$. No spinors $\bar\psi$ independent of $\psi$ appear in the action \eqref{A.47}.

For the analytic continuation to euclidean signature we do not change the real four-component spinor $\psi$, which will now describe the euclidean Majorana spinor. The analytic continuation only changes the action $S_E$ \eqref{A.47} by multiplication with an overall factor $-i$ from $e_E/e_M=-i$, and a multiplication of $\partial_0$ by a factor $i$ from $(e_0\ ^0)_E/(e_0\ ^0)_M=i$. The two factors $-i$ and $i$ cancel for the time derivative, such that the analytic continuation of the action \eqref{A.47} becomes for euclidean signature
\be\label{A.48}
S_E=\int d^4x\psi^T(\partial_0+i\sum_kT_k\partial_k)\psi.
\ee
The euclidean action \eqref{A.48}  is ``real'' in the sense of sect. \ref{Realfermionicactions}, i.e. $S_E$ is invariant under the involution $\theta$. However, the action \eqref{A.48} is no longer an element of a real Grassmann algebra since $T_k$ is multiplied by $i$. The Lorentz generators in the Majorana representation can be obtained from eq. \eqref{A.A2} of \eqref{eu3} by applying the similarity transformation \eqref{A.34}, or alternatively by computing the commutator \eqref{2.3} in the Majorana representation. They are given in sect. \ref{EuclideanMajoranaspinors} by eqs. \eqref{y} and \eqref{BB5N} for Minkowski and euclidean signature, respectively. 

We also can group the four Grassmann variables $\psi_\gamma$ into a two-component complex vector $\zeta=(\zeta_1,\zeta_2)$ according to
\ba\label{A.51}
\zeta_1&=&\frac{1}{\sqrt{2}}(\psi_1+i\psi_2)~,~\zeta_2=\frac{1}{\sqrt{2}}(\psi_3+i\psi_4),\nn\\
\zeta^*_1&=&\frac{1}{\sqrt{2}}(\psi_1-i\psi_2)~,~\zeta^*_2=\frac{1}{\sqrt{2}}(\psi_3-i\psi_4).
\ea
In terms of $\zeta$ the action becomes
\be\label{A.52}
S_E=2\int d^4x\zeta^\dagger (\partial_0+\rho(s)\sum_k\tau_k\partial_k)\zeta,
\ee
with $\rho(s=1)=-1$ for Minkowski signature and $\rho(s=0)=i$ for euclidean signature. In this form the invariance of the action under a global phase rotation of $\zeta$ is most easily visible. This phase rotation corresponds to the infinitesimal transformation \eqref{BB6N} in sect. \ref{EuclideanMajoranaspinors}. The spinors $\zeta$ transform as Weyl spinors, demonstrating the equivalence of Weyl and Majorana spinors in four dimensions.

\section*{Appendix B: Complex structures for real and \\complex Grassmann algebra}
\renewcommand{\theequation}{B.\arabic{equation}}
\setcounter{equation}{0}
A complex structure can be introduced for a real Grassmann algebra (real coefficients $a$ in eq. \eqref{3.1} or similar expressions) in a rather standard way. Let $\hat{\psi}$ be arbitrary vectors of a $2N$-dimensional real vector space such that $\lambda_1\hat{\psi}_1+\lambda_2\hat{\psi}_2$ is defined with real coefficients $\lambda_{1,2}$. For example, the components $\hat{\psi}_\alpha$ may be independent Grassmann variables or real numbers. A complex structure is defined by two  real $2N\times 2N$ matrices $K,I$ obeying 
\be\label{BB1}
K^2=1,~I^2=-1,~ \{K,I\}=0.
\ee
Then $K$ plays the role of complex conjugation, $\hat{\psi}^*=K\hat{\psi}$, and $I$ stands for the multiplication with $i$. Combining the conditions (\ref{BB1}) one infers 
\be\label{BB1A}
\Tr K=0.
\ee

A typical example is
{\renewcommand{\arraystretch}{1}
\be\label{BB2}
K=\left(\begin{array}{ccc}
1 &,& 0\\0 &,&-1\end{array}\right)
~,~
I=\left(\begin{array}{ccc}
0 &,& -1\\1 &,& 0\end{array}\right),
\ee}
with the representation
\be\label{BB3}
\hat{\psi}={\psi_R\choose\psi_I},~\hat{\psi}^*=K\hat{\psi}=
{\psi_R\choose-\psi_I}~,~I{\hat\psi}=
{-\psi_I\choose\psi_R}.
\ee
More generally, we may define the projections
\be\label{BB4}
\hat{\psi}_R=\frac{1+K}{2}\hat{\psi}~,~\hat{\psi}_I=\frac{1-K}{2}\hat{\psi}.
\ee
Since $\hat{\psi}_R$ and $\hat{\psi}_I$ obey constraints, i.e. $K\hat{\psi}_R=\hat{\psi}_R$, they have only $N$ independent components each. By a suitable change of basis we can always choose the representation (\ref{BB2}) and use $\psi_{R,I}$ according to (\ref{BB3}) as the independent components. An isomorphism maps the $2N$-component vector $\hat{\psi}$ to an $N$-component complex vector $\varphi$,
\be\label{BB5}
\varphi(\psi)=\psi_R+i\psi_I~,~\varphi (K\hat{\psi})~=~\varphi^*(\psi)~,~\varphi(I\hat{\psi})~{=}~i\varphi(\psi).
\ee
The anticommutation property $\{K,I\}=0$ guarantees $(i\varphi)^*=-i\varphi^*$ according to $KI\hat{\psi}=-IK\hat{\psi}$. 

Consider next linear transformations, $\hat{\psi}\rightarrow\hat{\psi}'=\hat{A}\hat{\psi}$, represented by real $2N\times2N$ matrices $\hat{A}$. For regular $\hat{A}$ they form a group. Only a subgroup is compatible with the complex structure, namely those obeying 
\be\label{BB5A}
[\hat{A},I]=0.
\ee
This condition guarantees that the multiplication with $i$ is represented by the same matrix $I$ for $\hat{\psi}'$ and $\hat{\psi}$ and commutes with all compatible linear transformations. One infers from eq. \eqref{BB5A} that $\hat{A}$ can be represented as 
\be\label{BB6}
\hat{A}=A_R+A_II,
\ee
with $A_R$ and $A_I$ real $N\times N$ matrices. Furthermore, we define the complex conjugate matrix $\hat{A}^*$ as a real $2N\times 2N$ matrix by
\be\label{BB7}
\hat{A}^*=K\hat{A}K.
\ee
For transformations which are compatible with the given complex structure one has
\be\label{BB7A}
\hat{A}^*=A_R-A_II,
\ee
and we see that the ``imaginary part'' $A_I$ is odd under complex conjugation. For the particular representation (\ref{BB2}) one has

{\renewcommand{\arraystretch}{1}

\be\label{BB8}
\hat{A}=\left(\begin{array}{rc}
A_R,& -A_I \\ A_I,& A_R
\end{array}\right)
~,~  
\hat{A}^*=\Bigg(
\begin{array}{rc}
A_R,&A_I\\-A_I,&A_R
\end{array}\Bigg).
\ee}

For matrices compatible with the complex structure the isomorphism (\ref{BB5}) reflects complex multiplication with a $N\times N$ matrix $A,\varphi\rightarrow A\varphi$, with 
\be\label{BB9}
\varphi(\hat A\hat\psi)=A\varphi(\hat\psi)~,~A=A_R+iA_I.
\ee
Here the definition of $\hat{A}^*$ guarantees
\be\label{BB10}
(A\varphi)^*=A^*\varphi^*(\hat\psi)~{=}~\varphi(K\hat{A}\hat\psi)=\varphi(\hat{A}^*K\hat\psi),
\ee
and the compatibility requirement (\ref{BB5A}) ensures 
\be\label{BB10A}
iA\varphi(\hat\psi)=Ai\varphi~=~\varphi( I\ha\hat\psi)=\varphi(\ha I\hat\psi).
\ee
It also implies that the real part of $\hat A$ is even under complex conjugation whereas the ``imaginary part'' is odd
\be\label{BB11}
\ha_R=\frac12(\ha+\ha^*)~,~\ha_I=\frac12(\ha-\ha^*)~,~\ha=\ha_R+\ha_I~,~[\ha_R,K]=0~,~\{\ha_I,K\}=0.
\ee
We note that transformations $\ha$ not obeying the compatibility condition (\ref{BB5A}) remain well defined - they simply cannot be represented by a complex matrix multiplication (\ref{BB9}). 

A special case is the multiplication with complex numbers. It is realized by diagonal real $N\times N$ matrices $A_R=\lambda_R,A_I=\lambda_I$, with $\hat A=\lambda_R+\lambda_II$. In the basis \eqref{BB2} this amounts to
\be\label{B.15X1}
\hat\psi\to\hat\Lambda\hat\psi~,~\hat\Lambda=\left(\begin{array}{lr}
\lambda_R,&-\lambda_I\\\lambda_I,&\lambda_R\end{array}\right),
\ee
with 
\be\label{B.15X2}
\varphi(\hat\Lambda\hat\psi)=\lambda\varphi(\hat\psi)~,~\lambda=\lambda_R+i\lambda_I.
\ee
Compatibility with the complex structure follows from 
\be\label{B.15X3}
K\hat\Lambda\hat\psi=\hat\Lambda^*K\hat\psi~,~\hat\Lambda^*=K\hat\Lambda K,
\ee
with
\be\label{B.15X4}
\hat\Lambda^*=
\left(\begin{array}{rl}
\lambda_R,&\lambda_I\\-\lambda_I,&\lambda_R\end{array}\right)
\ee
and 
\be\label{B.15X5}
\varphi(\hat\Lambda^*\hat\psi)=\lambda^*\varphi(\hat\psi).
\ee
If we express complex conjugation by an involution $\theta(\varphi)=\varphi^*,\theta(\varphi^*)=\varphi,$ one has 
\be\label{B.15X6}
\theta(\lambda\varphi)=\lambda^*\theta(\varphi)~,~\theta(A\varphi)=A^*\theta(\varphi),
\ee
corresponding to
\be\label{B.15X7}
\varphi(K\hat\Lambda\hat\psi)=\lambda^*\varphi(K\hat\psi)=\lambda^*\varphi^*(\hat\psi).
\ee

Each pair of matrices $(K,I)$ defines a map from the ``real'' spinors $\hat\psi$ to the ``complex'' spinors $\varphi,\varphi^*$ and corresponds to a particular complex structure. If we choose a given matrix $I$ the subclass of linear transformations \eqref{BB5A}, which are compatible with the complex structure, is uniquely defined. In particular, the multiplication of spinors by complex numbers is fixed. However, there remain still many different possibilities for the choice of the involution $K$ obeying eq. \eqref{BB1}. The complex structure is determined uniquely only once the action of the complex conjugation $K$ on the spinors is specified. In other words, the use of a complex Grassmann algebra (complex coefficients $a$ in eq. \eqref{3.1}) is not yet sufficient for fixing the complex structure. One has in addition to specify the action of the involution $K$ or $\theta$ on the Grassmann variables. 

As a first simple example, we consider the complex conjugation
\be\label{B.15A}
\tilde K=\cos \alpha K+\sin \alpha IK=\exp (\alpha I)K=
\left(\begin{array}{ccc}\cos\alpha&,&\sin\alpha\\\sin\alpha&,&-\cos\alpha \end{array}\right).
\ee
(For the last expression  we work in the basis where $(K,I)$ are given by eq. \eqref{BB2}.) The involution $\tilde K$ obeys eq. \eqref{BB1} and therefore defines a valid complex structure for arbitrary $\alpha$, differing from $K$ for $\alpha\neq 0$. The meaning of real and imaginary spinors $(\tilde \psi_R,\tilde\psi_I)$ with respect to $\tilde K$ differs from $(\psi_R,\psi_I)$ which are defined for $K$ or $\alpha=0$:
\be\label{B.15B}
\tilde \psi_{R,I}=\frac12(1\pm\tilde K)\hat\psi=\frac12\big(\hat\psi\pm\exp(\alpha I)K\hat\psi\big).
\ee
(Using eq. \eqref{BB3} one may express $\tilde \psi_R$ and $\tilde \psi_I$ in terms of $\psi_R$ and $\psi_I$.)

In the complex basis \eqref{BB5} defined by $K$ the action of $\tilde K$ is expressed by the isomorphism $\hat\psi\to\varphi$ as
\be\label{B.15C}
\varphi(\tilde K\hat\psi)=\varphi^{**}=e^{i\alpha}\varphi^*=\tilde\theta\varphi.
\ee
Here $\varphi^*$ refers to the complex structure based on $K$, and $\varphi^{**}$ denotes the complex conjugation which uses $\tilde K$. The isomorphism respects the division into real and imaginary parts\footnote{Note the different prefactors in the definitions of $\tilde \psi_I$ and $\tilde\varphi_I$ in the real and complex formulations \eqref{B.15B} and \eqref{B.15D}.}
\be\label{B.15D}
\tilde\varphi_R=\varphi(\tilde\psi_R)=\frac12(\varphi+\varphi^{**})~,~i\tilde\varphi_I=\varphi(\tilde\psi_I)=
\frac12(\varphi-\varphi^{**}).
\ee
Since $\theta$ and $\tilde\theta=e^{i\alpha}\theta$ both define a valid complex conjugation, the phase $e^{i\alpha}$ is arbitrary. It needs to be fixed for defining the complex structure uniquely. 

The multiplication with a phase in eq. \eqref{B.15C} generalizes to an arbitrary complex matrix $F=F_R+iF_I$ which obeys $FF^*=1$, such that 
\be\label{B.15E}
\tilde\theta=F\theta~,~\tilde K=\hat F K,
\ee
with 
\be\label{B.15F}
\hat F=F_R+F_II=\left(\begin{array}{lr}F_R,&-F_I\\F_I,&F_R\end{array}\right).
\ee
Indeed, the form \eqref{B.15E}, \eqref{B.15F} for $\tilde K$ is the most general choice obeying $\{\tilde K,I\}=0$, and the condition $\tilde K^2=1$ results in
\be\label{B.15G}
F^2_R+F^2_I=1~,~F_RF_I=F_IF_R~,~FF^*=1.
\ee
We conclude that two complex structures which share the same $I$ are related by
\be\label{B.15GA}
\varphi^{**}=F\varphi^*~,~F^*F=1.
\ee

In a language with $2N$-component spinors $\hat\varphi=(\varphi,\varphi^*)$ we may view the general form of the complex conjugation (with fixed $I$) as a similarity transformation of the spinors
\be\label{B.15H}
\tilde\theta(\hat\varphi)={\tilde\theta\varphi\choose\tilde\theta\varphi^*}=
{\varphi^{**}\choose \tilde\theta\varphi^*}=\hat F\hat\varphi,
\ee
with matrix
\be\label{B.15I}
\hat F=\left(\begin{array}{ccc}0&,&F\\F^*&,&0\end{array}\right)~,~
\hat F^2=1.
\ee
(In this language no additional complex conjugation of coefficients is involved.) The relations
\be\label{B.15J}
\tilde\theta \varphi=F\varphi^*~,~\tilde\theta\varphi^*=F^*\varphi
\ee
reflect the involutive property
\be\label{B.15K}
(\varphi^{**})^{**}=\tilde\theta\varphi^{**}=F\tilde\theta\varphi^*=FF^*\varphi=\varphi.
\ee

This general discussion applies to our discussion of spinors in a complex Grassmann algebra. Since the meaning of multiplication with complex numbers (in particular $i$) is fixed, the use of a complex Grassmann algebra corresponds to a fixed $I$ in a general framework of ``real'' spinors. Nevertheless, the meaning of the complex conjugation $K$ needs still to be specified, corresponding to the choice of $\theta$ in the main text. We use the $2N$ component vector $\hat{\psi}$ composed from the independent Grassmann variables $\psi$ and $\ps$
\be\label{BB12}
\hp={\psi\choose\ps}.
\ee
Since multiplication with complex numbers is already defined for $\psi$ and $\ps$, it will not be represented by a matrix $I$ in this case. General linear transformations are now represented by complex $2N\times 2N$ matrices $\ha,~\hp\rightarrow\ha\hp$. However, the possibility of multiplication with complex matrices does not yet specify the meaning of the complex conjugate of $\hp$. Only the definition of an involution $\hp\rightarrow\hp^{**}$ fully defines the complex structure.

A general complex structure is defined by the choice of a matrix $F$
\be\label{BB13}
\psi^{**}=F\bar\psi~,~FF^*=1.
\ee
Here we use eq. \eqref{B.15GA}, with $\varphi$ replaced by $\psi$ and $\varphi^*$ replaced by $\bar\psi$, such that $\hat\varphi$ is replaced by $\hat\psi$. Indeed, the involution $\psi\leftrightarrow \bar\psi$ can be associated with $\varphi\leftrightarrow\varphi^*$, and eq. \eqref{BB13} follows from the observation that two complex structures sharing the same $I$ are related by eq. \eqref{B.15GA}. For a complex structure based on $\theta_M$ one has $F=\epsilon D^*$, while for $\theta$ one replaces $\epsilon D^*$ by $H^*$ and combines this with the involution $q\to \theta q$ in momentum space. For $\hat\psi$ in eq. \eqref{BB12} one has
\be\label{B.39}
\hat\psi^{**}=\hat F\hat\psi,
\ee
with $\hat F$ given by eq. \eqref{B.15I}. 

For linear transformations compatible with complex matrix multiplication one infers from eq. \eqref{B.15X6} \be\label{B.38}
(A\psi)^{**}=\theta(A\psi)=A^*\theta\psi=A^*\psi^{**}.
\ee
Among the general complex linear transformations $\hp\rightarrow\hp'=\ha\hp$ only a subclass is compatible with a given complex structure $K$. For a transformation acting on $\psi,\psi\to A\psi$, a compatible matrix $\hat A$ reads
\be\label{B41}
\hat A=\left(\begin{array}{ccc}
A&,&0\\0&,&\bar A\end{array}\right),
\ee
where $\bar A$ is given by
\be\label{B.42}
\bar A=F^{-1}A^*F=F^*A^*F.
\ee
Indeed, for a compatible transformation $\psi\to A\psi$ we require according to eqs. \eqref{BB9}, \eqref{BB10} the rule $\psi^{**}\to A^*\psi^{**}$. Inserting $\psi^{**}=F\bar\psi$ yields $\bar\psi\to \bar A\bar\psi$, where $\bar A$ obeys eq. \eqref{B.42}. Obviously, the class of compatible transformations depends on the specific choice of a complex structure $K$. They form a subgroup of the group of regular complex linear transformations. If convenient, we will use the standard symbol $\psi^*$ instead of $\psi^{**}$ for the complex conjugate spinor.

\section*{Appendix C: Properties of Clifford algebra}
\renewcommand{\theequation}{C.\arabic{equation}}
\setcounter{equation}{0}
In this appendix we collect the relations for the matrices $B,C$ and $D$ for convenience of the reader. The defining relations are
\ba\label{C.1}
\begin{array}{lll}
B_1\gamma^\mu B^{-1}_1=-\gamma^{\mu*}&,&B_2\gamma^\mu B^{-1}_2=\gamma^{\mu *},\\ 
C_1\gamma^\mu C^{-1}_1=-\gamma^{\mu T}&,&C_2\gamma^\mu C^{-1}_2=\gamma^{\mu T},\\
D_1\gamma^\mu D^{-1}_1=\gamma^{\mu\dagger}&,&D_2\gamma^\mu D^{-1}_2=-\gamma^{\mu\dagger}.
\end{array}
\ea
In even dimensions all these matrices can be realized as unitary transformations
\be\label{C.1A}
B^\dagger_iB_i=C^\dagger_iC_i=D^\dagger_iD_i=1.
\ee
We concentrate first on even $d$ and choose conventions where 
\be\label{C.2}
B_2=B_1\bar\gamma~,~C_2=C_1\bar\gamma~,~D_2=i^{s-1}D_1\bar\gamma,
\ee
and
\be\label{C.3}
C_1=B_1D_1.
\ee
The involutive properties imply
\be\label{C.4}
D^\dagger_i=\alpha_i D_i~,~C^T_i=\delta_iC_i~,~B^*_iB_i=\epsilon_i,
\ee
with 
\be\label{C.4a}
|\alpha_i|=1~,~\epsilon^2_i=1,
\ee
and
\be\label{C.5}
\delta_1=\left\{\begin{array}{rl}
1&\text{for }d=6,8 \text{ mod }8\\
-1&\text{for }d=2,4 \text{ mod }8
\end{array}\right.~,~\delta_2=(-1)^{\frac d2}\delta_1.
\ee
We choose phase conventions for $\alpha_i$ such that 
\be\label{C.6}
D^\dagger_1=D_1~,~D^\dagger_2=-D_2.
\ee
This guarantees the hermiticity of $S_{kin}$ in eq. \eqref{2.9}. The appropriate values for $\epsilon_1$ and $\epsilon_2$ will be computed below.

For even $s$ we can choose\footnote{Here we use conventions where the matrices $\gamma^0\dots \gamma^{s-1}$ are antihermitean. For $s=d-1$ it is more common to denote the hermitean matrix by $\gamma^0$, which corresponds to $\gamma^{d-1}$ in the notation of this appendix.} 
\be\label{C.7}
D_1=\tau\gamma^0\gamma^1\dots \gamma^{s-1},
\ee
while for odd $s$ we take
\be\label{C.8}
D_1=\tau\gamma^0\gamma^1\dots\gamma^{s-1}\bar\gamma.
\ee
The phase $\tau$ obeys
\ba\label{C.9}
\begin{array}{lllll}
\tau=1& \text{for}& s=0,1 \text{ mod }8,&\tau=i&\text{for }s=2,3 \text{ mod } 8,\\ 
\tau=-1&\text{for}&s=4,5\text{ mod }8,&\tau=-i&\text{for }s=6,7 \text{ mod } 8,
\end{array}
\ea
guaranteeing eq. \eqref{C.6}. For even $d$ and $s=d$ this implies, with $\eta=i^{(d/2)}$, that $D_1=\bar\gamma$.) In particular, one has $D_1=1$ for $s=0$ and $D_1=\gamma^0\bar\gamma,D_2=\gamma^0$ for $s=1$. Eqs. \eqref{C.7}, \eqref{C.8} yield for $D_1$ the relation
\be\label{C.10}
C_1D_1C^{-1}_1=\rho_1 D^T_1,
\ee
with 
\be\label{C.11}
\rho_1=\left\{\begin{array}{lll}
(-1)^{\frac s2}&\text{for}&s\text{ even}\\
(-1)^{\frac{d-s+1}{2}}&\text{for}&s\text{ odd},
\end{array}\right.
\ee
and similar for $D_2$
\be\label{C.12}
C_1D_2C^{-1}_1=\rho_2D^T_2,
\ee
with 
\be\label{C.13}
\rho_2=\left\{\begin{array}{lll}
(-1)^{\frac{d-s}{2}}&\text{for}&s\text{ even}\\
(-1)^{\frac{s+1}{2}}&\text{for}&s\text{ odd}.
\end{array}\right.
\ee

These relations can be used for a computation of $\epsilon_i$. For 
\be\label{C.14}
B_1=C_1D^\dagger_1~,~B^*_1=C^*_1D^T_1,
\ee
one obtains
\ba\label{C.15}
\epsilon_1&=&B^*_1B_1=\rho_1C^*_1C_1D_1C^{-1}_1C_1D^\dagger_1\nn\\
&=&\rho_1C^*_1C_1=\rho_1\delta_1C^\dagger_1C_1=\rho_1\delta_1,
\ea
and therefore for even $d$
\be\label{C.16}
\epsilon_1=\left\{\begin{array}{lll}
(-1)^{\frac s2}\delta_1&\text{for}&s\text{ even}\\
(-1)^{\frac{d-s+1}{2}}\delta_1&\text{for}&s\text{ odd}.
\end{array}\right.
\ee
Employing $D=D_2$ one finds
\be\label{C.17}
B_2=B_1\bar\gamma=(-i)^{s-1}C_1D_2=-(-i)^{s-1}C_1D^\dagger_2,
\ee
such that 
\be\label{C.18}
\epsilon_2=B^*_2B_2=\rho_2\delta_1
\ee
is related to $\epsilon_1$ according to \cite{W1}
\be\label{C.19}
\epsilon_2=(-1)^{\frac d2-s}\epsilon_1.
\ee
In terms of $\delta_1$ and $\delta_2$ this yields
\be\label{C.20}
\begin{array}{lll}
\epsilon_i=(-1)^{\frac s2}\delta_i&\text{for}&s\text{ even}\\
\epsilon_1=(-1)^{\frac{s-1}{2}}\delta_2&\text{for}&s\text{ odd}\\
\epsilon_2=(-1)^{\frac{s+1}{2}}\delta_1&\text{for}&s\text{ odd}.
\end{array}
\ee
In particular, for $s=1$ one obtains
\be\label{C.21}
\epsilon_1=\delta_2~,~\epsilon_2=-\delta_1,
\ee
while for $s=0$ one has
\be\label{C.22}
\epsilon_i=\delta_i.
\ee
The different properties of $\epsilon_1$ and $\epsilon_2$ for $s$ even or odd arise from the presence of $\bar\gamma$ in the definition \eqref{C.8} for $D_1$ for $s$ odd. They are a manifestation of the modulo two periodicity in the signature. We also note the exchange of $\epsilon_1$ and $\epsilon_2$ for $s\to d-s$. 

Finally, the transpose of $\bar\gamma$ in eq. \eqref{2.4}  obeys for both $C_1$ and $C_2$
\be\label{C.23}
\bar\gamma^T=(-1)^{\frac d2}C\bar\gamma C^{-1}.
\ee
For the complex conjugate one finds the relation
\ba\label{C.24}
\bar\gamma^*&=&\bar\gamma^T=(-1)^{\frac d2}B_1D_1\bar\gamma D^{-1}_1B^{-1}_1\nn\\
&=&(-1)^{\frac d2-s}B_1\bar\gamma B^{-1}_1=(-1)^{\frac d2-s}B_2\bar\gamma B^{-1}_2,
\ea
such that for both $B_1$ and $B_2$ one finds
\be\label{C.25}
\bar\gamma^*=(-1)^{\frac d2-s}B\bar\gamma B^{-1}.
\ee

The Clifford algebra for odd $d$ obtains from the $d-1$-dimensional Clifford algebra with the same $s$ by adding $\gamma^{d-1}=\bar\gamma_{d-1}$, where $\bar\gamma$ is given by eq. \eqref{2.4} in $d-1$ dimensions. The matrices $B$ and $C$ for odd dimensions can therefore be inferred from the $d-1$ dimensional case. However, only one of the matrices $C_1$ or $C_2$ is compatible with the relations \eqref{C.23}, 
\be\label{C.26}
C\bar\gamma_{d-1}C^{-1}=(-1)^{\frac{d-1}{2}}\bar\gamma^T_{d-1}.
\ee
For $d=3,7$ mod $8$ we have to choose $C_1$, while for $d=5,9$ mod $8$ only $C_2$ is consistent with the definitions \eqref{C.1}. Similarly, the relation \eqref{C.25} restricts the choice of $B$ to
\be\label{C.27}
B=\left\{\begin{array}{lllll}
B_1&\text{for}&d-2s&=&3\text{ mod }4\\
B_2&\text{for}&d-2s&=&5\text{ mod }4.
\end{array}\right.
\ee
The corresponding relations for $\delta_i$ and $\epsilon_i$ are transported to odd dimensions from the values of the $d-1$-dimensional $B$ and $C$ matrices.

An overview of the values for $\delta_i$ and $\epsilon_i$ for $s=0,1$ is given in Table 1.

\bigskip
\begin{tabular}{|c|c|c|c|c|c|c|}
\hline 
$d$&$\delta_1$&$\delta_2$&$\epsilon_1(s=0)$&$\epsilon_2(s=0)$&$\epsilon_1(s=1)$&$\epsilon_2(s=1)$\\\hline 
2 mod 8&-1&+1&-1&+1&+1&+1\\
3 mod 8&-1&-&-1&-&-&+1\\
4 mod 8&-1&-1&-1&-1&-1&+1\\
5 mod 8&-&-1&-&-1&-1&-\\
6 mod 8&+1&-1&+1&-1&-1&-1\\
7 mod 8&+1&-&+1&-&-&-1\\
8 mod 8&+1&+1&+1&+1&+1&-1\\
9 mod 8&-&+1&-&+1&+1&-\\\hline
\end{tabular}

\medskip
Table 1: values of $\delta_i$ and $\epsilon_i$ for $s=0$ and $s=1$.

\end{document}